\documentclass{article}
\usepackage{amssymb,amsmath}
\usepackage[dvipdfmx]{graphicx}
\usepackage{bbm}
\usepackage{bm}
\usepackage{caption}
\usepackage{color}
\usepackage{comment}
\usepackage{comment}
\usepackage{enumerate}
\usepackage{here}
\usepackage{subfig}
\DeclareMathOperator*{\Tr}{{\rm Tr}}

\numberwithin{equation}{section}

\begin{document}

\thispagestyle{empty}
\begin{flushright}
Imperial/TP/18/AH/11


\end{flushright}
\vskip2cm
\begin{center}
{\Large \bf $(0,4)$ brane box models
}

\vskip1.5cm
Amihay Hanany\footnote{a.hanany@imperial.ac.uk}

\bigskip
{\it 
Theoretical Physics, The Blackett Laboratory, Imperial College London,\\
Prince Consort Road, London, SW7 2AZ United Kingdom}
\\
\bigskip
and 
\\
\bigskip
Tadashi Okazaki\footnote{tokazaki@perimeterinstitute.ca}

\bigskip
{\it Perimeter Institute for Theoretical Physics,\\
Waterloo, Ontario, N2L 2Y5 Canada}

\end{center}

\vskip2cm
\begin{abstract}
Two-dimensional $\mathcal{N}=(0,4)$ supersymmetric quiver gauge theories 
are realized as D3-brane box configurations (two dimensional intervals) which are bounded by NS5-branes and intersect with D5-branes.
The periodic brane configuration is mapped to D1-D5-D5$'$ brane system at orbifold singularity via T-duality. 
The matter content and interactions are encoded by the $\mathcal{N}=(0,4)$ quiver diagrams 
which are determined by the brane configurations. 
The Abelian gauge anomaly cancellation indicates the presence of Fermi multiplets at the NS-NS$'$ junction. 
We also discuss the brane construction of 
$\mathcal{N}=(0,4)$ supersymmetric boundary conditions in 3d $\mathcal{N}=4$ gauge theories 
involving two-dimensional boundary degrees of freedom that cancel gauge anomaly. 
\end{abstract}


\newpage
\setcounter{tocdepth}{2}
\tableofcontents

\section{Introduction}
\label{sec_intro}
$\mathcal{N}=(0,4)$ supersymmetric field theories are less 
understood due to the difficulty with their construction and quantization, 
however, they involve intriguing aspects and applications. 

$\mathcal{N}=(0,4)$ supersymmetric sigma model 
has the Yukawa couplings which can obey the ADHM equations of the instanton construction 
under certain assumptions \cite{Witten:1994tz}. 
This indicates that there exists a certain $\mathcal{N}=(0,4)$ sigma model for every instanton. 
Since then, there have been a lot of studies of $\mathcal{N}=(0,4)$ supersymmetric field theories. 
The classical aspects of $\mathcal{N}=(0,4)$ supermultiplets 
and off-shell formalism were studied in \cite{Gates:1994bu, Galperin:1994qn, Galperin:1995pq} 
and quantum properties were studied in \cite{Lambert:1995dp}. 
The elliptic genera or superconformal indices of $\mathcal{N}=(0,4)$ gauge theories 
have been computed in \cite{Putrov:2015jpa, Gadde:2015tra}.

$\mathcal{N}=(0,4)$ superconformal field theory also can play a significant role in string and M-theory 
to describe the dynamics of intersecting brane configurations 
and many features of the holographic dual supergravity solutions. 
One relevant example is the D1-D5-KK system which is a $1/8$ BPS configuration in Type IIB string theory. 
It is dual to the triple intersection of M5-branes \cite{Maldacena:1997de, Minasian:1999qn}. 
The dual $\mathcal{N}=(0,4)$ SCFT is studied in \cite{Sugawara:1999qp, Okuyama:2005gq}. 
Another relevant example is the D1-D5-D5$'$ system 
whose near horizon geometry is 
the geometry $AdS_{3}$ $\times$ $S^{3}$ $\times$ $S^{3}$ $\times$ $\mathbb{R}$ 
\cite{Cowdall:1998bu, Boonstra:1998yu, Gauntlett:1998kc}. 
The $\mathcal{N}=(0,4)$ gaugel theory which lives on the 
common world-volume of D1-branes and intersecting D5-branes is studied in \cite{Tong:2014yna}.

In this paper, we present a novel construction of $\mathcal{N}=(0,4)$ supersymmetric quiver gauge theories 
from brane box configuration of D3-branes which intersect with NS5-, D5-, NS5$'$- and D5$'$-branes. 
$\mathcal{N}=(0,4)$ supersymmetry can be preserved by half BPS boundary conditions 
in 3d $\mathcal{N}=4$ supersymmetric field theory \cite{Chung:2016pgt} 
and these boundary conditions can be realized in brane setup by starting with Hanany-Witten setup \cite{Hanany:1996ie} 
and introducing additional NS5$'$ and D5$'$-branes on which half-infinite D3-branes can end. 
We promote this setup to define $\mathcal{N}=(0,4)$ supersymmetric gauge theory 
by displacing D3-branes between NS5- and NS5$'$-branes. 
We call this brane configuration the D3-brane box model. 
The periodic D3-brane box model turns out to be T-dual to the D1-D5-D5$'$ brane system probing orbifold singularity, 
which is a generalization of the D1-D5-KK system and D1-D5-D5$'$ system.

In section \ref{sec_04theory} 
we start with reviewing $\mathcal{N}=(0,4)$ supersymmetry in two dimensions. 
We make use of $\mathcal{N}=(0,2)$ notation to formulate $\mathcal{N}=(0,4)$ theories. 
In section \ref{sec_04bcbrane} 
we realize $\mathcal{N}=(0,4)$ supersymmetric boundary conditions 
for 3d $\mathcal{N}=4$ gauge theories by adding extra 5-branes in Hanany-Witten construction 
and promoting to brane box configuration in Type IIB string theory. 
In section \ref{sec_d1d5kk} 
we study the D1-D5-D5$'$-KK-KK$'$ system 
which is T-dual of the brane box configuration. 
We determine the spectrum and interaction by using the techniques developed in \cite{Douglas:1996sw}. 
We propose $\mathcal{N}=(0,4)$ quiver which can be read from the brane box configuration 
and the D1-D5-D5$'$-KK-KK$'$ system. 
In section \ref{sec_box} 
we analyze the anomaly in brane box model. 
We discuss that 
the cancellation of the Abelian anomaly in brane box model 
requires the existence of tetravalent Fermi multiplet charged under the Abelian parts 
for quadrants of D3-branes which lives at the NS5-NS5$'$ junction. 
Furthermore, we argue the brane construction of 
$\mathcal{N}=(0,4)$ supersymmetric boundary conditions 
in 3d $\mathcal{N}=4$ supersymmetric gauge theory 
involving two-dimensional boundary degrees of freedom 
which cancel gauge anomaly.

\section{$\mathcal{N}=(0,4)$ supersymmetric theories}
\label{sec_04theory}

\subsection{$\mathcal{N}=(0,2)$ superspace}
\label{sec_02sp}
There are four types of supermultiplets for $\mathcal{N}=(0,4)$ supersymmetry in two dimensions. 
Let us start by introducing the notation and convention in $\mathcal{N}=(0,2)$ superspace 
\footnote{
The superspace formalism for $\mathcal{N}=(0,4)$ supersymmetry is given in \cite{Galperin:1994qn, Galperin:1995pq}. 
}. 
The $\mathcal{N}=(0,2)$ superspace is parametrized 
by bosonic coordinates $x^{\pm}$ and fermionic coordinates $\{ \theta^{+}, \overline{\theta}^{+}\}$ 
where the spinor conventions are as in \cite{Wess:1992cp, Witten:1993yc, Adams:2003zy}.  
The supercharges $Q_{+}$ and $\overline{Q}_{+}=Q_{+}^{\dag}$ of chiral $\mathcal{N}=(0,2)$ supersymmetry in the superspace is given by 
\begin{align}
\label{02_sch1}
Q_{+}&=\frac{\partial}{\partial \theta^{+}}+i\overline{\theta}^{+}\partial_{+},& 
\overline{Q}_{+}&=-\frac{\partial}{\partial \overline{\theta}^{+}}-i\theta^{+}\partial_{+}. 
\end{align}
They satisfy 
\begin{align}
\label{02_alg}
\left\{Q_{+}, Q_{+}\right\}&=\left\{\overline{Q}_{+}, \overline{Q}_{+}\right\}=0,& 
\left\{Q_{+},\overline{Q}_{+}\right\}&=-2i\partial_{+}. 
\end{align}
The superderivatives are 
\begin{align}
\label{02_sderiv1}
D_{+}&=\frac{\partial}{\partial \theta^{+}}-i\overline{\theta}^{+}\partial_{+},& 
\overline{D}_{+}&=-\frac{\partial}{\partial \overline{\theta}^{+}}+i\theta^{+}\partial_{+}
\end{align}
which obey 
\begin{align}
\label{02_sderiv2}
\left\{D_{+}, D_{+}\right\}&=\left\{\overline{D}_{+}, \overline{D}_{+}\right\}=0,& 
\left\{D_{+}, \overline{D}_{+}\right\}&=2i\partial_{+}
\end{align}
and anticommute with the supercharges
\begin{align}
\label{02_sderiv3}
\left\{D_{+}, Q_{+}\right\}&=0,& 
\left\{D_{+}, \overline{Q}_{+}\right\}&=0,& 
\left\{\overline{D}_{+}, Q_{+}\right\}&=0,& 
\left\{\overline{D}_{+}, \overline{Q}_{+}\right\}&=0. 
\end{align}
Note that superfields are functions of $(x^{\pm}, \theta^{+}, \overline{\theta}^{+})$ with constraints in terms of $\overline{D}_{+}$, 
which annihilates combinations of $y^{+}=x^{+}-i\theta^{+}\overline{\theta}^{+}$, $y^{-}=x^{-}$ and $\theta^{+}$.

\paragraph{$\mathcal{N}=(0,2)$ gauge multiplets}
For simplicity, we focus on Abelian gauge theory. 
The $\mathcal{N}=(0,2)$ gauge multiplet consists of a real  adjoint valued superfield $\mathcal{A}_{+}$, 
whose lowest component is the right-moving component of gauge field, and $\mathcal{A}_{-}$ whose lowest component is the left-moving component of gauge field. 
A supergauge transformation is 
\begin{align}
\label{sgauge1}
\mathcal{A}_{+}&\rightarrow \mathcal{A}_{+}-i\left(\Lambda-\overline{\Lambda}\right),& 
\mathcal{A}_{-}&\rightarrow \mathcal{A}_{-}-i\left(\Lambda+\overline{\Lambda}\right)
\end{align}
where $\Lambda$ is a chiral superfield $\overline{D}_{+}\Lambda=\overline{D}_{+}\overline{\Lambda}=0$. 
In the Wess-Zumino gauge, the real superfields $\mathcal{A}_{+}$ and $\mathcal{A}_{-}$ have the component expansions
\begin{align}
\label{02_gmult1}
\mathcal{A}_{+}&=\theta^{+}\overline{\theta}^{+}\left(A_{0}+A_{1}\right),& 
\mathcal{A}_{-}&=\left(A_{0}-A_{1}\right)-2i\theta^{+}\overline{\lambda}_{-}-2i\overline{\theta}^{+}\lambda_{-}+2\theta^{+}\overline{\theta}^{+}D
\end{align}
Here the left-moving component $A_{-}=A_{0}-A_{1}$ of the gauge field has two real left-moving fermionic partners 
$\lambda_{-}$ and $\overline{\lambda}_{-}$ while the right-moving one $A_{+}=A_{0}+A_{1}$ has none. 
$D$ is a real auxiliary field. 

In gauge theories, the superspace derivatives $D_{+}$ and $\overline{D}_{+}$ 
are extended to gauge covariant superderivatives 
$\mathcal{D}_{+}=e^{-\mathcal{A}_{+}}D_{+}e^{\mathcal{A}_{+}}$ 
and $\overline{\mathcal{D}}_{+}=e^{\overline{\mathcal{A}}_{+}}\overline{D}_{+}e^{-\overline{\mathcal{A}}_{+}}$. 
In the Wess-Zumino gauge they are expressed as
\begin{align}
\label{02_gcov1}
\mathcal{D}_{0}+\mathcal{D}_{1}&=\partial_{0}+\partial_{1}+iA_{+},& 
\mathcal{D}_{0}-\mathcal{D}_{1}&=\partial_{0}-\partial_{1}+i\mathcal{A}_{-}, \\
\label{02_gcov2}
\mathcal{D}_{+}&=\frac{\partial}{\partial \theta^{+}}
-i\overline{\theta}^{+}\left(\mathcal{D}_{0}+\mathcal{D}_{1}\right),& 
\overline{\mathcal{D}}_{+}&=-\frac{\partial}{\partial \overline{\theta}^{+}}
+i\theta^{+}\left(\mathcal{D}_{0}+\mathcal{D}_{1}\right). 
\end{align}

The field strength is given by the uncharged Fermi multiplet
\begin{align}
\label{02_fstrengh}
\Upsilon&=\left[\overline{\mathcal{D}}_{+},\mathcal{D}_{0}-\mathcal{D}_{1}\right]
=\overline{D}_{+}\left(\partial_{-}\mathcal{A}_{+}+i\mathcal{A}_{-}\right)\nonumber\\
&=-2\left[
\lambda_{-}(y)-i\theta^{+}\left(D-iF_{01}\right)
\right]\nonumber\\
&=-2\left[
\lambda_{-}(x)-i\theta^{+}\left(D-iF_{01}\right)-i\theta^{+}\overline{\theta}^{+}
\left(
\mathcal{D}_{0}+\mathcal{D}_{1}
\right)\lambda_{-}
\right]
\end{align}
where $F_{01}=\partial_{0}A_{1}-\partial_{1}A_{0}$. 
The kinetic terms for the $\mathcal{N}=(0,2)$ gauge multiplets are given 
by integration over all of superspace $d^{2}\theta=d\theta^{+}d\overline{\theta}^{+}$
\begin{align}
\label{02_gauge_A}
S_{\mathrm{gauge}}^{(0,2)}&=\frac{1}{8e^{2}}
\int d^{2}x d^{2}\theta 
\overline{\Upsilon}\Upsilon\nonumber\\
&=\frac{1}{e^{2}}
\int d^{2}x \left[
\frac12 F_{01}^{2}+i\overline{\lambda}_{-}\left(\partial_{0}+\partial_{1}\right)\lambda_{-}+\frac12 D^{2}
\right]. 
\end{align}

\subsubsection{$\mathcal{N}=(0,2)$ chiral multiplets}
The $\mathcal{N}=(0,2)$ chiral superfield $\Phi$ satisfies the chirality constraint 
\begin{align}
\label{02_ch1}
\overline{D}_{+}\Phi&=0. 
\end{align}
It is expanded in the (super)coordinates $y^{+}=x^{+}-i\theta^{+}\overline{\theta}^{+}$, $y^{-}=x^{-}$ and $\theta^{+}$ as
\begin{align}
\label{02_ch2}
\Phi&=\phi(y)+\sqrt{2}\theta^{+}\psi_{+}(y)\nonumber\\
&=\phi(x)+\sqrt{2}\theta^{+}\psi_{+}(x)-i\theta^{+}\overline{\theta}^{+}\partial_{+}\phi(x)
\end{align} 
where $\phi$ is complex scalar and $\psi_{+}$ is its right-moving fermionic partner. 
The kinetic terms for the $\mathcal{N}=(0,2)$ chiral superfield are given by
\begin{align}
\label{02_ch_A1}
S_{\mathrm{chiral}}^{(0,2)}&=-\frac{i}{2}\int d^{2}x d^{2}\theta 
\overline{\Phi}\partial_{-}\Phi. 
\end{align}
For a field $\Phi^{(Q)}$ with $U(1)$ charge $Q$, the covariant chirality constraint $\overline{\mathcal{D}}_{+}\Phi^{(Q)}=0$ 
can be solved by $\Phi=e^{-Q\mathcal{A}}\Phi^{(Q)}$. 
In components we have 
\begin{align}
\label{02_ch2a}
\Phi&=\phi(x)+\sqrt{2}\theta^{+}\psi_{+}(x)-i\theta^{+}\overline{\theta}^{+}(D_{0}+D_{1})\phi(x)
\end{align}
where $D_{\alpha}=\partial_{\alpha}+iQu_{\alpha}$. 
The kinetic terms for the $\mathcal{N}=(0,2)$ charged chiral superfield are given by
\begin{align}
\label{02_ch_A2}
S_{\mathrm{chiral}}^{(0,2)}
&=-\frac{i}{2}\int d^{2}x d^{2}\theta \overline{\Phi}^{(Q)}\left(\mathcal{D}_{0}-\mathcal{D}_{1}\right)\Phi^{(Q)}
\nonumber\\
&=\int d^{2}x \left[
-|D_{\alpha}\phi|^{2}+i\overline{\psi}_{+}\left(D_{0}-D_{1}\right)\psi_{+}
-iQ\sqrt{2}\overline{\phi}\lambda_{-}\psi_{+}+iQ\sqrt{2}\phi\overline{\psi}_{+}\overline{\lambda}_{-}+QD|\phi|^{2}
\right]. 
\end{align}

\subsubsection{$\mathcal{N}=(0,2)$ Fermi multiplets}
The $\mathcal{N}=(0,2)$ Fermi multiplets satisfy the conditions 
\begin{align}
\label{02_f1}
\overline{D}_{+}\Gamma&=\sqrt{2}E,& \overline{D}_{+}E&=0. 
\end{align}
Here $E$ determines the potential term and 
it can be solved by assuming that $E$ is a holomorphic function of chiral superfields $\Phi$ \cite{Witten:1993yc}.  
The $\mathcal{N}=(0,2)$ Fermi multiplet is expanded in the coordinates as
\begin{align}
\label{02_f2}
\Gamma&=\chi_{-}(x)-\sqrt{2}\theta^{+}G(x)-i\theta^{+}\overline{\theta}^{+}(D_{0}+D_{1})\chi_{-}(x)-\sqrt{2}\overline{\theta}^{+}E
\end{align}
where $\chi_{-}$ is a left-moving fermion and $G$ is a complex auxiliary field. 
In general $E$ will have an expansion 
\begin{align}
\label{02_f3}
E(\Phi)&=E(\phi_{i})+\sqrt{2}\theta^{+}\frac{\partial E}{\partial \phi^{i}}\psi_{+i}-i\theta^{+}\overline{\theta}^{+}(D_{0}+D_{1})E(\phi_{i}). 
\end{align}
The $\mathcal{N}=(0,2)$ Fermi multiplets may also transform as some representation $R$ of the gauge group. 
For a field $\Gamma^{(Q)}$ with $U(1)$ charge $Q$, 
the covariant constraints $\overline{\mathcal{D}}_{+}\Gamma^{(Q)}=E$ and $\overline{\mathcal{D}}_{+}E^{(Q)}=0$ can be solved 
by taking $\Gamma=e^{-Q\mathcal{A}}\Gamma^{(Q)}$ and $E=e^{-Q\mathcal{A}}E^{(Q)}$. 
The kinetic terms for the $\mathcal{N}=(0,2)$ Fermi multiplet are given by 
\begin{align}
\label{02_f_A1}
S_{\mathrm{ferm}}^{(0,2)}
&=-\frac12 \int d^{2}x d^{2}\theta \overline{\Gamma}\Gamma\nonumber\\
&=\int d^{2}x
\left[ i\overline{\chi}_{-}(D_{0}+D_{1})\chi_{-}+|G|^{2}-|E^{(Q)}(\phi)|^{2}-\overline{\chi}_{-}\frac{\partial E^{(Q)}(\phi)}{\partial \phi^{i}}\psi_{+i}
+\overline{\psi}_{+i}\frac{\partial \overline{E}^{(Q)}(\overline{\phi})}{\partial \overline{\phi}^{i}}\chi_{-}
\right]. 
\end{align}
We see that the holomorphic function $E^{(Q)}(\phi)$ appears as a potential term for chiral multiplet in (\ref{02_f_A1}), 
which we call an $E$-term potential. 
By definition (\ref{02_f1}),  
$E$-term transforms in the same way as the Fermi multiplet $\Gamma$.

%
%
%
\subsubsection{$\mathcal{N}=(0,2)$ superpotential}
Let $J^{a}(\Phi)$ be a superpotential which is a holomorphic function of chiral superfields $\Phi$ 
for a set of Fermi multiplets $\{\Gamma^{a}\}$. 
Then a supersymmetric action can be also constructed by integrating over half of superspace as
\begin{align}
\label{02_spot1}
S_{J}^{(0,2)}&=-\frac{1}{\sqrt{2}}\int d^{2}x d\theta^{+} \sum_{a}\Gamma_{a}J^{a}(\Phi)\Bigl|_{\overline{\theta}^{+}=0}-\mathrm{h.c.}\nonumber\\
&=-\sum_{a}\int d^{2}x \left[G_{a}J^{a}(\phi)+\sum_{i}\chi_{-a}\frac{\partial J^{a}}{\partial \phi_{i}}\psi_{+i}\right] - \mathrm{h.c.}
\end{align}
By integrating out the auxiliary field $G_{a}$, one obtains a potential term $\sim |J^{a}(\phi)|^{2}$. 
We shall call this a $J$-term potential. 
It follows from gauge invariance of $\Gamma_{a}J^{a}$ that 
$Q_{\Gamma_{a}}=-Q_{J^{a}}$. 
Thus $J$-term transforms in the conjugate representation, namely as $\widetilde{\Gamma}$. 
The bosonic potential terms specified by holomorphic functions $E(\phi)_{a}$ and $J^{a}(\phi)$ 
are associated to the Fermi multiplet $\Gamma_{a}$. 
It is important to note that in $\mathcal{N}=(0,2)$ theories, there is a symmetry 
between Fermi multiplet $\Gamma$ and its conjugate $\widetilde{\Gamma}$ 
under an exchange of $E$- and $J$-terms.

Since $\mathcal{N}=(0,2)$ Fermi multiplet $\Gamma_{a}$ is not a genuine chiral superfield 
obeying $\overline{\mathcal{D}}_{+}\Gamma=\sqrt{2}E$, 
one needs to impose the condition 
\begin{align}
\label{02_constraint}
E\cdot J&=\sum_{a}E_{a}J^{a}=0
\end{align}
to ensure that the $J$-term potential $\Gamma_{a}J^{a}(\Phi)$ is chiral, 
i.e. $\overline{\mathcal{D}}_{+}(\Gamma_{a}J^{a})=0$.

It is important to note that 
as 3d $\mathcal{N}=2$ supersymmetric theories with a superpotential $W_{3d}(\Phi)$ admits 
$\mathcal{N}=(0,2)$ supersymmetric boundary conditions with $W(\Phi)$ being constant \cite{Okazaki:2013kaa}, 
the condition (\ref{02_constraint}) can be relaxed so that 
\begin{align}
\label{factor_W}
E\cdot J=W_{3d}(\Phi)
\end{align}
if $\mathcal{N}=(0,2)$ theories live on a boundary of 3d $\mathcal{N}=2$ theories \cite{Gadde:2013sca, Dimofte:2017tpi}. 
This is a 3d analogue of the Warner problem \cite{Warner:1995ay} so that 
(\ref{factor_W}) exhibits holomorphic factorization of 3d bulk superpotential. 

One simple example of the superpotential is an FI and $\theta$ term 
\begin{align}
\label{02_FI}
S_{\mathrm{FI}}&=\frac{t}{4}\mathrm{Tr}\int d^{2}x d\theta^{+} \Upsilon\Bigl|_{\overline{\theta}^{+}=0}
 + \mathrm{h.c.}
\nonumber\\
&=\mathrm{Tr}
\int d^{2}x \left[
-rD+\frac{\theta}{2\pi}F_{01}
\right]
\end{align}
where $t=ir+\frac{\theta}{2\pi}$ is a complex combination of a FI parameter $r$ and $\theta$ angle. 

In total, there are three contributions to potential energy, i.e. $D$-, $E$- and $J$-terms. 
$\mathcal{N}=(0,2)$ gauge theories are expected to flow in the IR 
to the non-linear sigma model whose target space is determined 
by the vanishing $D$-, $E$- and $J$-terms.

\subsection{$\mathcal{N}=(0,4)$ supersymmetric theories}
\label{sssec_04}
Let us discuss the $\mathcal{N}=(0,4)$ supersymmetric gauge theories. 
There are four kinds of supermultiplets; 
vector multiplets, hypermultiplets, twisted hypermultiplets and Fermi multiplets. 
We construct them by using $\mathcal{N}=(0,2)$ supermultiplets 
which have an enhanced $SO(4)\cong SU(2)_{C}\times SU(2)_{H}$ R-symmetry of $\mathcal{N}=(0,4)$ supersymmetry.

\subsubsection{$\mathcal{N}=(0,4)$ vector multiplets}
The $\mathcal{N}=(0,4)$ vector multiplet $V$ consist of an $\mathcal{N}=(0,2)$ vector multiplet $\Upsilon$ 
and an adjoint-valued $\mathcal{N}=(0,2)$ Fermi multiplet $\Gamma$
\begin{align}
\label{04_v}
V&=(\Upsilon,\Gamma). 
\end{align}
In components it contains 
a gauge field, 
a pair of left-moving complex fermions $\lambda_{-}^{A\tilde{A}}$ 
which transform as $(\bm{2},\bm{2})_{-}$ under the $SU(2)_{C}\times SU(2)_{H}$ R-symmetry 
and auxiliary fields transforming as $(\bm{1}, \bm{3})$. 
The $\mathcal{N}=(0,4)$ vector multiplet does not contain scalar fields, hence there is no Coulomb branch. 
The adjoint-valued $\mathcal{N}=(0,2)$ Fermi multiplet $\Gamma$ obeys 
\begin{align}
\label{04_v_f1}
\overline{\mathcal{D}}_{+}\Gamma&=E_{\Gamma}
\end{align}
where $E_{\Gamma}$ is a holomorphic function of chiral superfields 
which transforms in the adjoint representation of the gauge group. 
It can be expressed as (\ref{04_v_f2}) in terms of chiral fields of $\mathcal{N}=(0,4)$ twisted hypermultiplets.

%
%
%
\subsubsection{$\mathcal{N}=(0,4)$ hypermultiplets}
The $\mathcal{N}=(0,4)$ hypermultiplet consists of 
a pair of $\mathcal{N}=(0,2)$ chiral multiplets $L$ and $R$ which transform 
in conjugate representations of the gauge group
\begin{align}
\label{04_hm}
H&=(R,L). 
\end{align}
Under the R-symmetry $SU(2)_{C}\times SU(2)_{H}$ the chiral field and the anti chiral field transform in the $(\bm{1}, \bm{2})$ 
and the right-moving fermions transform as $(\bm{2},\bm{1})_{+}$. 
The kinetic terms for $\mathcal{N}=(0,4)$ hypermultiplet are 
a sum of those for $\mathcal{N}=(0,2)$ chiral multiplets given in (\ref{02_ch_A2}). 
In addition, $\mathcal{N}=(0,4)$ hypermultiplet can couple to $\mathcal{N}=(0,4)$ vector multiplet 
through the $\mathcal{N}=(0,2)$ Fermi multiplet $\Gamma$ as a $J$-type potential \cite{Tong:2014yna}
\begin{align}
\label{04h_coupling}
J^{\Gamma}&=RL. 
\end{align}

%
%
%
\subsubsection{$\mathcal{N}=(0,4)$ twisted hypermultiplets}
\label{sec_04thm}
In a similar fashion as in $\mathcal{N}=(0,4)$ hypermultiplet, 
$\mathcal{N}=(0,4)$ twisted hypermultiplet consists of a pair of 
$\mathcal{N}=(0,2)$ chiral multiplets $U$ and $D$ transforming 
in conjugate representations under the gauge group
\begin{align}
\label{04_hm}
T&=(U,D). 
\end{align}
In contrast to the hypermultiplet, 
a pair of complex scalars transform as $(\bm{2},\bm{1})$ 
and the right-moving fermions transform as $(\bm{1},\bm{2})_{+}$ under the R-symmetry $SU(2)_{C}\times SU(2)_{H}$. 
Again the kinetic terms for $\mathcal{N}=(0,4)$ twisted hypermultiplet are given by a sum of 
(\ref{02_ch_A2}) for both $U$ and $D$. 
The $\mathcal{N}=(0,4)$ twisted hypermultiplet can couple to a vector multiplet via the $E$-term potential 
(\ref{04_v_f2}).

It is important to note that 
the condition (\ref{02_constraint}) cannot be satisfied 
if a single $\mathcal{N}=(0,4)$ vector multiplet is coupled to 
both $\mathcal{N}=(0,4)$ hypermultiplet and $\mathcal{N}=(0,4)$ twisted hypermultiplet 
since $E_{\Gamma}J^{\Gamma}=\Tr UDRL\neq 0$. 
This can be solved by introducing $\mathcal{N}=(0,4)$ Fermi multiplets which 
couple to both the $\mathcal{N}=(0,4)$ hypermultiplet and the twisted hypermultiplet. 

Since the $\mathcal{N}=(4,4)$ vector multiplet decomposes into $\mathcal{N}=(0,4)$ vector multiplet 
and adjoint valued $\mathcal{N}=(0,4)$ twisted hypermultiplet, 
$E_{\Gamma}$ in (\ref{04_v_f1}) can be expressed as a holomorphic function of adjoint valued chiral multiplets $U$ and $D$ 
which constitute the $\mathcal{N}=(0,4)$ twisted hypermultiplet \cite{Tong:2014yna}: 
\begin{align}
\label{04_v_f2}
E_{\Gamma}&=UD. 
\end{align}

\subsubsection{$\mathcal{N}=(0,4)$ Fermi multiplets}
\label{sec_04fm}
%
The $\mathcal{N}=(0,4)$ Fermi multiplet $\Xi$ consists of a pair of $\mathcal{N}=(0,2)$ Fermi multiplets $\Gamma$ and $\Gamma'$
\begin{align}
\label{04_fm}
\Xi&=(\Gamma,\Gamma'). 
\end{align}
A pair of left-moving fermions $\xi_{-}$ and $\tilde{\xi}$ 
in the multiplet transform as $(\bm{1},\bm{1})_{-}$ under the $SU(2)_{C}\times SU(2)_{H}$ R-symmetry. 

Although they are trivial under the R-symmetry, 
they play a key role in defining consistent $\mathcal{N}=(0,4)$ gauge theories. 
As discussed in \ref{sec_04thm}, when both $\mathcal{N}=(0,4)$ hyper and twisted hypermultiplets couple to a gauge field, 
it is required to introduce neutral Fermi multiplet so that the supersymmetric condition (\ref{02_constraint}) is satisfied. 
The simplest situation would be the case with 
one hyper multiplet and one twisted hyper multiplet. 
In addition, as we discuss in section \ref{sec_04anomaly}, 
when there are enough hyper and twisted hypermultiplets in the gauge theory, 
the charged Fermi multiplets are required to cancel a gauge anomaly.

\subsubsection{Anomaly}
\label{sec_04anomaly}
$\mathcal{N}=(0,4)$ supersymmetric gauge theory can be anomalous  
because left- and right-moving fermions are not necessarily paired together.

Let $G$ be a simple compact group of which a system of right- and left-handed chiral fermions 
transform under a unitary representation $R$ 
through coupling to a (background) gauge field whose field strength is ${\bf f}$. 
We define the quadratic index $C(R)$ of $R$ as 
a sum of length-squared of weights $\lambda$
\begin{align}
\label{quadratic_index}
C(R)&=\frac{1}{\mathrm{rank} G}\sum_{\lambda\in R}
\|\lambda \|^{2}
\end{align}
where $\|\alpha \|^{2}=2$ for long roots $\alpha$. 
This is normalized so that $C(\textrm{adjoint})=2h$ 
where $h$ is the dual Coxeter number.

The contribution to anomaly the 4-form is summarized as follows:
\begin{align}
\label{t_Anom}
\begin{array}{c|c|c}
\textrm{2d $\mathcal{N}=(0,2)$ multiplet}&\textrm{$R$}&{\bf f}^2 \\  \hline 
\textrm{chiral $\Phi$}&\textrm{$\Box$ or $\overline{\Box}$} &-C(R) \\
& \textrm{adjoint}& -2h\\ \hline
\textrm{Fermi $\Gamma$}&\textrm{$\Box$ or $\overline{\Box}$}& C(R) \\ 
&\textrm{adjoint}& 2h\\ \hline
\textrm{gauge $\Upsilon$}&\textrm{adjoint}& 2h\\
\end{array}
\end{align}
The left- and right-moving fermions have the opposite contributions to the anomaly 
whereas the fundamental and anti-fundamental representations have the same contributions. 
For $SU(N)$ we have $C(\square)$ $=$ $1$ 
and $C(\textrm{adjoint})=2h=2N$ 
and the anomaly contributions are summarized as 
\begin{align}
\label{t_Anom2a}
\begin{array}{c|c|c}
\textrm{2d $\mathcal{N}=(0,2)$ multiplet}&R&{\bf f}_{\mathfrak{su}(N)}^2 \\  \hline 
\textrm{chiral $\Phi$}&\textrm{$\Box$ or $\overline{\Box}$} &-1 \\
& \textrm{adjoint}& -2N\\ \hline
\textrm{Fermi $\Gamma$}&\textrm{$\Box$ or $\overline{\Box}$}& 1 \\ 
&\textrm{adjoint}& 2N\\ \hline
\textrm{gauge $\Upsilon$}&\textrm{adjoint}& 2N\\
\end{array}
\end{align}

In particular, the gauge anomaly is required to be cancelled for a consistent quantum field theory, 
which leads to an important constraint. 
Unlike the gauge anomaly, the global anomaly may remain in the theory. 
If the global anomaly remains in the IR, the current of the global symmetry of Lie algebra $\mathfrak{h}$ 
can be holomorphic or anti-holomorphic, i.e. left- or right-moving. 
Then  the corresponding global symmetry can be enhanced to 
the affine Lie algebra $\widehat{\mathfrak{h}}$ of level $|2\mathcal{A}_{\mathfrak{h}}|$ 
where $\mathcal{A}_{\mathfrak{h}}$ is the anomaly coefficient. 
The affine Lie algebra $\widehat{\mathfrak{h}}$ acts in the holomorphic or anti-holomorphic sector of the associated CFT 
depending on the sign of the anomaly coefficient $\mathcal{A}_{\mathfrak{h}}$.

\section{$\mathcal{N}=(0,4)$ boundary conditions}
\label{sec_04bcbrane}

\subsection{Brane construction}
\label{sec_04bcbrane_1}
We consider Type IIB superstring theory in Minkowski spacetime 
with time coordinate $x^{0}$ and space coordinates $x^{1},\cdots, x^{9}$ \cite{Hanany:1996ie}. 
Let $\mathcal{Q}_{L}$ and $\mathcal{Q}_{R}$ be the supercharges 
generated by left- and right-moving world-sheet degrees of freedom. 
They satisfy the chirality conditions of Type IIB superstring theory:  
$\overline{\Gamma}\mathcal{Q}_{L}=\mathcal{Q}_{L}$, $\overline{\Gamma}\mathcal{Q}_{R}=\mathcal{Q}_{R}$ 
where $\overline{\Gamma}=\Gamma_{0}\cdots \Gamma_{9}$. 

We introduce NS5-branes with world-volumes in $(x^{0}$, $x^{1}$, $x^{2}$, $x^{3}$, $x^{4}$, $x^{5})$ directions, 
D5-branes with world-volumes in $(x^{0}$, $x^{1}$, $x^{2}$, $x^{7}$, $x^{8}$, $x^{9})$ directions, 
NS5$'$-branes with world-volumes in $(x^{0}$, $x^{1}$, $x^{6}$, $x^{7}$, $x^{8}$, $x^{9})$ directions, 
D5$'$-branes with world-volumes in $(x^{0}$, $x^{1}$, $x^{3}$, $x^{4}$, $x^{5}$, $x^{6})$ directions, 
and D3-branes in $(x^{0}, x^{1}, x^{2}, x^{6})$ directions: 
\begin{align}
\label{04_brane1}
\begin{array}{ccccccccccc}
&0&1&2&3&4&5&6&7&8&9\\
\textrm{D3}
&\circ&\circ&\circ&-&-&-&\circ&-&-&- \\
\textrm{NS5}
&\circ&\circ&\circ&\circ&\circ&\circ&-&-&-&- \\
\textrm{D5}
&\circ&\circ&\circ&-&-&-&-&\circ&\circ&\circ \\
\textrm{NS5$'$}
&\circ&\circ&-&-&-&-&\circ&\circ&\circ&\circ \\
\textrm{D5$'$}
&\circ&\circ&-&\circ&\circ&\circ&\circ&-&-&- \\
\end{array}
\end{align}
All the branes share the $(x^0, x^1)$ directions. 
We consider the case in which 
the D3-branes are bounded by all the 5-branes in the $(x^{2}, x^{6})$ directions. 
According to the Kaluza-Klein reduction in these two directions, 
the world-volume theories on the D3-branes therefore are macroscopically two dimensional. 

The presence of all branes breaks down the space-time $SO(1,9)$ Lorentz symmetry to 
$SO(1,1)_{01}\times SO(3)_{345}\times SO(3)'_{789}$ 
where $SO(1,1)$ acts on $x^{0}$, $x^{1}$, 
$SO(3)_{345}$ acts on $x^{3}, x^{4}, x^{5}$ 
and $SO(3)_{789}$ acts on $x^{7}, x^{8}, x^{9}$. 
The brane configuration (\ref{04_brane1}) preserves linear combinations of supercharges 
$\epsilon_{L}\mathcal{Q}_{L}+\epsilon_{R}\mathcal{Q}_{R}$ 
with $\epsilon_{L}$ and $\epsilon_{R}$ being spinors such that 
\begin{align}
\label{NS_SUSY1}
\Gamma_{012345}\epsilon_{L}&=\epsilon_{L}, & 
\Gamma_{012345}\epsilon_{R}&=-\epsilon_{R}, \\
\label{D5_SUSY2}
\Gamma_{012789}\epsilon_{R}&=\epsilon_{L}\\
\label{NS'_SUSY3}
\Gamma_{016789}\epsilon_{L}&=\epsilon_{L}, &
\Gamma_{016789}\epsilon_{R}&=-\epsilon_{R}, \\
\label{D5'_SUSY}
\Gamma_{013456}\epsilon_{R}&=\epsilon_{L}, \\
\label{D3_SUSY}
\Gamma_{0126}\epsilon_{R}&=\epsilon_{L}. 
\end{align}
These conditions give rise to three non-trivial projection conditions 
so that the presence of branes breaks supersymmetry to $1/8$ of the original supersymmetry 
and four supercharges are preserved. 
Besides, these conditions lead to 
$\Gamma_{01}\epsilon_{L}=\epsilon_{L}$, 
$\Gamma_{01}\epsilon_{R}=\epsilon_{R}$, 
which implies that chiral $\mathcal{N}=(0,4)$ supersymmetry exists in the $(x^{0}, x^{1})$ directions 
and thereby $SO(3)_{345}\times SO(3)_{789}$ is identified with the $SO(4)_{R}\cong SU(2)_{C}\times SU(2)_{H}$ R-symmetry 
of the $\mathcal{N}=(0,4)$ supersymmetry.

\subsection{$\mathcal{N}=(0,4)$ boundary conditions}
\label{sec_04bc}
When the D3-branes are finite only in $x^{6}$ direction and are semi-infinite in the region $x^2\ge 0$, 
the configuration of D3-, NS5- and D5-branes leads to 3d $\mathcal{N}=4$ supersymmetric field theories 
and two extra 5-branes, NS5$'$- and D5$'$-branes break further half of supersymmetry 
to give rise to $\mathcal{N}=(0,4)$ boundary conditions at $x^2=0$ in these theories 
\cite{Chung:2016pgt}. 
%
%
%
%
%

\subsubsection{3d $\mathcal{N}=4$ vector multiplet}
\label{subsec_04BC1}
When a D3-brane is stretched between two parallel NS5-branes along $x^{6}$ direction, 
the low-energy effective theory is that of 3d $\mathcal{N}=4$ Abelian vector multiplet. 
It contains a three-dimensional gauge fields $A_{\mu}$, $\mu=0,1,2$, 
three real scalar fields $\phi^{i}$, $i=3,4,5$ which transforming as $(\bm{3},\bm{1})$ of $SU(2)_{C}\times SU(2)_{H}$, 
an auxiliary field $D$ and a fermionic fields $\lambda$. 
The irreducible 3d $\mathcal{N}=4$ vector multiplet $\mathbb{V}$ 
decomposes into a sum of $\mathcal{N}=(0,4)$ vector multiplet $V$ 
and $\mathcal{N}=(0,4)$ twisted hypermultiplet $T$:
\begin{align}
\label{3d4_v_half}
\mathbb{V}&=(V, T). 
\end{align}

\begin{enumerate}

\item \textbf{NS5$'$ and $(0,4)$ vector multiplet}

When the D3-brane further ends on an NS5$'$-brane, 
it fixes the motion of the D3-brane in $(x^{3}, x^{4}, x^{5})$ 
while the two-dimensional gauge field $A_{\alpha}$, $\alpha=0,1$ is free to fluctuate 
and normal component $A_{2}$ of gauge field obeys the Dirichlet boundary condition. 
Correspondingly, 
the field theory admits the half BPS boundary conditions 
\begin{align}
\label{half_v1}
F_{2\alpha} |_{\partial}&=0,& D_{\alpha}\phi^{i}|_{\partial}&=0, 
\end{align}
along with the massless left-moving fermions.  
The boundary massless modes which are 
two-dimensional gauge fields $A_{\alpha}$ and the left-moving fermions form the irreducible $\mathcal{N}=(0,4)$ vector multiplet. 
Therefore the NS5$'$-brane allows boundary conditions 
in which $(0,4)$ vector multiplet $V$ on the boundary is free 
while $\mathcal{N}=(0,4)$ twisted hypermultiplet $T$ vanishes. 

\item \textbf{D5$'$ and $(0,4)$ twisted hypermultiplet}

Conversely, when the D3-brane terminates on a D5$'$-brane, 
the three scalar fields $\phi^{i}$ are free to fluctuate 
as the D5$'$-brane spans in the $(x^{3}, x^{4}, x^{5})$ directions. 
On the other hand, the gauge field $A_{\alpha}$ 
satisfies the Dirichlet boundary condition and 
the scalar field $A_{2}$ is free to move at the boundary. 
In field theory analysis this corresponds to the half of BPS boundary conditions in 
3d $\mathcal{N}=4$ Abelian vector multiplet 
\begin{align}
\label{half_v2}
F_{\alpha\beta}|_{\partial}&=0,& D_{2}\phi^{i}|_{\partial}&=0, 
\end{align}
when the massless right-moving fermions survive at the boundary. 
The three scalar fields $\phi^{i}$ and the scalar field $A_{2}$ can form a pair of 
complex scalars transforming as $(\bm{2},\bm{1})$. 
They eventually combine with the right-moving fermions into $\mathcal{N}=(0,4)$ twisted hypermultiplet. 
Hence the D5$'$-brane sets $\mathcal{N}=(0,4)$ vector multiplet $V$ to zero 
and leaves $\mathcal{N}=(0,4)$ twisted hypermultiplet $T$ at the boundary. 

When $N_{c}$ D3-branes end on a D5$'$-brane, 
the half BPS boundary conditions are generalized to the Nahm pole boundary condition \cite{Chung:2016pgt}:
\begin{align}
\label{nahm_BC}
F_{\alpha\beta} |_{\partial}&=0,& 
D_{2}\phi^{i}|_{\partial}&=\epsilon^{ijk}[\phi^{j},\phi^{k}]|_{\partial}.
\end{align}
The pole governed by the Nahm equation would  
describe the D3-branes which polarize into a fuzzy funnel configuration  
\cite{Callan:1997kz,Constable:1999ac}.

\end{enumerate}

\subsubsection{3d $\mathcal{N}=4$ hypermultiplet}
\label{subsec_04BC2}
The dynamics of a D3-brane ending on two parallel D5-branes 
would be described by a theory of 3d $\mathcal{N}=4$ hypermultiplet. 
The bosonic massless modes in the theory are 
the fluctuations of the D3-brane in the $(x^{7}, x^{8}, x^{9})$ directions and the scalar field $A_{6}$. 
The three real scalar fields transform as 
$(\bm{1},\bm{3})$ under the $SU(2)_{C}\times SU(2)_{H}$ R-symmetry 
while the scalar field $A_{6}$ transform as $(\bm{1},\bm{1})$. 
The irreducible 3d $\mathcal{N}=4$ hypermultiplet $\mathbb{H}$ 
decomposes as the sum of $\mathcal{N}=(0,4)$ hypermultiplet $H$ and $\mathcal{N}=(0,4)$ Fermi multiplet $\Xi$:
\begin{align}
\label{3d4_h_half}
\mathbb{H}&=(H, \Xi). 
\end{align}

\begin{enumerate}

\item \textbf{NS5$'$ and $(0,4)$ hypermultiplet}

When the D3-brane terminating on two D5-branes further ends on an NS5$'$-brane, 
the three scalar fields describing the fluctuations in the  $(x^{7}, x^{8}, x^{9})$ directions 
and the scalar field $A_{6}$ still remain. 
The corresponding half BPS boundary conditions in 
3d $\mathcal{N}=4$ hypermultiplet can be found in field theory analysis 
\begin{align}
\label{half_h1}
\partial_{2}q |_{\partial}&=0,& 
\partial_{2}\widetilde{q} |_{\partial}&=0, 
\end{align}
when the massless right-moving fermions are left at the boundary. 
The pair of complex scalar fields and the right-moving fermions form $\mathcal{N}=(0,4)$ hypermultiplet. 
Thus the NS5$'$-brane keeps $\mathcal{N}=(0,4)$ hypermultiplet $H$ 
and sets $\mathcal{N}=(0,4)$ Fermi multiplet $\Xi$ to zero. 

\item \textbf{D5$'$ and $(0,4)$ Fermi multiplet}

When the D3-brane between the two D5-branes further attach on an D5$'$-brane, 
all the bosonic massless modes are set to zero. 
These boundary conditions correspond to the half of BPS boundary conditions in 
3d $\mathcal{N}=4$ hypermultiplet 
\begin{align}
\label{half_h2}
\partial_{\alpha}q |_{\partial}&=0,&
\partial_{\alpha}\widetilde{q} |_{\partial}&=0, 
\end{align}
when the massless left-moving fermions are kept at the boundary. 
The left-moving fermions without any bosonic degrees of freedom 
consistently form $\mathcal{N}=(0,4)$ Fermi multiplet. 
So the D5$'$-brane kills $\mathcal{N}=(0,4)$ hypermultiplet $H$ 
and leaves $\mathcal{N}=(0,4)$ Fermi multiplet $\Xi$.

\end{enumerate}

Altogether there are four types of boundary conditions which 
correspond to the four types of junctions of branes; 
D3-NS-NS$'$, D3-NS-D5$'$, D3-D5-NS$'$ and D3-D5-D5$'$. 
These intersections respectively admit boundary local operators 
which are involved in $\mathcal{N}=(0,4)$ vector, twisted hyper, hyper and Fermi multiplets.

\subsection{Boundary anomaly and linking number}
\label{sec_bdyanomaly}
When we consider chiral supersymmetric boundary conditions in 3d supersymmetric gauge theory, 
there also exists an anomaly contribution from bulk fields. 
It is argued in the analysis \cite{Dimofte:2017tpi} 
of $(0,2)$ boundary conditions for 3d $\mathcal{N}=2$ theory,
that the bulk fields may have half of the contributions as those from boundary fields.  
For $SU(N)$ the anomaly contribution is given by
\begin{align}
\label{t_Anom_bc1}
\begin{array}{c|c|c}
\textrm{3d $\mathcal{N}=2$ multiplet with $1/2$ BPS b.c.}&R&{\bf f}^{2}_{\mathfrak{su}(N)} \\  \hline 
\textrm{chiral multiplet with $N$ b.c.}&\textrm{$\Box$ or $\overline{\Box}$} &-\frac12 \\
& \textrm{adjoint}& -N\\ 
\textrm{chiral multiplet with $D$ b.c.}&\textrm{$\Box$ or $\overline{\Box}$}& \frac12 \\ 
&\textrm{adjoint}& N\\ \hline
\textrm{gauge multiplet with $\mathcal{N}$ b.c.}&\textrm{adjoint}& N\\ 
\textrm{gauge multiplet with $\mathcal{D}$ b.c.}&\textrm{adjoint}& -N\\
\end{array}
\end{align}
where $N, D$ b.c. stand for the Neumann and Dirichlet boundary condition for 
3d $\mathcal{N}=2$ chiral multiplet scalar fields and 
$\mathcal{N}, \mathcal{D}$ b.c. imply the Neumann and Dirichlet  boundary conditions for 3d gauge field. 

As discussed in section \ref{sec_04anomaly}, gauge anomaly needs to be cancelled. 
Otherwise, the Neumann boundary condition for gauge field is not consistent. 
For example, 
the boundary gauge anomaly polynomial for 
3d $\mathcal{N}=4$  $U(N_{c})$ gauge theory with $N_{f}$ fundamental hypermultiplets obeying 
$(\mathcal{N}, N)$ boundary condition is given by 
\begin{align}
\label{ncnf_AN}
\mathcal{I}_{(\mathcal{N},N)}^{(N_{c})-[N_{f}]}
&=
-(N_{f}-2N_{c})\Tr ({\bf s}^{2})
\end{align}
where ${\bf s}$ is the field strength of $U(N)$ gauge field. 
The boundary anomaly polynomial for 3d $\mathcal{N}=4$ $\prod_{i}^{n} U(N_{i})$ 
linear quiver gauge theory with bi-fundamental hypermultiplets 
obeying $(\mathcal{N},N)$ boundary conditions is 
\
\begin{align}
\label{quiver_N_AN}
\mathcal{I}^{(N_{1})-(N_{2})- \cdots (N_{n})}_{(\mathcal{N}, N)}
&=(-N_{2}+2N_{1})
\Tr ({\bf s}_{1}^{2})
+\sum_{i=2}^{n-1}(2N_{i}-N_{i-1}-N_{i+1})\Tr ({\bf s}_{i}^{2})
+(-N_{n-1}+2N_{n})\Tr ({\bf s}_{n}^{2})
\end{align}
where $\mathfrak{s}_{i}$ is the field strength of $U(N_{i})$ gauge field. 

\begin{figure}
\begin{center}
\includegraphics[width=15cm]{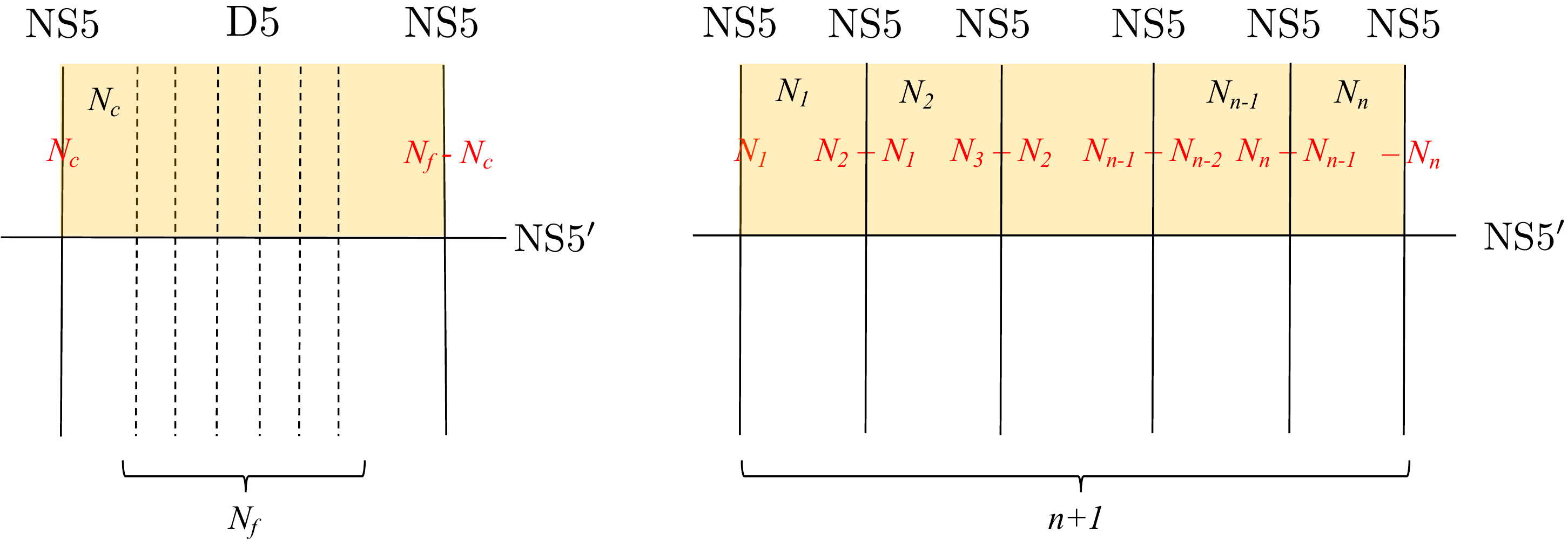}
\caption{Brane construction of 3d $\mathcal{N}=4$ $U(N_{c})$ gauge theory with $N_{f}$ hypermultiplets and 
that of 3d $\mathcal{N}=4$ $\prod_{i=1}^{n} U(N_{i})$ 
linear quiver gauge theory with bi-fundamental hypermultiplets. 
The numbers with red color indicate the linking numbers. }
\label{figanomaly}
\end{center}
\end{figure}

As shown in Figure \ref{figanomaly}, 
the 3d $\mathcal{N}=4$ $U(N_{c})$ gauge theory with $N_{f}$ hypers 
can be constructed as a world-volume theory of stack of $N_{c}$ D3-branes 
intersecting with $N_{f}$ D5-branes stretched between two NS5-branes. 
Also the 3d $\mathcal{N}=4$ $\prod_{i=1}^{n} U(N_{i})$ 
linear quiver gauge theory with bi-fundamental hypermultiplets can be 
realized as $n$ sets of $N_{i}$ D3-branes suspended between 
$(n+1)$ NS5-branes. 
We observe that 
\begin{quote}
\textit{
the boundary non-Abelian gauge anomaly coefficient   
is encoded as the difference of linking numbers of 
the corresponding right and left NS5-branes.}
\end{quote}

The gauge anomaly cancellation can be achieved by introducing two-dimensional boundary degrees of freedom 
which are charged under the gauge symmetry. 
It can be seen that this is naturally achieved by considering consistent linking number 
of boundary NS5$'$-branes and making use of our identification of matter multiplets in brane box configuration.

\subsection{Brane box configuration}
\label{sec_box}
Furthermore, we introduce several NS5- and NS5$'$-branes which bound the D3-branes 
at a finite distance in both $x^{6}$ and $x^{2}$ directions. 
This configuration leads to two-dimensional $\mathcal{N}=(0,4)$ quiver gauge theory in which 
each box of $N$ D3-branes defines $U(N)$ gauge symmetry factor. 
The gauge coupling of the gauge theory is given by 
\begin{align}
\label{coupling}
\frac{1}{e^{2}}&=
\frac{\Delta x^{2} \Delta x^{6}}{g_{s}}
\end{align}
where $g_{s}$ is the string coupling.  
$\Delta x^{6}$ and $\Delta x^{2}$ are the distance of two NS5-branes along $x^{6}$ and that of two NS5$'$-branes along $x^{2}$. 
We call this brane box model. 
In the following section, we study the resulting $\mathcal{N}=(0,4)$ quiver gauge theory 
by identifying the matter content and interactions.

\section{D1-D5-KK system}
\label{sec_d1d5kk}
Now we would like to study $\mathcal{N}=(0,4)$ quiver gauge theory 
which is constructed from a periodic array of D3-branes in $(x^2, x^6)$ plane. 
In this section we analyze the spectrum of brane tiling model 
by taking the T-dual configuration \cite{Hanany:1998it}. 

The T-duality along $x^{6}$ turns the $k$ NS5-branes into $k$ Kaluza-Klein (KK) monopoles 
which can be described by a multi-centered Taub-NUT metric with non-trivial geometry along the directions $(x^{6'}, x^{7}, x^{8}, x^{9})$ 
where $x^{6'}$ is the T-dual coordinate of $x^{6}$. 
Since the original $k$ NS5-branes coincide in the directions $(x^{7}, x^{8}, x^{9})$, 
the centers of the corresponding $k$ KK monopoles also coincide. 
This means that the geometry contains $A_{k-1}$ singularities. 
Also the T-duality along $x^{6}$ translates the D3-branes, D5-branes and D5$'$-branes into 
D2-branes, D6-branes and the D4$'$-branes respectively. 

Similarly the T-duality along $x^{2}$ directions translates  
the $k'$ NS5$'$-branes into the $k'$ KK monopoles extending along $(x^{0}, x^{1}, x^{6'}, x^{7}, x^{8}, x^{9})$ 
with non-trivial geometry along $(x^{2'}, x^{3}, x^{4}, x^{5})$. 
Again since the centers of KK monopoles coincide, the space-time contains $A_{k'-1}$ singularities. 
Under the T-duality along $x^{2}$, 
the D2-branes, D6-branes and D4$'$-branes turn into D1-branes, D5-branes and D5$'$-branes respectively. 
Consequently the T-dual configuration is the D1-D5-KK system:
\begin{align}
\label{d1d5d5_ale}
\begin{array}{ccccccccccc}
&0&1&2'&3&4&5&6'&7&8&9\\
\textrm{D1}
&\circ&\circ&-&-&-&-&-&-&-&- \\
\textrm{KK}
&\circ&\circ&\circ&\circ&\circ&\circ&-&-&-&- \\
\textrm{D5}
&\circ&\circ&-&-&-&-&\circ&\circ&\circ&\circ \\
\textrm{KK$'$}
&\circ&\circ&-&-&-&-&\circ&\circ&\circ&\circ \\
\textrm{D5$'$}
&\circ&\circ&\circ&\circ&\circ&\circ&-&-&-&- \\
\end{array}
\end{align}
Before orbifolding, the configuration (\ref{d1d5d5_ale}) is invariant under 
the space-time symmetry $SO(1,1)_{0,1}$ $\times$ $SO(4)_{2'345}$ $\times$ $SO(4)_{6'789}$. 
Let $(\tilde{A}, \tilde{A}')$ and $(A, A')$ represent the $\bm{2}$'s of the 
 $SU(2)_{C}$ $\times SU(2)_{C}'$ $\cong$ $SO(4)_{2'345}$ 
 and those of the $SU(2)_{H}$ $\times$ $SU(2)_{H}'$ $\cong$ $SO(4)_{6'789}$. 
 Here the $SU(2)_{C}$ $\times$ $SU(2)_{H}$ is the R-symmetry group in the $\mathcal{N}=(0,4)$ gauge theories 
 in such a way that $SU(2)_{C}$ rotates the twisted hypermultiplet scalars 
 while the $SU(2)_{H}$ rotates the hypermultiplet scalars.

\subsection{D1-branes on $\mathbb{C}^{2}/\mathbb{Z}_{k}\times \mathbb{C}^{2}/\mathbb{Z}_{k'}$}
\label{subsec_d1zkzk}

\subsubsection{$\mathcal{N}=(0,2)$ quiver}
\label{subsec_02quiver}
Let us firstly consider the configuration in which D1-branes probing 
$\mathbb{C}^{2}/\mathbb{Z}_{k}$ $\times$ $\mathbb{C}^{2}/\mathbb{Z}_{k'}$. 
This corresponds to the T-dual D3-brane box model where neither D5- nor D5$'$-branes exist. 
Generically the two-dimensional low-energy effective world-volume theory would preserve $\mathcal{N}=(0,4)$ supersymmetry. 

To identify the gauge theory on D1-branes at singularity, 
we start from the $N$ D1-branes over $\mathbb{C}^{4}$ 
and use the technique developed in \cite{Franco:2015tna}. 
The world-volume theory is 2d $\mathcal{N}=(8,8)$ supersymmetric $U(N)$ gauge theory. 
In terms of $\mathcal{N}=(0,2)$ supermultiplets, 
it consists of $\mathcal{N}=(0,2)$ gauge multiplet $\Upsilon$, 
four $\mathcal{N}=(0,2)$ adjoint chiral multiplets $R$, $L$, $U$ and $D$, 
and three $\mathcal{N}=(0,2)$ adjoint Fermi multiplets $\Delta$, $\nabla$ and $\Lambda$. 
We introduce four complexifed coordinates $z_{1}=x^{6'}+ix^{7}$, $z_{2}=x^{8}+ix^{9}$, 
$z_{3}=x^{2'}+ix^{3}$, $z_{4}=x^{4}+ix^{5}$. 
In terms of $\mathcal{N}=(0,2)$ supermultiplets, 
$J$- and $E$ terms are given by 
\begin{align}
\label{je_88}
\begin{array}{ccc}
&J&E \\
\Delta: &
L\cdot U-U\cdot L=0&
D\cdot R-R\cdot D=0\\
\nabla: &
U\cdot R-R\cdot U=0&
D\cdot L-L\cdot D=0\\
\Lambda: &
R\cdot L-L\cdot R=0&
D\cdot U-U\cdot D=0.\\
\end{array}
\end{align}
Here and in the following $\cdot$ stands for appropriate index contraction 
which includes gauge and global symmetry groups.

\begin{figure}
\begin{center}
\includegraphics[width=5cm]{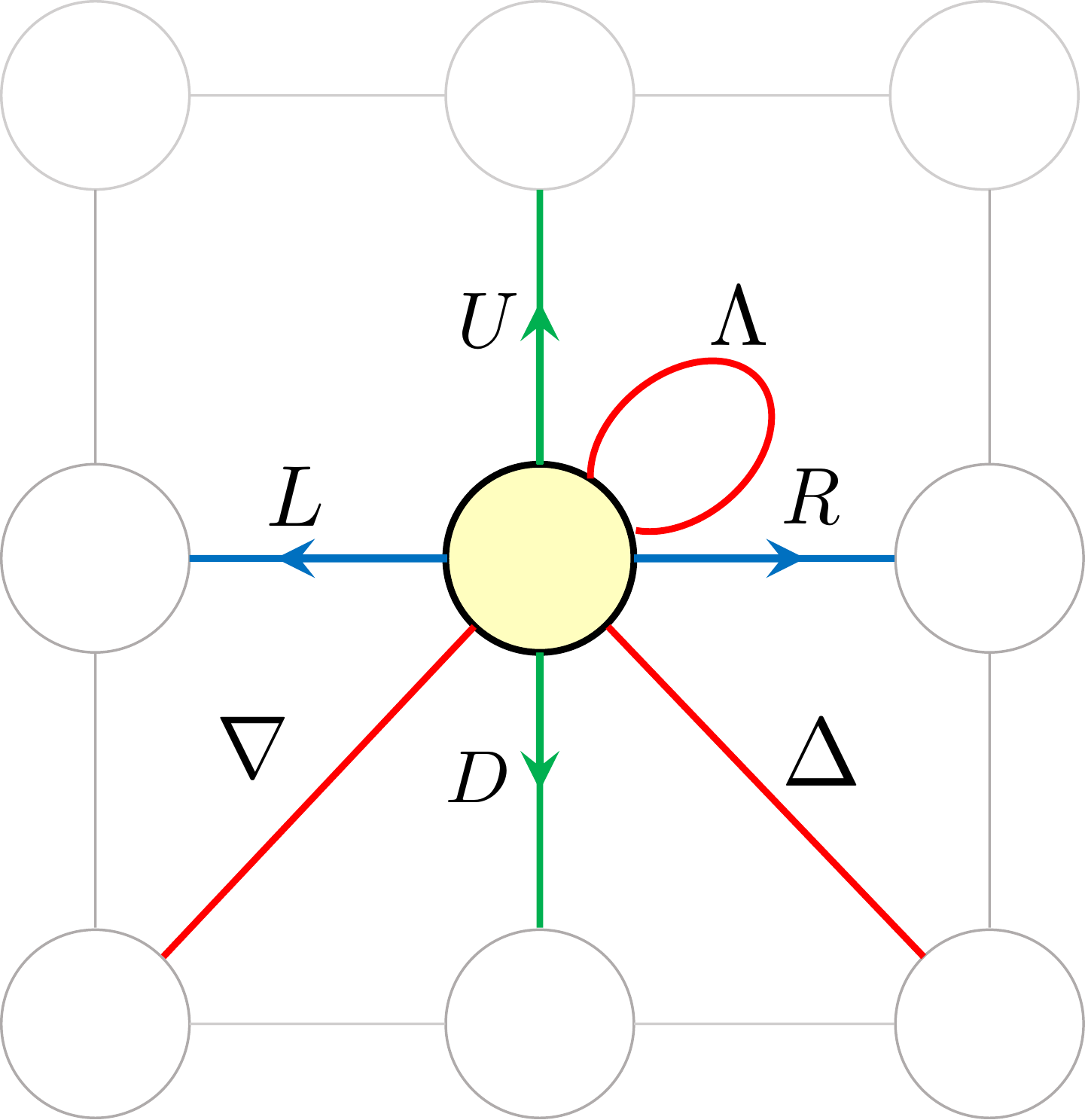}
\caption{$\mathcal{N}=(0,2)$ quiver for basic building block of $\mathcal{N}=(0,4)$ quiver gauge theory. 
A circular node represents the $\mathcal{N}=(0,2)$ gauge multiplet. 
A pair of blue arrows of $\mathcal{N}=(0,2)$ chiral multiplets $(R,L)$ form $\mathcal{N}=(0,4)$ hypermultiplet. 
A pair of green arrows of the chiral multiplets $(U,D)$ form $\mathcal{N}=(0,4)$ twisted hypermultiplet. 
Red diagonal lines $\Delta$ and $\nabla$ are the $\mathcal{N}=(0,2)$ Fermi multiplets. 
A red loop of adjoint $\mathcal{N}=(0,2)$ Fermi multiplet $\Lambda$ is involved in $\mathcal{N}=(0,4)$ vector multiplets. }
\label{fig04quiver}
\end{center}
\end{figure}

Let us take generators of orbifold group labeled by $(a,k-a,0,0)$ and $(0,0,b,k'-b)$ 
where $a\le k$, $b\le k'$ are positive integers. 
They act on the complex coordinates as
\begin{align}
\label{orb1a}
z_{1}&\mapsto e^{\frac{2\pi ai}{k}}z_{1},&  
z_{2}&\mapsto e^{-\frac{2\pi ai}{k}}z_{2},& 
z_{3}&\mapsto e^{\frac{2\pi bi}{k'}}z_{3},&  
z_{4}&\mapsto e^{-\frac{2\pi bi}{k'}}z_{4}. 
\end{align}
Since the action of orbifold on the Chan-Paton factor is 
$\mathcal{R}_{CP}$ $=$ $N (\bigoplus_{I} \mathcal{R}_{I})$ 
where $\mathcal{R}_{I}$ is one-dimensional unitary representation of orbifold, 
the orbifold turns the gauge symmetry group into \cite{Douglas:1996sw}
\begin{align}
\label{zkzk_g}
G&=\prod_{i}U(N)_{i}
=\prod_{i_{1}=1}^{k}\prod_{i_{2}=1}^{k'}U(N)_{i_{1}, i_{2}}
\end{align}
where $i=(i_{1}, i_{2})$, $i_{1}$ $=$ $1,\cdots, k$, $i_{2}$ $=$ $1,\cdots, k'$ label the gauge nodes. 
As the chiral multiplets $R$, $L$, $U$ and $D$ describe 
the motions of D1-branes in $z_{1}$, $z_{2}$, $z_{3}$ and $z_{4}$ respectively, 
it follows that 
under the orbifold action (\ref{orb1a}), 
they are represented as
\begin{align}
\label{orb1b}
{R^{i}}_{j}: \qquad j&=i+(a,0)& \mod& (k, k')  \nonumber\\
{L^{i}}_{j}: \qquad j&=i+(-a,0)& \mod& (k,k') \nonumber\\
{U^{i}}_{j}: \qquad j&=i+(0,b)& \mod& (k,k') \nonumber\\
{D^{i}}_{j}: \qquad j&=i+(0,-b)& \mod& (k,k'). 
\end{align}
According to the form of interaction terms (\ref{je_88}) and chiral fields (\ref{orb1b}) in the presence of orbifold, 
three $\mathcal{N}=(0,2)$ Fermi multiplets transform under the 
gauge group (\ref{zkzk_g}) as
\begin{align}
\label{orb1c}
\Delta_{i,j}: \qquad j&=i+(a,-b) &\mod& (k,k') \nonumber\\
\nabla_{i,j}: \qquad j&=i+(-a,-b) &\mod& (k,k')\nonumber\\
\Lambda_{i,j}: \qquad j&=i &\mod& (k,k').
\end{align}
Each of $kk'$ boxes involves four $\mathcal{N}=(0,2)$ chiral multiplets and three $\mathcal{N}=(0,2)$ Fermi multiplets 
so that in total there exist $4kk'$ $\mathcal{N}=(0,2)$ chiral multiplets and $3kk'$ $\mathcal{N}=(0,2)$ Fermi multiplets in the quiver gauge theory. 
The basic building block of the $\mathcal{N}=(0,4)$ quiver gauge theories in terms of $\mathcal{N}=(0,2)$ quiver 
is depicted in Figure \ref{fig04quiver}. 
%
%
%
%
%

\subsubsection{$\mathcal{N}=(0,4)$ quiver}
\label{subsec_04quiver}
Now we would like to construct $\mathcal{N}=(0,4)$ quiver. 
From the discussion in section \ref{sssec_04}, 
the $\mathcal{N}=(0,2)$ supermultiplets appearing in $\mathcal{N}=(0,2)$ quiver of Figure \ref{fig04quiver} should be promoted to 
$\mathcal{N}=(0,4)$ supermultiplets in a consistent way. 

\begin{figure}
\begin{center}
\includegraphics[width=9.5cm]{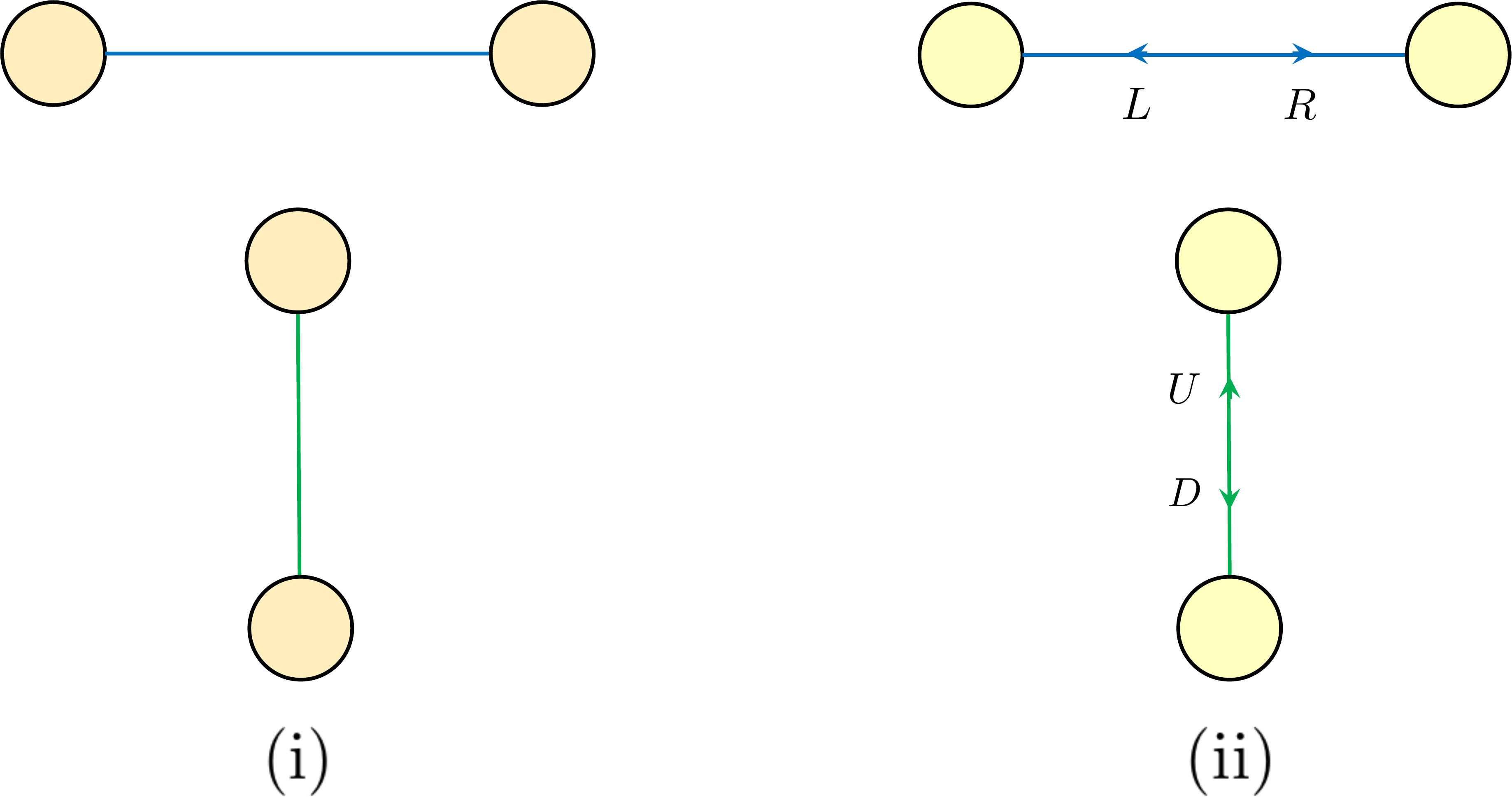}
\caption{
(i) Blue line of $\mathcal{N}=(0,4)$ hyper and green line of $\mathcal{N}=(0,4)$ twisted hypermultiplets.  
(ii) The corresponding pairs of $\mathcal{N}=(0,2)$ chiral multiplets $(R,L)$ and $(U,D)$.}
\label{fig04hmth_dec}
\end{center}
\end{figure}

The $\mathcal{N}=(0,2)$ chiral multiplets always appear as a pair of arrows with opposite orientations between adjacent gauge nodes. 
The pair of $\mathcal{N}=(0,2)$ chiral multiplets ${R^{i}}_{j}$ and ${L^{i}}_{j}$ corresponding to horizontal arrows 
will form the bi-fundamental $\mathcal{N}=(0,4)$ hypermultiplets 
while the pair of $\mathcal{N}=(0,2)$ chiral multiplets ${U^{i}}_{j}$ and ${D^{i}}_{j}$ corresponding to vertical arrows  
will combine into the bi-fundamental $\mathcal{N}=(0,4)$ twisted hypermultiplets 
(see Figure \ref{fig04hmth_dec}). 

\begin{figure}
\begin{center}
\includegraphics[width=9.5cm]{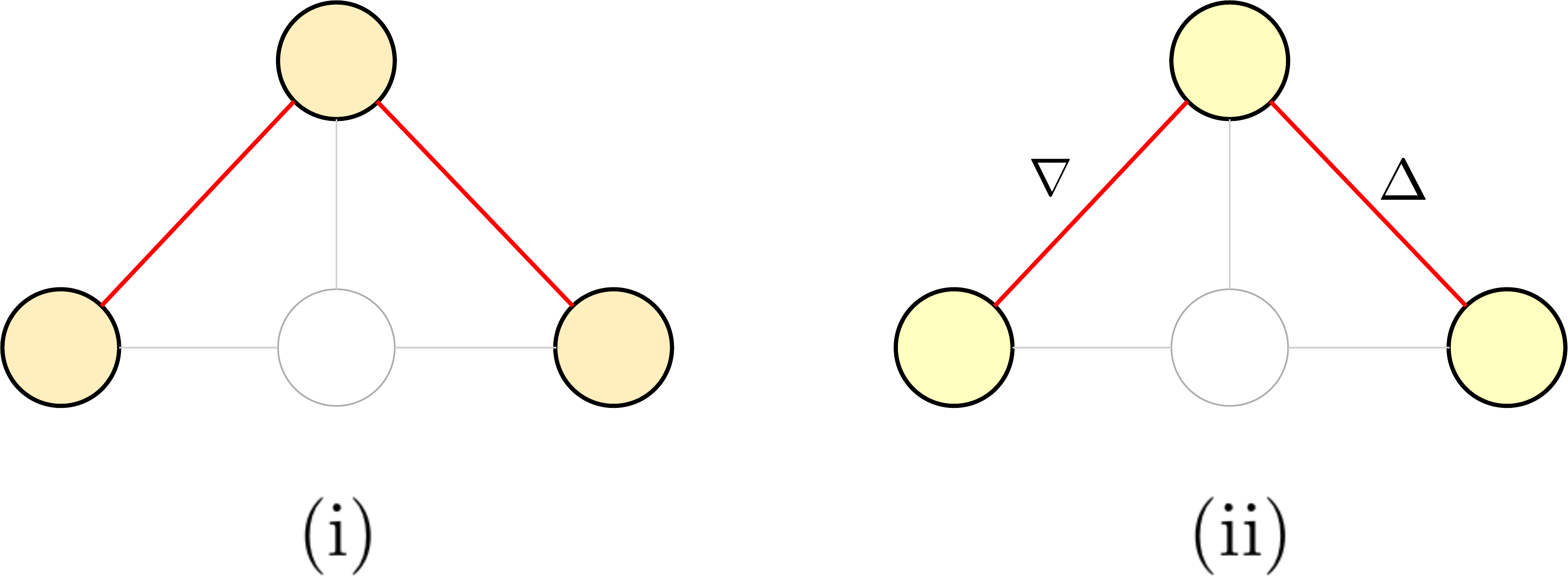}
\caption{
(i) A V-shaped configuration of Fermi multiplet in the $\mathcal{N}=(0,4)$ quiver diagram. 
(ii) The corresponding pair of $\mathcal{N}=(0,2)$ Fermi multiplets $(\Delta, \nabla)$ in the $\mathcal{N}=(0,2)$ quiver diagram.}
\label{fig04fm_dec}
\end{center}
\end{figure}

Meanwhile, the three $\mathcal{N}=(0,2)$ Fermi multiplets split into two types. 
Two $\mathcal{N}=(0,2)$ multiplets $\Delta_{ij}$ and $\nabla_{ij}$ represented as 
diagonal red edges produce the following $J$- and $E$-terms: 
\begin{align}
\label{je_bifund}
J:\qquad & 
\Delta_{i,i+(a,-b)}\left(
{L^{i+(a,-b)}}_{i+(0,-b)}{U^{i+(0,-b)}}_{i}
-{U^{i+(a,-b)}}_{i+(a,0)}{L^{i+(a,0)}}_{i}
\right),
\nonumber\\
E:\qquad &
\Delta_{i+(a,-b),i}
\left(
{D^{i}}_{i+(0,-b)}{R^{i+(0,-b)}}_{i+(a,-b)}
-{R^{i}}_{i+(a,0)}{D^{i+(a,0)}}_{i+(a,-b)}
\right),
\nonumber\\
J:\qquad &
\nabla_{i,i-(a,b)}\left(
{U^{i-(a,b)}}_{i-(a,0)}{R^{i-(a,0)}}_{i}
-{R^{i-(a,b)}}_{i-(0,b)}{U^{i-(0,b)}}_{i}
\right),
\nonumber\\
E:\qquad &
\nabla_{i-(a,b),i}
\left(
{D^{i}}_{i-(0,b)}{L^{i-(0,b)}}_{i-(a,b)}
-{L^{i}}_{i-(a,0)}{D^{i-(a,0)}}_{i-(a,b)}
\right). 
\end{align}
They give rise to the interaction terms between $\mathcal{N}=(0,4)$ hypermultiplets 
and $\mathcal{N}=(0,4)$ twisted hypermultiplets. 
Unlike the $\mathcal{N}=(0,2)$ chiral multiplets, 
the $\mathcal{N}=(0,2)$ Fermi multiplets do not appear as a pair of arrows 
but rather a pair of links forming a V-shaped configuration as shown in Figure \ref{fig04fm_dec}. 
We identify them with the \textit{Fermi multiplet} in $\mathcal{N}=(0,4)$ quiver. 
%
%
%
%
%

\begin{figure}
\begin{center}
\includegraphics[width=10.5cm]{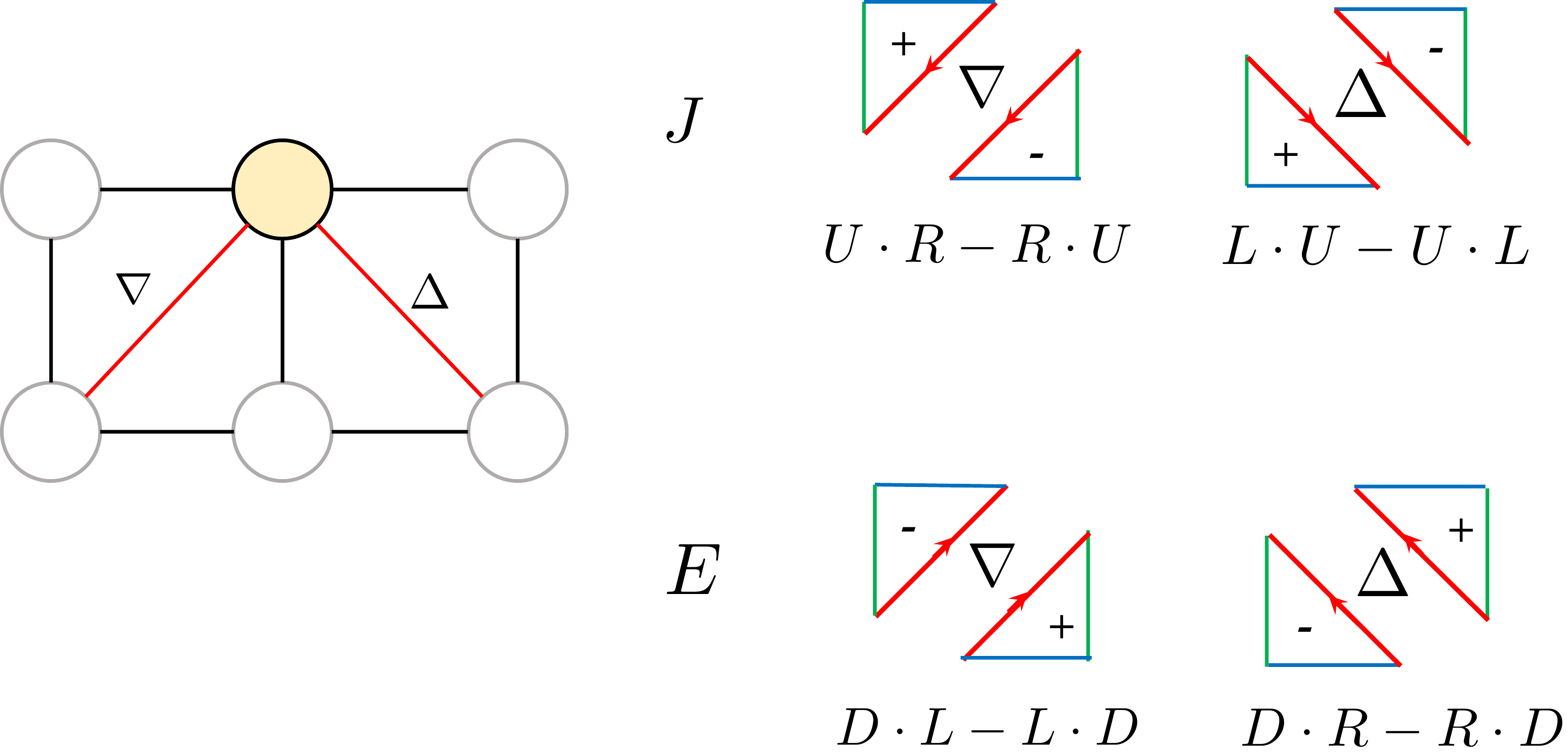}
\caption{
Triangular loops of the $E$- and $J$-terms in quiver diagram. 
The positive and negative signs in the triangles represent 
the clockwise and anti-clockwise orientations which determine the contributions to the $E$- and $J$-terms. 
}
\label{figje_bifund}
\end{center}
\end{figure}

The four types of interactions (\ref{je_bifund}) can be easily read from closed loops in quiver diagram. 
For each of red edges $\Delta_{i,i+(a,-b)}$ and $\nabla_{i,i-(a,b)}$ in the Fermi multiplet, 
one can draw two triangles sharing the corresponding edge. 
Given the orientation of the Fermi edge, the triangles lead to two closed loops. 
If the orientation is directed from $i$ to $i+(a,-b)$ or to $i-(a,b)$, 
the two loops contribute to $J$-terms associated to the corresponding Fermi multiplet. 
If the orientation is opposite, they contribute to the $E$-terms. 
Assigning positive and negative signs for clockwise and anti-clockwise loops respectively, 
we obtain the $E$- and $J$-terms (\ref{je_bifund})
for $\Delta_{i,i+(a,-b)}$ and $\nabla_{i,i-(a,b)}$ (see Figure \ref{figje_bifund}). 

\begin{figure}
\begin{center}
\includegraphics[width=5cm]{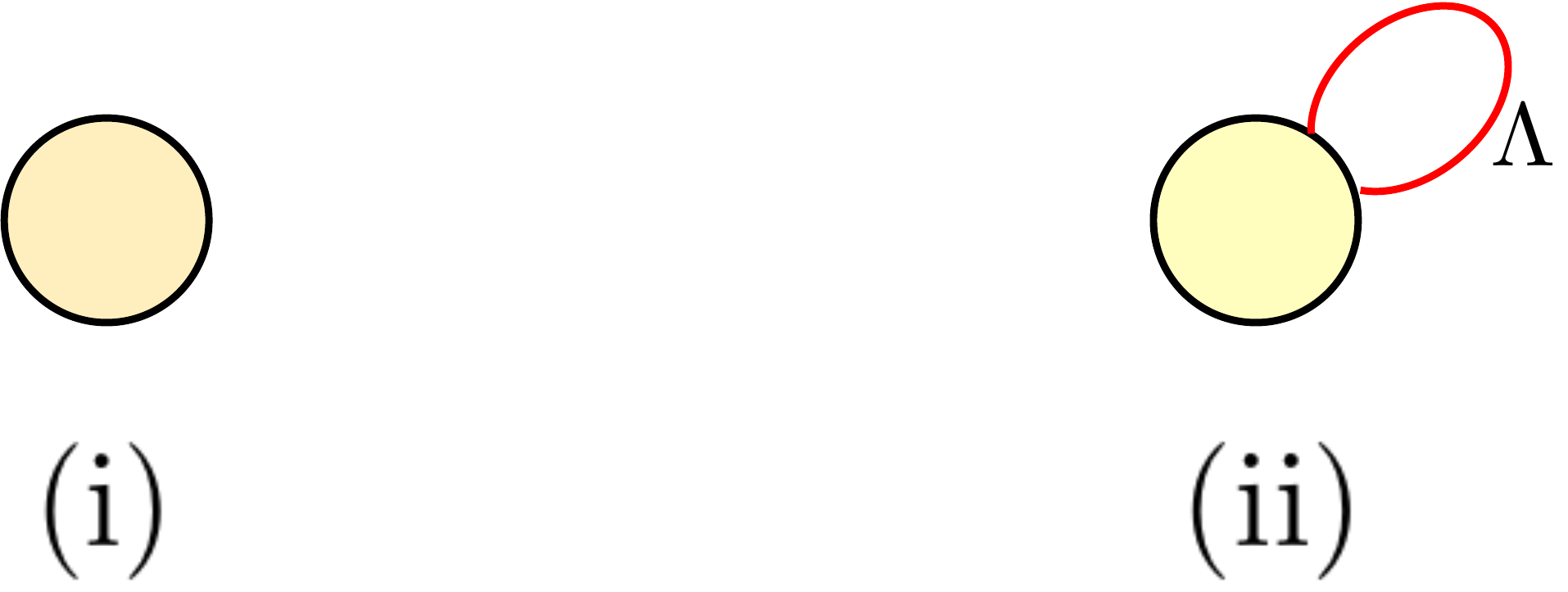}
\caption{(i) An orange node of vector multiplet in $\mathcal{N}=(0,4)$ quiver diagram. 
(ii) The corresponding $\mathcal{N}=(0,2)$ gauge node and adjoint Fermi multiplet $\Lambda$ in $\mathcal{N}=(0,2)$ quiver diagram. 
}
\label{fig04vm_dec}
\end{center}
\end{figure}

The remaining $\mathcal{N}=(0,2)$ Fermi multiplets $\Lambda_{ij}$ sketched as the circular red edges 
transform as adjoint representation under the corresponding gauge group. 
They lead to $J$- and $E$-terms 
\begin{align}
\label{j_adj02}
J:\qquad & 
\Lambda_{i, i}\left(
{R^{i}}_{i+(a,0)}{L^{i+(a,0)}}_{i}
-{L^{i}}_{i+(-a,0)}{R^{i+(-a,0)}}_{i}
\right),\\
\label{e_adj02}
E:\qquad &
\Lambda_{i,i}
\left(
{D^{i}}_{i+(0,-b)}{U^{i+(0,-b)}}_{i}
-{U^{i}}_{i+(0,b)}{D^{i+(0,b)}}_{i}
\right).
\end{align}
The $J$-terms (\ref{j_adj02}) describe the interaction between hypermultiplets 
while the $E$-terms (\ref{e_adj02}) describe the interaction between twisted hypermultiplets. 
These Fermi multiplets will combine with the $\mathcal{N}=(0,2)$ gauge multiplets to form the $\mathcal{N}=(0,4)$ vector multiplets. 
We represent the $\mathcal{N}=(0,4)$ vector multiplet as an orange node (see Figure \ref{fig04vm_dec}). 

Here we can summarize a simple dictionary between the 
$\mathcal{N}=(0,2)$ quiver and $\mathcal{N}=(0,4)$ quiver as follows: 

\begin{itemize}

\item A circular orange node of $\mathcal{N}=(0,4)$ vector multiplet consists of 
$\mathcal{N}=(0,2)$ gauge node and $\mathcal{N}=(0,2)$ adjoint Fermi multiplet $\Lambda$. 
The map is shown in Figure \ref{fig04vm_dec}.

\item Blue edge of $\mathcal{N}=(0,4)$ hyper 
and green edge of twisted hypermultiplets in $\mathcal{N}=(0,4)$ quiver diagram translates into  
pairs of arrows of $\mathcal{N}=(0,2)$ chiral multiplets $(R,L)$ and $(U,D)$ respectively in $\mathcal{N}=(0,2)$ quiver diagram. 
This is shown in Figure \ref{fig04hmth_dec}.

\item V-shaped configuration of the Fermi multiplet appears in $\mathcal{N}=(0,4)$ quiver diagram. 
This is shown in Figure \ref{fig04fm_dec}.

\end{itemize}

\begin{figure}
\begin{center}
\includegraphics[width=9.5cm]{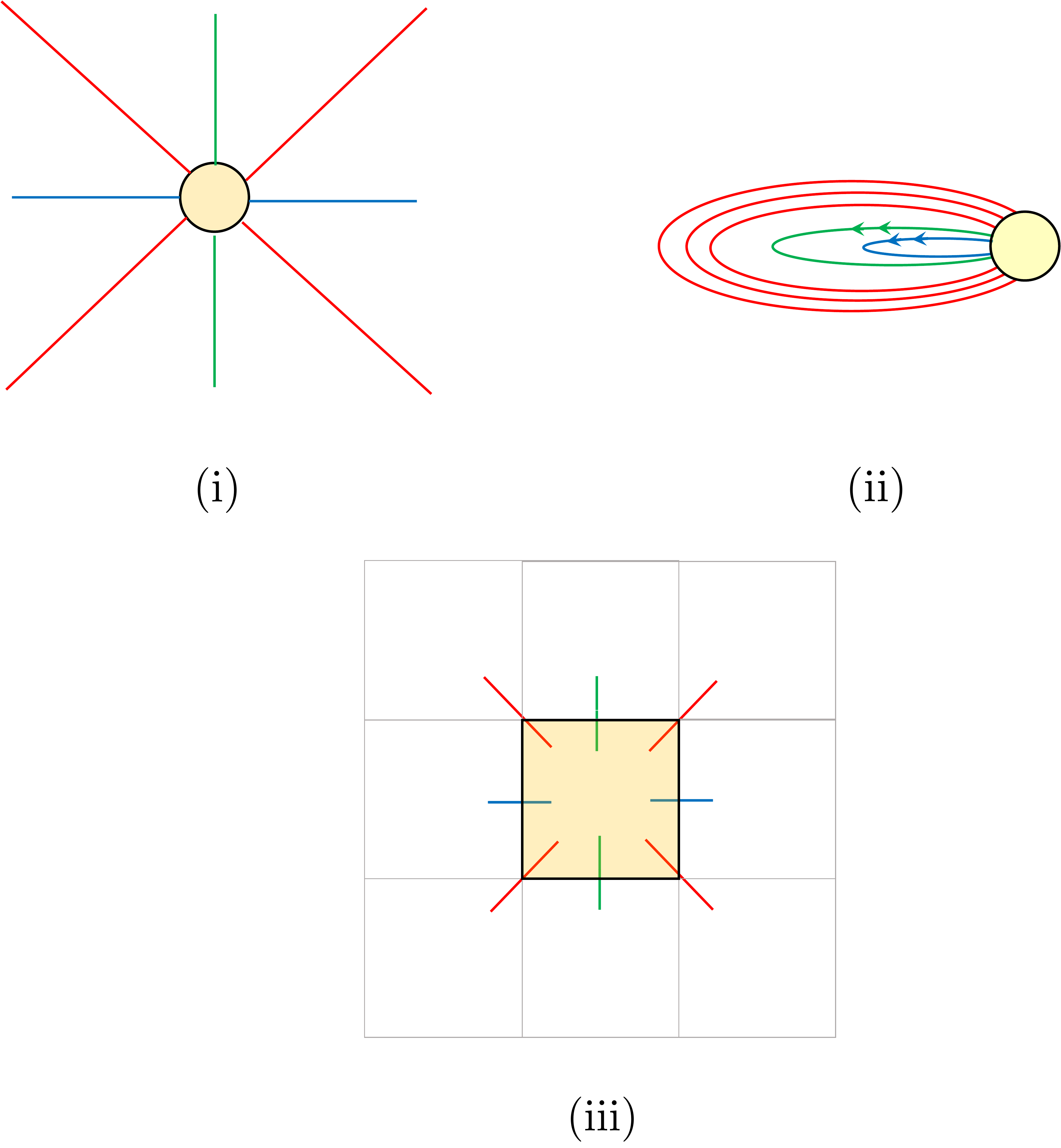}
\caption{
(i) $\mathcal{N}=(0,4)$ quiver diagram for D1-branes probing $\mathbb{C}^{4}$. 
(ii) The corresponding $\mathcal{N}=(0,2)$ quiver. 
(iii) The T-dual configuration as a single periodic box of D3-branes. 
}
\label{figc4}
\end{center}
\end{figure}

It would be instructive to present both $\mathcal{N}=(0,2)$ quiver and $\mathcal{N}=(0,4)$ quiver for 
2d $\mathcal{N}=(8,8)$ supersymmetric gauge theory of D1-branes probing $\mathbb{C}^{4}$. 
They are shown in Figure \ref{figc4}. 
The $\mathcal{N}=(0,4)$ quiver contains a single $\mathcal{N}=(0,4)$ gauge node, 
a blue line of $\mathcal{N}=(0,4)$ adjoint hypermultiplet forming a single loop, 
a green line of $\mathcal{N}=(0,4)$ adjoint twisted hypermultiplet forming a single loop 
and a red V-shaped configuration adjoint Fermi multiplet consisting of two loops. 

At this stage we can go back to the brane box configuration. 
For the case of D1-branes probing $\mathbb{C}^{4}$, 
the T-dual configuration is a periodic single box of D3-branes as shown in Figure \ref{figc4}. 
It is now easy to read off the field content in the D3-brane box model. 
Gauge nodes correspond to a boxes of D3-branes. 
To each of gauge nodes there are 
bivalent blue links as a pair of edges of $\mathcal{N}=(0,4)$ hypermultiplets, 
bivalent green links of $\mathcal{N}=(0,4)$ twisted hypermultiplets and 
a pair of V-shaped configuration of Fermi multiplets.

Let us see further examples in the following.

\subsubsection{$\mathbb{C}^{2}/\mathbb{Z}_{2}\times \mathbb{C}^{2}$}
\label{subsec_d1zkzk}
In this case the gauge group is $U(N)_{1,0}$ $\times$ $U(N)_{2,0}$. 
There are eight $\mathcal{N}=(0,2)$ chiral multiplets 
\begin{align}
\label{z2_cm}
\{ {R^{i_{1}}}_{i_{1}+1}, {L^{i_{1}}}_{i_{1}-1}, {U^{i_{1}}}_{i_{1}}, {D^{i_{1}}}_{i_{1}} \}_{i_{1}=1,2}. 
\end{align}
${R^{i_{1}}}_{i_{1}+1}$ and 
${L^{i_{1}}}_{i_{i-1}}$ transform as bi-fundamental representation 
under the gauge group $U(N)_{1,0}$ $\times$ $U(N)_{2,0}$ 
and they form two bi-fundamental $\mathcal{N}=(0,4)$ hypermultiplets: 
$({R^{1}}_{2}, {L^{2}}_{1})$ $\oplus$ $({R^{2}}_{1}, {L^{1}}_{2})$. 
${U^{i_{1}}}_{ i_{1}}$ and ${D^{i_{1}}}_{i_{1}}$ transform as adjoint representation under the gauge group 
and they combine into two adjoint $\mathcal{N}=(0,4)$ twisted hypermultiplets: 
$({U^{1}}_{1}, {D^{1}}_{1})$ $\oplus$ $({U^{2}}_{2}, {D^{2}}_{2})$. 

There are also six $\mathcal{N}=(0,2)$ Fermi multiplets 
\begin{align}
\label{z2_fm}
\{ \Delta_{i_{1}, i_{1}+1}, \nabla_{i_{1}, i_{1}-1}, \Lambda_{i_{1}, i_{1}}\}_{i_{1}=1,2}. 
\end{align}
$\Delta_{i_{1}, i_{1}+1}$ and $\nabla_{i_{1},i_{1}-1}$ transform as bi-fundamental representation 
under $U(N)_{1,0}$ $\times$ $U(N)_{2,0}$. 
$\Lambda_{i_{1},i_{1}}$ are adjoint under $U(N)_{1,0}$ $\times$ $U(N)_{2,0}$ 
and they combine $\mathcal{N}=(0,2)$ gauge multiplet to form two $\mathcal{N}=(0,4)$ vector multiplets.

\begin{figure}
\begin{center}
\includegraphics[width=12.5cm]{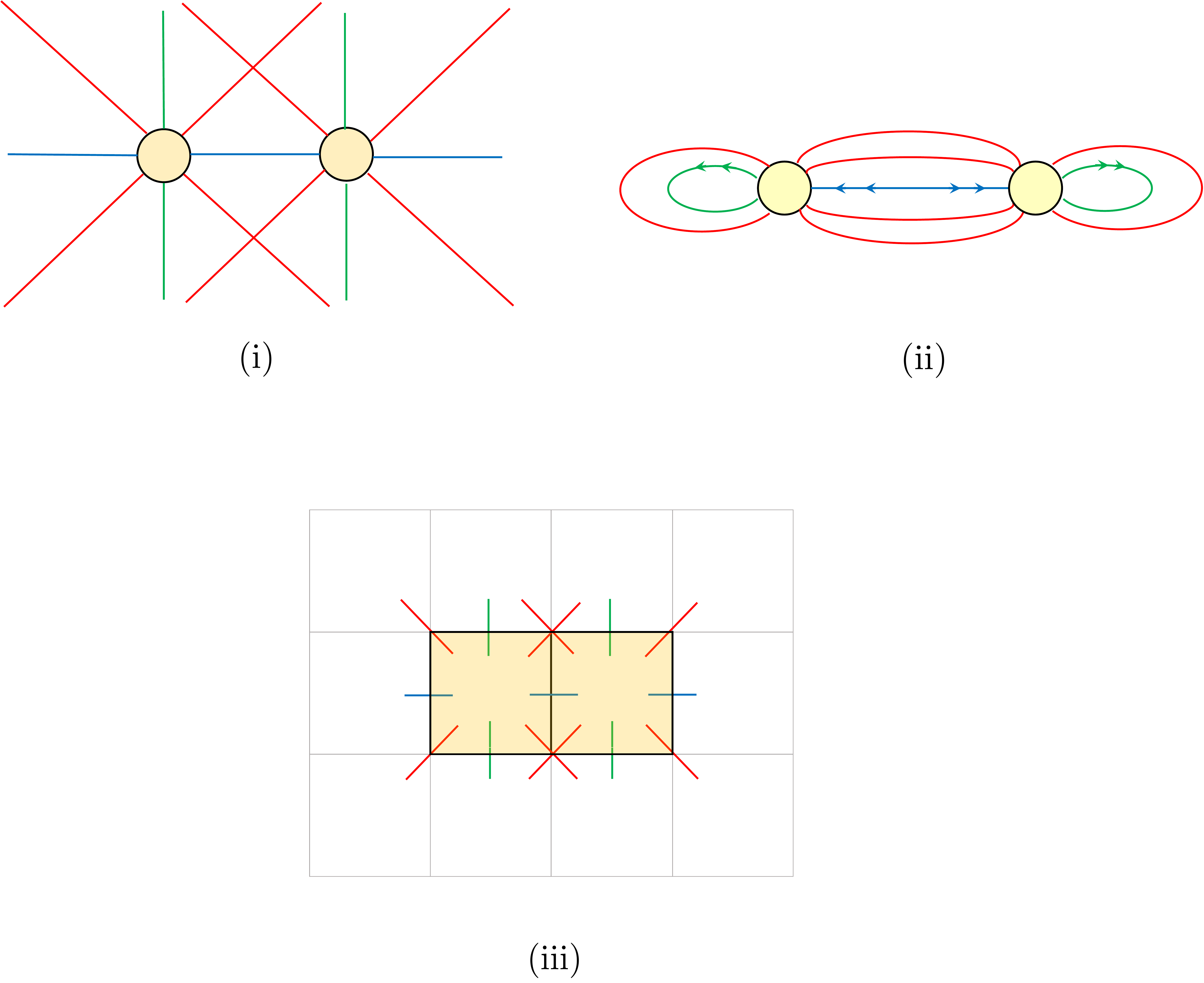}
\caption{
(i) $\mathcal{N}=(0,4)$ quiver diagram for D1-branes on $\mathbb{C}^{2}/\mathbb{Z}_{2}\times \mathbb{C}^{2}$. 
(ii) The corresponding $\mathcal{N}=(0,2)$ quiver diagram. 
(iii) D3 brane box configuration which is T-dual to D1-branes on $\mathbb{C}^{2}/\mathbb{Z}_{2}\times \mathbb{C}^{2}$. }
\label{fig44quiver}
\end{center}
\end{figure}

The theory can be encoded by $\mathcal{N}=(0,4)$ quiver 
or $\mathcal{N}=(0,2)$ quiver diagram shown in Figure \ref{fig44quiver}. 
The orange node corresponds to $\mathcal{N}=(0,4)$ vector multiplet for each factor of gauge symmetry groups. 
The blue lines describe bi-fundametnal $\mathcal{N}=(0,4)$ hypermultiplets and 
the blue arrows represent $R$ and $L$ forming these $\mathcal{N}=(0,4)$ hypermultiplets. 
The green loops represent $\mathcal{N}=(0,4)$ twisted hypermultiplet consisting of 
pairs of $\mathcal{N}=(0,2)$ chiral multiplets $U$ and $D$. 
Four red edges between two nodes describe $\Delta$ and $\nabla$ 
while two red loops are adjoint $\Lambda$.

In this case $\mathcal{N}=(0,4)$ supersymmetry is enhanced to $\mathcal{N}=(4,4)$ 
in such a way that  bi-fundamental $\mathcal{N}=(0,4)$ hypermultiplets 
$(R,L)$ and bi-fundamental Fermi multiplets $(U, D)$ further combine into $\mathcal{N}=(4,4)$ hypermultiplets 
and adjoint $\mathcal{N}=(0,4)$ twisted hypermultiplets $(U,D)$ and 
$\mathcal{N}=(0,4)$ vector multiplets are promoted to $\mathcal{N}=(4,4)$ vector multiplets. 

The T-dual brane box configuration is shown in Figure \ref{fig44quiver}. 
This is obtained by the dimensional reduction of the 3d $\mathcal{N}=4$ quiver gauge theory 
which has the same gauge group and bi-fundamental hypermultiplets.

\subsubsection{$\mathbb{C}^{2}/\mathbb{Z}_{2}\times \mathbb{C}^{2}/\mathbb{Z}_{2}$ $(1,1,0,0)$, $(0,0,1,1)$}
\label{subsec_d1zkzk}
\begin{figure}
\begin{center}
\includegraphics[width=12.5cm]{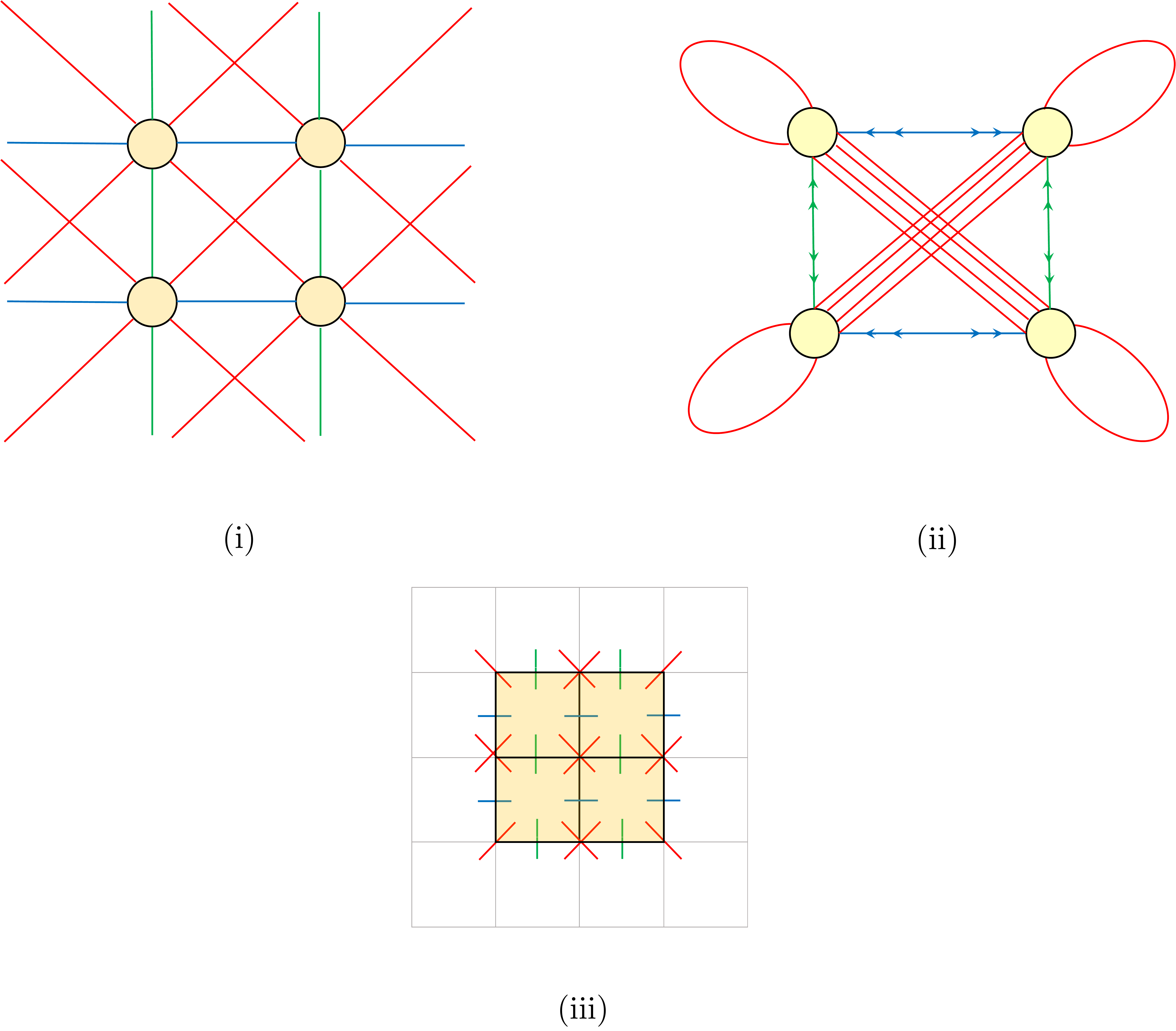}
\caption{
(i) $\mathcal{N}=(0,4)$ quiver diagram for D1-branes on $\mathbb{C}^{2}/\mathbb{Z}_{2}\times \mathbb{C}^{2}/\mathbb{Z}_{2}$. 
(ii) The corresponding $\mathcal{N}=(0,2)$ quiver diagram. 
(iii) D3-brane box configuration which is T-dual to D1-branes on 
$\mathbb{C}^{2}/\mathbb{Z}_{2}$ $\times$ $\mathbb{C}^{2}/\mathbb{Z}_{2}$. 
Horizontal, vertical and diagonal edges represent hyper, twisted hyper and Fermi multiplets. 
}
\label{figz2z2}
\end{center}
\end{figure}

There are two generators of $\mathbb{C}^{2}/\mathbb{Z}_{2}$ 
$\times$ $\mathbb{C}^{2}/\mathbb{Z}_{2}$ singularity. 
One of them labeled by $(1,1,0,0)$ acts on the two complex coordinates as 
$z_{1}$ $\mapsto$ $e^{\pi i}z_{1}$, $z_{2}$ $\mapsto$ $e^{\pi i}z_{2}$ 
while the other labeled by $(0,0,1,1)$ acts on the others as
$z_{3}$ $\mapsto$ $e^{\pi i}z_{3}$, $z_{4}$ $\mapsto$ $e^{\pi} z_{4}$. 
The gauge group of the world-volume theory of D1-branes at $\mathbb{C}^{2}/\mathbb{Z}_{2}$ $\times$ 
$\mathbb{C}^{2}/\mathbb{Z}_{2}$ is given by 
\begin{align}
\label{z2z2_G}
G&=\prod_{i_{1}=1}^{2}\prod_{i_{2}=1}^{2}U(N)_{i_{1},i_{2}}
\end{align}
The matter content consists of sixteen $\mathcal{N}=(0,2)$ chiral multiplets
\begin{align}
\label{z2z2content}
\left\{
{R^{i}}_{i+(1,0)}, {L^{i}}_{i+(-1,0)}, {U^{i}}_{i+(0,1)}, {D^{i}}_{i+(0,-1)}
\right\}
\end{align}
and twelve $\mathcal{N}=(0,2)$ Fermi multiplets 
\begin{align}
\label{z2z2content}
\left\{
\Delta_{i,i-(1,1)}, \nabla_{i,i-(1,1)}, \Lambda_{i,i}
\right\}. 
\end{align}
where $i=(i_{1},i_{2})$ are pairs of modulo 2 gauge indices with $i_{1}=1,2$ and $i_{2}=1,2$. 
The quiver diagram is shown in Figure \ref{figz2z2}. 

The blue and green arrows describe bi-fundamental $\mathcal{N}=(0,4)$ hyper and twisted hypermultiplets respectively. 
The red lines correspond to bi-fundamental Fermi multiplets 
while the red loop $\mathcal{N}=(0,2)$ Fermi multiplets combine 
with $\mathcal{N}=(0,2)$ gauge multiplets into $\mathcal{N}=(0,4)$ vector multiplets. 

There are twelve pairs of vanishing $J$- and $E$-terms
\begin{align}
\label{z2z2_JE}
\begin{array}{lcc}
&J&E \\
\Delta_{(1,1)(2,2)}:&
{L^{(2,2)}}_{(1,2)}{U^{(1,2)}}_{(1,1)}-{U^{(2,2)}}_{(2,1)}{L^{(2,1)}}_{(1,1)}=0&
{D^{(1,1)}}_{(1,2)}{R^{(1,2)}}_{(2,2)}-{R^{(1,1)}}_{(2,1)}{D^{(2,1)}}_{(2,2)}=0\\
\Delta_{(1,2)(2,1)}:&
{L^{(2,1)}}_{(1,1)}{U^{(1,1)}}_{(1,2)}-{U^{(2,1)}}_{(2,2)}{L^{(2,2)}}_{(1,2)}=0&
{D^{(1,2)}}_{(1,1)}{R^{(1,1)}}_{(2,1)}-{R^{(1,2)}}_{(2,2)}{D^{(2,2)}}_{(2,1)}=0\\
\Delta_{(2,1)(1,2)}:&
{L^{(1,2)}}_{(2,1)}{U^{(2,2)}}_{(2,1)}-{U^{(1,2)}}_{(1,1)}{L^{(1,1)}}_{(2,1)}=0&
{D^{(2,1)}}_{(2,2)}{R^{(2,2)}}_{(1,2)}-{R^{(2,1)}}_{(1,1)}{D^{(1,1)}}_{(1,2)}=0\\
\Delta_{(2,2)(1,1)}:&
{L^{(2,2)}}_{(1,2)}{U^{(1,2)}}_{(1,1)}-{U^{(1,1)}}_{(1,2)}{L^{(1,2)}}_{(2,2)}=0&
{D^{(2,2)}}_{(2,1)}{R^{(2,1)}}_{(1,1)}-{R^{(2,2)}}_{(1,2)}{D^{(1,2)}}_{(1,1)}=0\\
\nabla_{(1,1)(2,2)}:&
{U^{(2,2)}}_{(2,1)}{R^{(2,1)}}_{(1,1)}-{R^{(2,2)}}_{(1,2)}{U^{(1,2)}}_{(1,1)}=0&
{D^{(1,1)}}_{(1,2)}{L^{(1,2)}}_{(2,2)}-{L^{(1,1)}}_{(2,1)}{D^{(2,1)}}_{(2,2)}=0\\
\nabla_{(1,2)(2,1)}:&
{U^{(2,1)}}_{(2,2)}{R^{(2,2)}}_{(1,2)}-{R^{(2,1)}}_{(1,1)}{U^{(1,1)}}_{(1,2)}=0& 
{D^{(1,2)}}_{(1,1)}{L^{(1,1)}}_{(2,1)}-{L^{(1,2)}}_{(2,2)}{D^{(2,2)}}_{(2,1)}=0\\
\nabla_{(2,1)(1,2)}:&
{U^{(1,2)}}_{(1,1)}{R^{(1,1)}}_{(2,1)}-{R^{(1,2)}}_{(2,2)}{U^{(2,2)}}_{(2,1)}=0&
{D^{(2,1)}}_{(2,2)}{L^{(2,2)}}_{(1,2)}-{L^{(2,1)}}_{(1,1)}{D^{(1,1)}}_{(1,2)}=0\\
\nabla_{(2,2)(1,1)}:&
{U^{(1,1)}}_{(1,2)}{R^{(1,2)}}_{(2,2)}-{R^{(1,1)}}_{(2,1)}{U^{(2,1)}}_{(2,2)}=0&
{D^{(2,2)}}_{(2,1)}{L^{(2,1)}}_{(1,1)}-{L^{(2,2)}}_{(1,2)}{D^{(1,2)}}_{(1,1)}=0\\
\Lambda_{(1,1)(1,1)}:&
{R^{(1,1)}}_{(2,1)}{L^{(2,1)}}_{(1,1)}-{L^{(1,1)}}_{(2,1)}{R^{(2,1)}}_{(1,2)}=0&
{D^{(1,1)}}_{(1,2)}{U^{(1,2)}}_{(1,1)}-{U^{(1,1)}}_{(1,2)}{D^{(1,2)}}_{(1,1)}=0\\
\Lambda_{(1,2)(1,2)}:&
{R^{(1,2)}}_{(2,2)}{L^{(2,2)}}_{(2,1)}-{L^{(1,2)}}_{(2,2)}{R^{(2,2)}}_{(1,2)}=0&
{D^{(1,2)}}_{(1,1)}{U^{(1,1)}}_{(1,2)}-{U^{(1,2)}}_{(1,1)}{D^{(1,1)}}_{(1,2)}=0\\
\Lambda_{(2,1)(2,1)}:&
{R^{(2,1)}}_{(1,1)}{L^{(1,1)}}_{(2,1)}-{L^{(2,1)}}_{(1,1)}{R^{(1,1)}}_{(2,1)}=0&
{D^{(2,1)}}_{(2,2)}{U^{(2,2)}}_{(2,1)}-{U^{(2,1)}}_{(2,2)}{D^{(2,2)}}_{(2,1)}=0\\
\Lambda_{(2,2)(2,2)}:&
{R^{(2,2)}}_{(1,1)}{L^{(1,1)}}_{(2,2)}-{L^{(2,2)}}_{(1,1)}{R^{(1,2)}}_{(2,2)}=0&
{D^{(2,2)}}_{(2,1)}{U^{(2,1)}}_{(2,2)}-{U^{(2,2)}}_{(2,1)}{D^{(2,1)}}_{(2,2)}=0\\
\end{array}
\end{align}

The T-dual configuration is $2\times 2$ brane box model illustrated in Figure \ref{figz2z2}. 
In this case, two NS5-branes and two NS5$'$-branes are 
trivially identified going around $x^{6}$ and $x^{2}$ directions respectively.

\subsubsection{$\mathbb{C}^{2}/\mathbb{Z}_{2}\times \mathbb{C}^{2}/\mathbb{Z}_{3}$ $(1,1,0,0)$, $(0,0,1,2)$}
\label{sec_z2z3}
The $\mathcal{N}=(0,4)$ supersymmetric quiver gauge theory of 
regular $N$ D1-branes on $\mathbb{C}^{2}/\mathbb{Z}_{2}$ $\times$ $\mathbb{C}^{2}/\mathbb{Z}_{3}$ 
has gauge group 
\begin{align}
\label{z2z3_G}
G&=\prod_{i_{1}=1}^{2}\prod_{i_{2}=1}^{3}
U(N)_{i_{1}, i_{2}}. 
\end{align}
The matter content consists of 24 chiral multiplets
\begin{align}
\label{z2z3content1}
\left\{
{R^{i}}_{i+(1,0)}, {L^{i}}_{i+(-1,0)}, {U^{i}}_{i+(0,1)}, {D^{i}}_{i+(0,-1)}
\right\}
\end{align}
and 18 Fermi multiplets 
\begin{align}
\label{z2z2content}
\left\{
\Delta_{i,i-(1,1)}, \nabla_{i,i-(1,1)}, \Lambda_{i,i}
\right\}. 
\end{align}
where $i=(i_{1},i_{2})$ are pairs of gauge indices with $i_{1}=1,2$ and $i_{2}=1,2,3$.

The vanishing $E$- and $J$ terms are given by
\begin{align}
\label{z2z3_JE}
\begin{array}{lcc}
&J&E \\
\Delta_{(1,1)(2,3)}:&
{L^{(2,3)}}_{(1,3)}{U^{(1,3)}}_{(1,1)}-{U^{(2,3)}}_{(2,1)}{L^{(2,1)}}_{(1,1)}=0&
{D^{(1,1)}}_{(1,3)}{R^{(1,3)}}_{(2,3)}-{R^{(1,1)}}_{(2,1)}{D^{(2,1)}}_{(2,3)}=0\\
\Delta_{(1,2)(2,1)}:&
{L^{(2,1)}}_{(1,1)}{U^{(1,1)}}_{(1,2)}-{U^{(2,1)}}_{(2,2)}{L^{(2,2)}}_{(1,2)}=0&
{D^{(1,2)}}_{(1,1)}{R^{(1,1)}}_{(2,1)}-{R^{(1,2)}}_{(2,2)}{D^{(2,2)}}_{(2,1)}=0\\
\Delta_{(1,3)(2,2)}:&
{L^{(2,2)}}_{(1,2)}{U^{(1,2)}}_{(1,3)}-{U^{(2,2)}}_{(2,3)}{L^{(2,3)}}_{(1,3)}=0&
{D^{(1,3)}}_{(1,2)}{R^{(1,2)}}_{(2,2)}-{R^{(1,3)}}_{(2,3)}{D^{(2,3)}}_{(2,2)}=0\\
\Delta_{(2,1)(1,3)}:&
{L^{(1,3)}}_{(2,3)}{U^{(2,3)}}_{(2,1)}-{U^{(1,3)}}_{(1,1)}{L^{(1,1)}}_{(2,1)}=0&
{D^{(2,1)}}_{(2,3)}{R^{(2,3)}}_{(1,3)}-{R^{(2,1)}}_{(1,1)}{D^{(1,1)}}_{(1,3)}=0\\
\Delta_{(2,2)(1,1)}:&
{L^{(1,1)}}_{(2,1)}{U^{(2,1)}}_{(2,2)}-{U^{(1,1)}}_{(1,2)}{L^{(1,2)}}_{(2,2)}=0&
{D^{(2,2)}}_{(2,1)}{R^{(2,1)}}_{(1,1)}-{R^{(2,2)}}_{(1,2)}{D^{(1,2)}}_{(1,1)}=0\\
\Delta_{(2,3)(1,2)}:&
{L^{(1,2)}}_{(2,2)}{U^{(2,2)}}_{(2,3)}-{U^{(1,2)}}_{(1,3)}{L^{(1,3)}}_{(2,3)}=0&
{D^{(2,3)}}_{(2,2)}{R^{(2,2)}}_{(1,2)}-{R^{(2,3)}}_{(1,3)}{D^{(1,3)}}_{(1,2)}=0\\
\nabla_{(1,1)(2,3)}:&
{U^{(2,3)}}_{(2,1)}{R^{(2,1)}}_{(1,1)}-{R^{(2,3)}}_{(1,3)}{U^{(1,3)}}_{(1,1)}=0&
{D^{(1,1)}}_{(1,3)}{L^{(1,3)}}_{(2,3)}-{L^{(1,1)}}_{(2,1)}{D^{(2,1)}}_{(2,3)}=0\\
\nabla_{(1,2)(2,1)}:&
{U^{(2,1)}}_{(2,2)}{R^{(2,2)}}_{(1,2)}-{R^{(2,1)}}_{(1,1)}{U^{(1,1)}}_{(1,2)}=0& 
{D^{(1,2)}}_{(1,1)}{L^{(1,1)}}_{(2,1)}-{L^{(1,2)}}_{(2,2)}{D^{(2,2)}}_{(2,1)}=0\\
\nabla_{(1,3)(2,2)}:&
{U^{(2,2)}}_{(2,3)}{R^{(2,3)}}_{(1,3)}-{R^{(2,2)}}_{(1,2)}{U^{(1,2)}}_{(1,3)}=0& 
{D^{(1,3)}}_{(1,2)}{L^{(1,2)}}_{(2,2)}-{L^{(1,3)}}_{(2,3)}{D^{(2,3)}}_{(2,2)}=0\\
\nabla_{(2,1)(1,2)}:&
{U^{(1,2)}}_{(1,1)}{R^{(1,1)}}_{(2,1)}-{R^{(1,2)}}_{(2,2)}{U^{(2,2)}}_{(2,1)}=0&
{D^{(2,1)}}_{(2,2)}{L^{(2,2)}}_{(1,2)}-{L^{(2,1)}}_{(1,1)}{D^{(1,1)}}_{(1,2)}=0\\
\nabla_{(2,2)(1,1)}:&
{U^{(1,1)}}_{(1,2)}{R^{(1,2)}}_{(2,2)}-{R^{(1,1)}}_{(2,1)}{U^{(2,1)}}_{(2,2)}=0&
{D^{(2,2)}}_{(2,1)}{L^{(2,1)}}_{(1,1)}-{L^{(2,2)}}_{(1,2)}{D^{(1,2)}}_{(1,1)}=0\\
\nabla_{(2,3)(1,2)}:&
{U^{(1,2)}}_{(1,3)}{R^{(1,3)}}_{(2,3)}-{R^{(1,2)}}_{(2,2)}{U^{(2,2)}}_{(2,3)}=0& 
{D^{(2,3)}}_{(2,2)}{L^{(2,2)}}_{(1,2)}-{L^{(2,3)}}_{(1,3)}{D^{(1,3)}}_{(1,2)}=0\\
\Lambda_{(1,1)(1,1)}:&
{R^{(1,1)}}_{(2,1)}{L^{(2,1)}}_{(1,1)}-{L^{(1,1)}}_{(2,1)}{R^{(2,1)}}_{(1,2)}=0&
{D^{(1,1)}}_{(1,2)}{U^{(1,2)}}_{(1,1)}-{U^{(1,1)}}_{(1,2)}{D^{(1,2)}}_{(1,1)}=0\\
\Lambda_{(1,2)(1,2)}:&
{R^{(1,2)}}_{(2,2)}{L^{(2,2)}}_{(2,1)}-{L^{(1,2)}}_{(2,2)}{R^{(2,2)}}_{(1,2)}=0&
{D^{(1,2)}}_{(1,1)}{U^{(1,1)}}_{(1,2)}-{U^{(1,2)}}_{(1,1)}{D^{(1,1)}}_{(1,2)}=0\\
\Lambda_{(1,3)(1,3)}:&
{R^{(1,3)}}_{(2,3)}{L^{(2,3)}}_{(1,3)}-{L^{(1,3)}}_{(2,3)}{R^{(2,3)}}_{(1,3)}=0&
{D^{(1,3)}}_{(1,2)}{U^{(1,2)}}_{(1,3)}-{U^{(1,3)}}_{(1,1)}{D^{(1,1)}}_{(1,3)}=0\\
\Lambda_{(2,1)(2,1)}:&
{R^{(2,1)}}_{(1,1)}{L^{(1,1)}}_{(2,1)}-{L^{(2,1)}}_{(1,1)}{R^{(1,1)}}_{(2,1)}=0&
{D^{(2,1)}}_{(2,2)}{U^{(2,2)}}_{(2,1)}-{U^{(2,1)}}_{(2,2)}{D^{(2,2)}}_{(2,1)}=0\\
\Lambda_{(2,2)(2,2)}:&
{R^{(2,2)}}_{(1,1)}{L^{(1,1)}}_{(2,2)}-{L^{(2,2)}}_{(1,1)}{R^{(1,2)}}_{(2,2)}=0&
{D^{(2,2)}}_{(2,1)}{U^{(2,1)}}_{(2,2)}-{U^{(2,2)}}_{(2,1)}{D^{(2,1)}}_{(2,2)}=0\\
\Lambda_{(2,3)(2,3)}:&
{R^{(2,3)}}_{(1,3)}{L^{(1,3)}}_{(2,3)}-{L^{(2,3)}}_{(1,3)}{R^{(1,3)}}_{(2,3)}=0&
{D^{(2,3)}}_{(2,2)}{U^{(2,2)}}_{(2,3)}-{U^{(2,3)}}_{(2,1)}{D^{(2,1)}}_{(2,3)}=0\\
\end{array}
\end{align}

\begin{figure}
\begin{center}
\includegraphics[width=14.5cm]{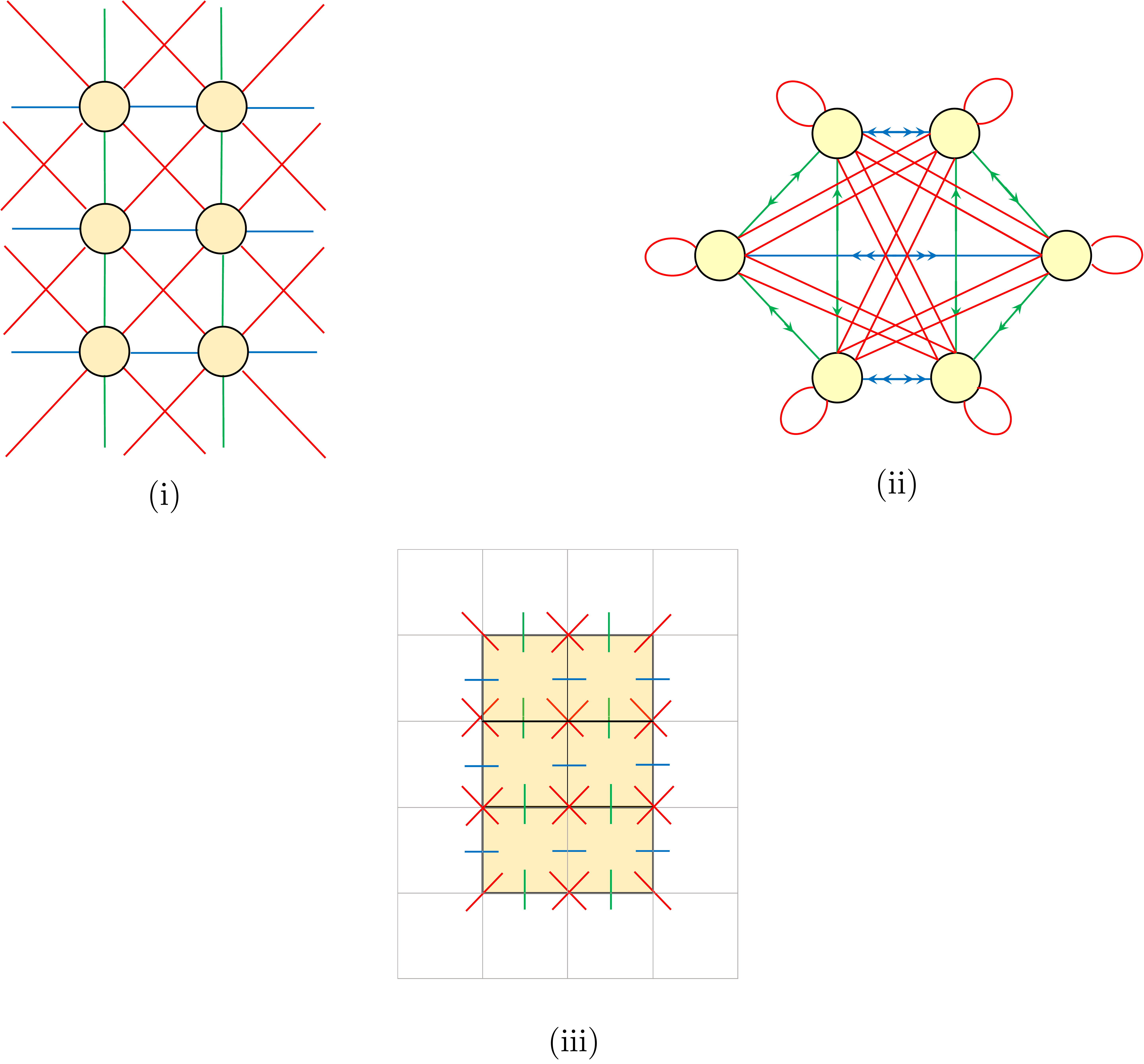}
\caption{
(i) $\mathcal{N}=(0,4)$ quiver diagram for D1-branes on $\mathbb{C}^{2}/\mathbb{Z}_{2}\times \mathbb{C}^{2}/\mathbb{Z}_{3}$. 
(ii) The corresponding $\mathcal{N}=(0,2)$ quiver diagram. 
(iii) D3-brane box configuration which is T-dual to D1-branes on 
$\mathbb{C}^{2}/\mathbb{Z}_{2}$ $\times$ $\mathbb{C}^{2}/\mathbb{Z}_{3}$. }
\label{figz2z3}
\end{center}
\end{figure}

The quiver diagram is shown in Figure \ref{figz2z3}. 
The T-dual brane box model is $2\times 3$ D3-brane box model, 
as depicted in Figure \ref{figz2z3}.

\subsubsection{$\mathbb{C}^{2}/\mathbb{Z}_{3}\times \mathbb{C}^{2}/\mathbb{Z}_{3}$ $(1,2,0,0)$, $(0,0,1,2)$}
\label{sec_z3z3}
\begin{figure}
\begin{center}
\includegraphics[width=15.5cm]{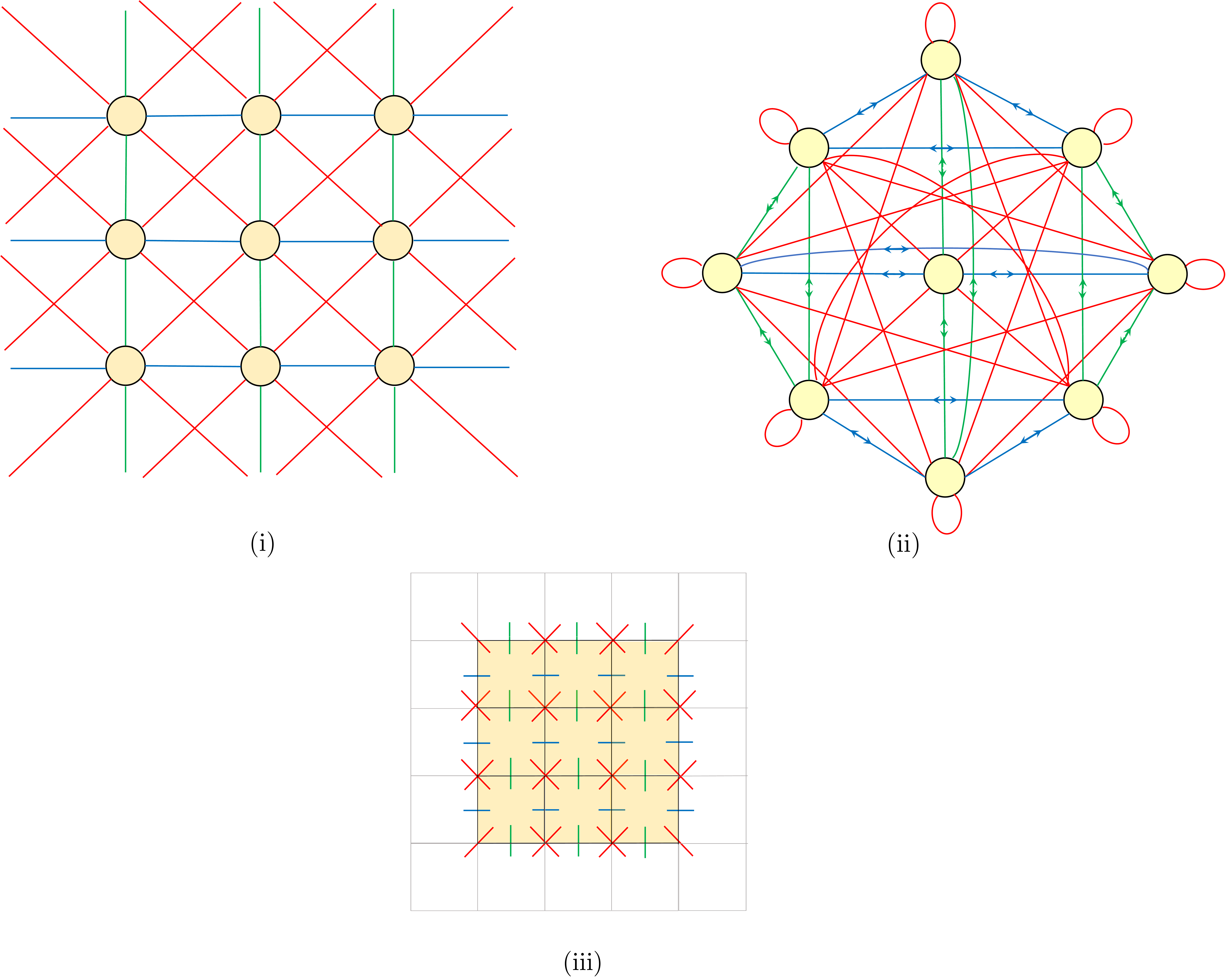}
\caption{
(i) $\mathcal{N}=(0,4)$ quiver diagram for D1-branes on $\mathbb{C}^{2}/\mathbb{Z}_{3}\times \mathbb{C}^{2}/\mathbb{Z}_{3}$. 
(ii) The corresponding $\mathcal{N}=(0,2)$ quiver diagram. 
(iii) $3\times 3$ D3-brane box configuration which is T-dual to D1-branes on 
$\mathbb{C}^{2}/\mathbb{Z}_{3}$ $\times$ $\mathbb{C}^{2}/\mathbb{Z}_{3}$. }
\label{figz3z3}
\end{center}
\end{figure}

In this case gauge group is
\begin{align}
\label{z3z3_G}
G&=\prod_{i_{1}=1}^{3}
\prod_{i_{2}=1}^{3}U(N)_{i_{1}, i_{2}}
\end{align}
and there are 36 chiral multiplets and 27 Fermi multiplets. 
The quiver diagram is shown in Figure \ref{figz3z3}. 
Note that unlike the above examples, 
the Fermi multiplets cannot be drawn as a bi-fundamental single line 
since each of $\mathcal{N}=(0,2)$ Fermi multiplets connects to distinct four gauge nodes. 
This always happens when both $k$ and $k'$ are larger than two. 
Therefore the Fermi multiplets generically take the V-shaped or X-shaped configuration.

The T-dual configuration is $3\times 3$ boxes of D3-branes shown in Figure \ref{figz3z3}.

\subsection{D1-D5-D5$'$ branes}
\label{sec_d1d5d5'}
Now consider the dynamics of $N$ D1-branes intersecting with $N_{f}$ D5-branes and $N_{f}'$ D5$'$-branes. 
The space-time symmetry is $SO(1,1)_{01}$ $\times$ $SO(4)_{2'345}$ $\times$ $SO(4)_{6'789}$. 
The low-energy effective world-volume theory is 
$\mathcal{N}=(0,4)$ supersymmetric $U(N)$ gauge theory with $SU(N_{f})$ $\times$ $SU(N_{f}')$ global symmetry. 
In addition to the massless modes from D1-D1 string, 
there are three sets of multiplets appear from D1-D5, D1-D5$'$ and D5-D5$'$ strings. 
This brane configuration is studied in \cite{Tong:2014yna}. 
It is shown that the $\mathcal{N}=(0,4)$ gauge theory flows to the conformal field theory with the central charge 
\begin{align}
\label{d1d5d5_cc}
c&=6N\frac{N_{f}N_{f}'}{N_{f}+N_{f}'}
\end{align}
which is evaluated from the near horizon geometry \cite{Gukov:2004ym} of 
$ AdS_{3}\times S^{3}\times S^{3}\times \mathbb{R}$ \cite{Cowdall:1998bu, Boonstra:1998yu, Gauntlett:1998kc}. 
Let us firstly give a brief review of the $\mathcal{N}=(0,4)$ gauge theory 
that arises from the intersection of D1-D5-D5$'$ brane system.

\paragraph{D1-D1 strings}
As discussed in the previous section, 
the world-volume theory of $N$ D1-branes is $\mathcal{N}=(8,8)$ supersymmetric $U(N)$ gauge theory. 
The theory contains a $\mathcal{N}=(0,2)$ gauge multiplet, 
four $\mathcal{N}=(0,2)$ adjoint chiral multiplets and three $\mathcal{N}=(0,2)$ adjoint Fermi multiplets. 
The matter content is summarized as 
\begin{align}
\label{d1d5b_modes}
\begin{array}{c|c}
\textrm{$\mathcal{N}=(0,4)$ supermultiplets}& SU(2)_{C}\times SU(2)_{C}'\times SU(2)_{H}\times SU(2)_{H}' \\ \hline 
\textrm{adjoint hypermultiplet}:\quad ({R^{i}}_{i}, {L^{i}}_{i}) 
&\textrm{bosons}: (\bm{1}, \bm{1},\bm{2},\bm{2}) \\ 
&\textrm{fermions}: (\bm{2},\bm{1},\bm{1},\bm{2}) \\ \hline
\textrm{adjoint twisted hypermultiplet}:\quad ({U^{i}}_{i}, {D^{i}}_{i})
&\textrm{bosons}: (\bm{2}, \bm{2},\bm{1},\bm{1}) \\ 
&\textrm{fermions}: (\bm{1},\bm{2},\bm{2},\bm{1}) \\ \hline
\textrm{adjoint Fermi multiplets}:\quad (\Delta_{ii}, \nabla_{ii}) 
&\textrm{fermions}: (\bm{1},\bm{2},\bm{1},\bm{2}) \\ 
\end{array}
\end{align}
There is also the $\mathcal{N}=(0,4)$ $U(N)$ vector multiplet 
consisting of the $\mathcal{N}=(0,2)$ gauge multiplet and the $\mathcal{N}=(0,2)$ adjoint Fermi multiplet. 
As in the brane setup (\ref{d1d5d5_ale}), 
the hypermultiplet scalar fields describe the dynamics of D1-branes 
along $6'7898$ in which D5-branes extend, 
while the twisted hypermultiplet scalar fields 
capture the motions of D1-branes along $2'345$ in which D5$'$-branes span. 
The $E$- and $J$-terms are given by (\ref{je_88}).

\paragraph{D1-D5 strings}
The massless modes coming from the open strings stretched between 
D1-branes and D5-branes give rise to $\mathcal{N}=(4,4)$ hypermultiplets. 
In our brane construction (\ref{d1d5d5_ale}), 
they decompose into $\mathcal{N}=(0,4)$ hypermultiplets and $\mathcal{N}=(0,4)$ Fermi multiplets 
transforming as follows: 
\begin{align}
\label{d1d5a_modes}
\begin{array}{c|c}
\textrm{$\mathcal{N}=(0,4)$ supermultiplets}& SU(2)_{C}\times SU(2)_{C}'\times SU(2)_{H}\times SU(2)_{H}' \\ \hline 
\textrm{fundamental hypermultiplets}:\quad (H^{i}_{a}, \widetilde{H}_{i}^{a})
&\textrm{bosons}: (\bm{1}, \bm{1},\bm{2},\bm{1}) \\
&\textrm{fermions}: (\bm{2}, \bm{1},\bm{1},\bm{1}) \\ \hline 
\textrm{fundamental Fermi multiplets}:\quad (\xi^{i}_{a},\widetilde{\xi}_{i}^{a})
&\textrm{fermions}: (\bm{1}, \bm{2},\bm{1},\bm{1}) \\
\end{array}
\end{align}
Here $H^{i}_{a}$, $\widetilde{H}_{i}^{a}$ are $\mathcal{N}=(0,2)$ chiral multiplets 
and $\xi^{i}_{a}$, $\widetilde{\xi}_{i}^{a}$ are $\mathcal{N}=(0,2)$ Fermi multiplets 
where $i=1,\cdots, N$ are gauge indices and $a=1,\cdots, N_{f}$ are flavor indices.

Each of the $\mathcal{N}=(0,4)$ supermultiplets transforms 
as the fundamental representation under the $U(N)$ gauge group 
and transform as the anti-fundamental representation under the $SU(N_{f})$ global symmetry. 
The fundamental hypermultiplets $(H,\widetilde{H})$ couple to the $U(N)$ vector multiplets through $J$-term 
\begin{align}
\label{j_d1d5}
\begin{array}{cc}
&J\\
{\Lambda^{i}}_{i}:& \qquad \widetilde{H}_{i}^{a}\cdot H^{i}_{a}=0\\
\end{array}
\end{align}
associated to the adjoint $\mathcal{N}=(0,2)$ Fermi multiplet $\Lambda_{ii}$ in the $\mathcal{N}=(0,4)$ vector multiplet.

The fundamental hypermultiplet $(H,\widetilde{H})$ can also couple to 
the adjoint twisted hypermultiplet $(U,D)$ appearing in the D1-D1 string spectrum (\ref{d1d5b_modes}) through 
the $E$- and $J$-terms associated to the Fermi multiplet $(\xi, \widetilde{\xi})$ 
\begin{align}
\label{je_d1d5}
\begin{array}{lcc}
&J&E \\
\xi^{i}_{a}:&
\widetilde{H}_{i}^{a}\cdot {D^{i}}_{i}=0&
{U^{i}}_{i}\cdot H^{i}_{a}=0\\
\widetilde{\xi}_{i}^{a}:&
 {D^{i}}_{i}\cdot H^{i}_{a}=0&
-\widetilde{H}_{i}^{a}\cdot {U^{i}}_{i}=0\\
\end{array}. 
\end{align}
These interactions reflect the fact that 
D1-D5 strings become massive when the D1-branes move along 
the directions parallel to D5$'$-branes.

\paragraph{D1-D5$'$ strings}
Similarly to the D1-D5 strings, the massless modes from the open strings stretched between 
D1-branes and D5$'$-branes give $\mathcal{N}=(4,4)$ hypermultiplets. 
They transform under the space-time symmetry 
so that they decompose into $\mathcal{N}=(0,4)$ twisted hypermultiplets 
$(T,\widetilde{T})$ and $\mathcal{N}=(0,4)$ Fermi multiplets $(\zeta,\widetilde{\zeta})$: 
\begin{align}
\label{d1d5b_modes}
\begin{array}{c|c}
\textrm{$\mathcal{N}=(0,4)$ supermultiplets}& SU(2)_{C}\times SU(2)_{C}'\times SU(2)_{H}\times SU(2)_{H}' \\ \hline 
\textrm{fundamental twisted hypermultiplets}:\quad (T^{i}_{\tilde{a}}, \widetilde{T}_{i}^{\tilde{a}})
&\textrm{bosons}: (\bm{2}, \bm{1},\bm{1},\bm{1}) \\
&\textrm{fermions}: (\bm{1}, \bm{1},\bm{2},\bm{1}) \\ \hline 
\textrm{fundamental Fermi multiplets}:\quad (\zeta^{i}_{\tilde{a}}, \widetilde{\zeta}_{i}^{\tilde{a}})
&\textrm{fermions}: (\bm{1}, \bm{1},\bm{1},\bm{2}) \\
\end{array}
\end{align}
where $T^{i}_{\tilde{a}}$, $\widetilde{T}_{i}^{\tilde{a}}$ are $\mathcal{N}=(0,2)$ chiral multiplets 
and $\zeta^{i}_{\tilde{a}}$, $\widetilde{\zeta}_{i}^{\tilde{a}}$ are $\mathcal{N}=(0,2)$ Fermi multiplets 
with $\tilde{a}=1,\cdots, N_{f}'$ being flavor indices.

They transform as the fundamental representations under the $U(N)$ gauge group. 
Under the $SU(N_{f}')$ global symmetry, 
they transform as the anti-fundamental representation. 
The $\mathcal{N}=(0,4)$ twisted hypermultiplet is coupled to $U(N)$ vector multiplet through 
$E$-term 
\begin{align}
\label{e_d1d5'}
\begin{array}{cc}
&E\\
{\Lambda^{i}}_{i}:& \qquad T^{i}_{a}\cdot \widetilde{T}_{i}^{a}=0\\
\end{array}
\end{align}

They also couple to the adjoint hypermultiplets $(R,L)$ in (\ref{d1d5a_modes}) 
through the $E$- and $J$-term potentials for 
the Fermi multiplets $(\zeta, \widetilde{\zeta})$
\begin{align}
\label{je_d1d5'}
\begin{array}{lcc}
&J&E \\
\zeta^{i}_{\tilde{a}}:&
\widetilde{T}_{i}^{\tilde{a}}\cdot {L^{i}}_{i}=0&
{R^{i}}_{i}\cdot T^{i}_{\tilde{a}}=0\\
\widetilde{\zeta}_{i}^{\tilde{a}}:&
{L^{i}}_{i}\cdot T^{i}_{\tilde{a}}=0&
-\widetilde{T}_{i}^{\tilde{a}}\cdot {R^{i}}_{i}=0\\
\end{array}. 
\end{align}
The presence of these interactions implies that 
the D1-D5$'$ strings become massive 
when the D1-branes move in the directions where the D5-branes span.

\paragraph{D5-D5$'$ strings}
The quantization of open strings stretched between D5- and D5$'$-branes leads to 
the $\mathcal{N}=(0,8)$ Fermi multiplets, as a pair of $\mathcal{N}=(0,4)$ Fermi multiplets 
\footnote{This is T-dual configuration of D0-D8 brane system \cite{Banks:1997zs, Bachas:1997kn}.}.
\begin{align}
\label{d5d5_modes}
\begin{array}{c|c}
\textrm{$\mathcal{N}=(0,4)$ supermultiplets}& SU(2)_{C}\times SU(2)_{C}'\times SU(2)_{H}\times SU(2)_{H}' \\ \hline 
\textrm{neutral Fermi multiplets}:\quad (\gamma^{a}_{\tilde{a}}, 
\widetilde{\gamma}_{a}^{\tilde{a}})&(\bm{1}, \bm{1},\bm{1},\bm{1}) \\
\textrm{neutral Fermi multiplets}:\quad (\gamma_{a}^{'\tilde{a}}, 
\widetilde{\gamma}^{'a}_{\tilde{a}})&(\bm{1}, \bm{1},\bm{1},\bm{1}) \\
\end{array}
\end{align}
The $\mathcal{N}=(0,4)$ Fermi multiplets are neutral under the gauge group. 
So they have no interaction with the $\mathcal{N}=(0,4)$ vector multiplet. 
Meanwhile, they transform as bi-fundamental representation under the $SU(N_{f})\times SU(N_{f}')$ global symmetry.

Without any $E$- and $J$-potential terms 
of these Fermi multiplets $(\gamma,\widetilde{\gamma})$, 
when both D5- and D5$'$-branes exist, it follows that 
$E\cdot J$ $=$ $E_{\Lambda}\cdot J_{\Lambda}$ 
$=$ $T\widetilde{T}H\widetilde{H}$ $\neq$ $0$. 
The supersymmetric constraint (\ref{02_constraint}) can hold 
if one of the $\mathcal{N}=(0,4)$ Fermi multiplets, which we take $(\gamma,\widetilde{\gamma})$, 
are couple to both fundamental hypermultiplets $(H,\widetilde{H})$ 
and fundamental twisted hypermultiplets $(T,\widetilde{T})$ via $E$- and $J$-terms 
\begin{align}
\label{je_d5d5'}
\begin{array}{lcc}
&J&E \\
\gamma^{a}_{\tilde{a}}:&
\frac{1}{\sqrt{2}}H^{i}_{a}\cdot \widetilde{T}_{i}^{\tilde{a}}=0&
-\frac{1}{\sqrt{2}}T^{i}_{\tilde{a}}\cdot \widetilde{H}_{i}^{a}=0\\
\widetilde{\gamma}_{a}^{\tilde{a}}:&
\frac{1}{\sqrt{2}}T^{i}_{\tilde{a}}\cdot \widetilde{H}_{i}^{a}=0&
-\frac{1}{\sqrt{2}}H_{a}^{i}\cdot \widetilde{T}_{i}^{\tilde{a}}=0\\
\end{array}
\end{align}

\begin{figure}
\begin{center}
\includegraphics[width=13.5cm]{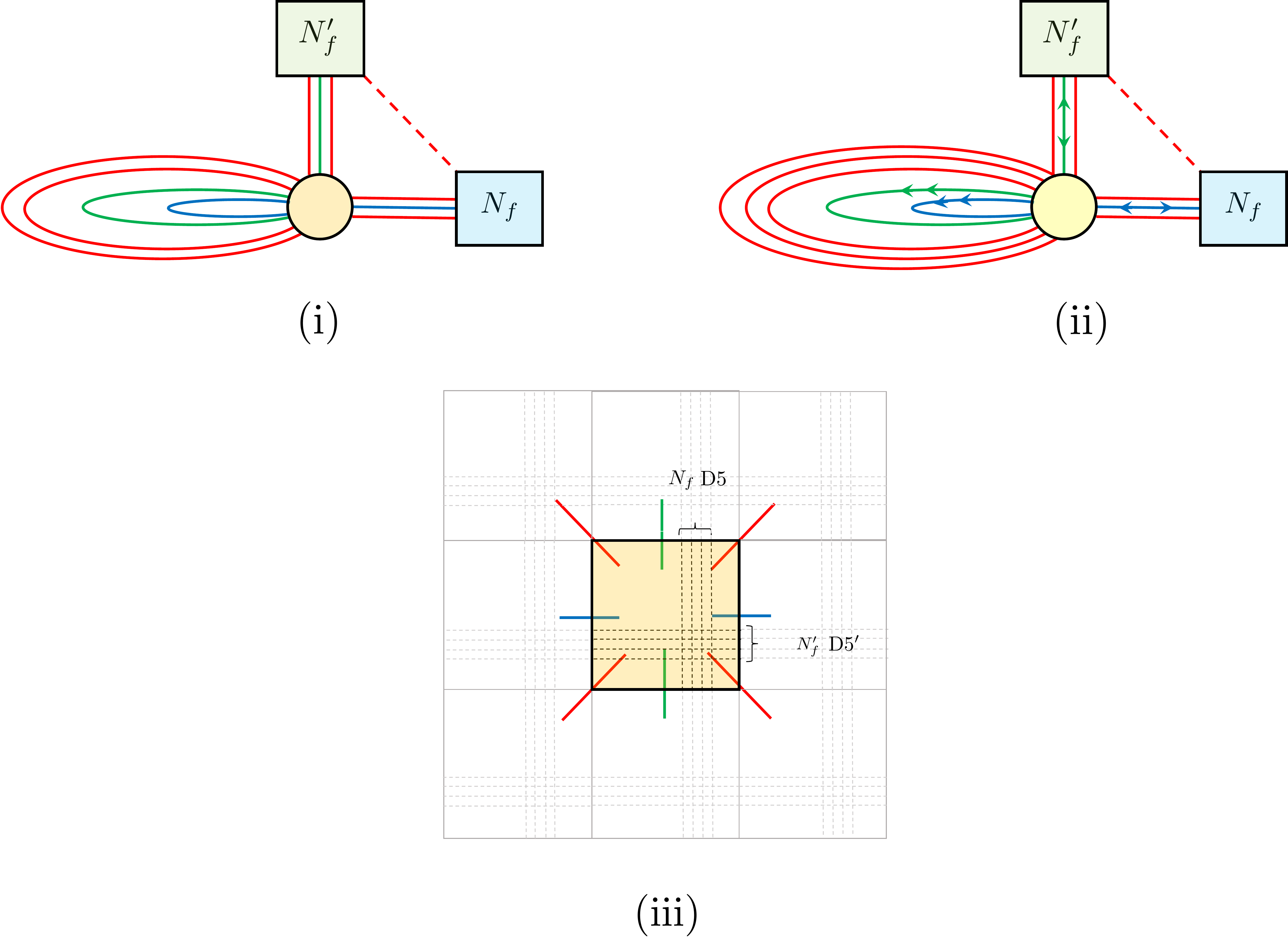}
\caption{(i) $\mathcal{N}=(0,4)$ quiver of D1-D5-D5$'$ brane system. 
The dotted red line represents neutral Fermi multiplets. 
(ii) The corresponding $\mathcal{N}=(0,2)$ quiver. 
(iii) A single brane box of D3-branes with flavor D5- and D5$'$-branes. 
This is T-dual configuration of D1-D5-D5$'$ brane system. 
}
\label{figd1d5d5quiver}
\end{center}
\end{figure}

To sum, the $\mathcal{N}=(0,4)$ gauge theory of D1-D5-D5$'$ brane system is encoded 
in the quiver diagram of Figure \ref{figd1d5d5quiver}.  
The two square nodes represent $SU(N_{f})$ and $SU(N_{f}')$ flavor symmetries.

The T-dual brane configuration is depicted in Figure \ref{figd1d5d5quiver}. 
NS5- and D5-branes are depicted as vertical bold and dotted lines 
while NS5$'$- and D5$'$-branes are depicted as horizontal bold and dotted lines. 
A single box of D3-branes surrounded by a NS5-brane and NS5$'$-brane correspond to the gauge node. 
The D3-branes intersect with $N_{f}$ D5-branes and $N_{f}'$ D5$'$-branes which 
are associated to the $SU(N)_{f}$ and $SU(N)_{f}'$ flavor symmetry.

\subsection{D1-D5-D5$'$ branes on $\mathbb{C}^{2}/\mathbb{Z}_{k}$ $\times \mathbb{C}^{2}/\mathbb{Z}_{k'}$}
\label{sec_d1d5d5zkzk}
\begin{figure}
\begin{center}
\includegraphics[width=6.5cm]{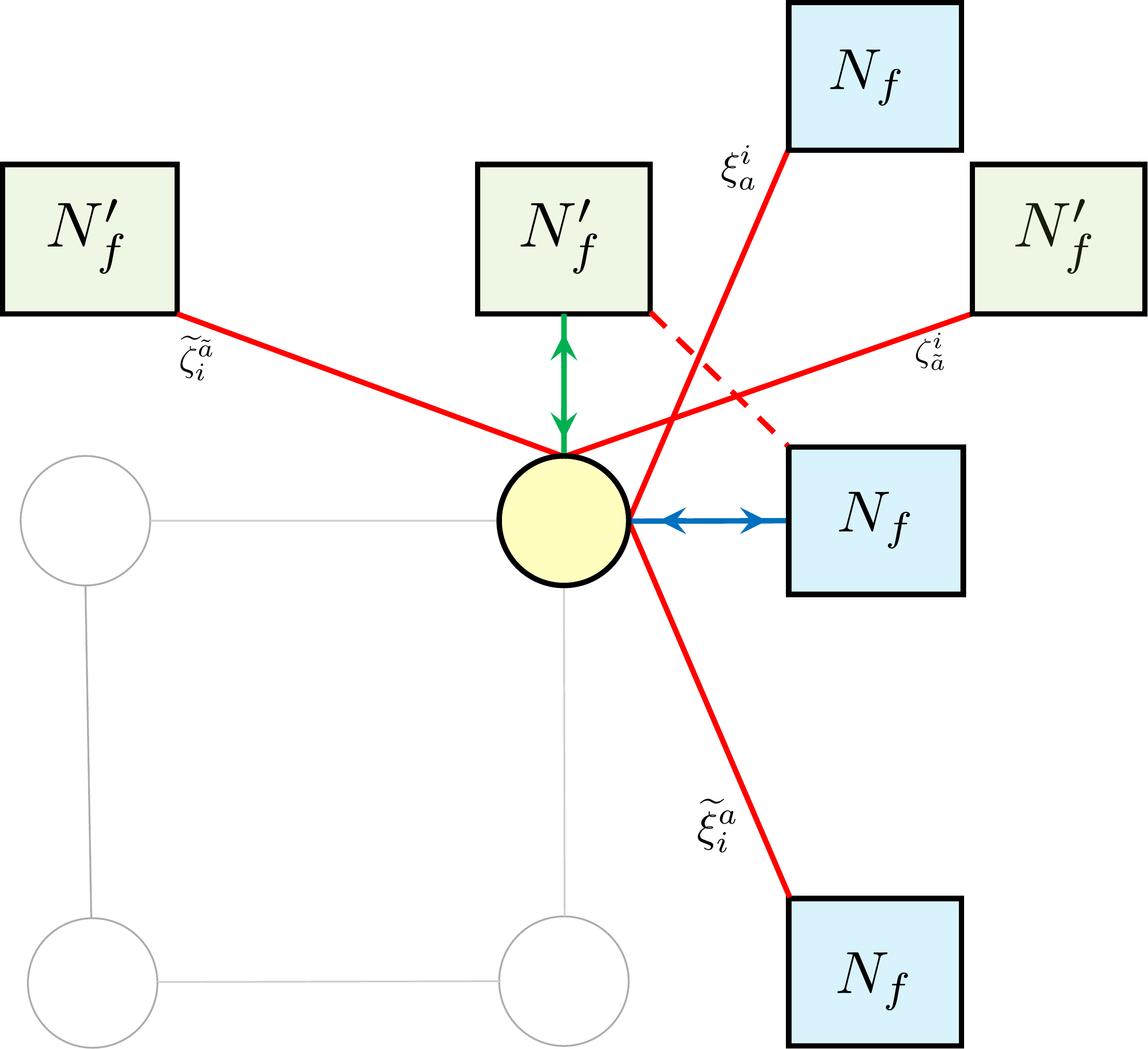}
\caption{A quiver diagram of adding $N_{f}$ flavor D5- and $N_{f}'$ D5$'$-branes 
to D1-branes in the presence of the singularity. 
Each of blue square boxes represents $SU(N_{f})$ flavor symmetry 
while each of green square boxes represents $SU(N_{f}')$ flavor symmetry. 
The blue and green edges are fundamental $\mathcal{N}=(0,4)$ hyper and twisted hypermultiplets. 
The red bold lines are Fermi multiplets. 
The red dotted line is neutral $\mathcal{N}=(0,4)$ Fermi multiplets. }
\label{fig04quiver_1}
\end{center}
\end{figure}

Generic brane box configuration 
consisting of a grid of $k$ NS5-branes and $k'$ NS5$'$-branes 
is T-dual of D1-D5-D5$'$ brane system on the singularity 
$\mathbb{C}^{2}/\mathbb{Z}_{k}$ $\times$ $\mathbb{C}^{2}/\mathbb{Z}_{k'}$. 
For simplicity let us concentrate on the generator of orbifold group labeled by $(1,1,0,0)$ and $(0,0,1,1)$. 
In order for the $SU(2)_{C}\times SU(2)_{H}$ R-symmetry of $\mathcal{N}=(0,4)$ supersymmetry to be preserved, 
the orbifold groups $\mathbb{Z}_{k}$ and $\mathbb{Z}_{k'}$ should be embedded 
in $SU(2)_{H}'$ and $SU(2)_{C}'$ respectively
\begin{align}
\label{emb_orb}
\mathbb{Z}_{k}&\subset SU(2)_{H}'\subset SO(4)_{6'789},\nonumber\\
\mathbb{Z}_{k'}&\subset SU(2)_{C}'\subset SO(4)_{2'345}. 
\end{align}
In addition to the restriction on fields from D1-D1 strings, which we have already seen, 
there is a further restriction on fields from 
D1-D5 strings, D1-D5$'$ strings and D5-D5$'$ strings. 
The action of $\mathbb{Z}_{k'}$ should be embedded in the Chan Paton factors of $N_{f}$ D5-branes 
and that of $\mathbb{Z}_{k}$ should be embedded in the Chan Paton factors of $N_{f}'$ D5$'$-branes 
so that flavor symmetry group is 
\begin{align}
\label{zkzk_gf}
G_{F}&=\prod_{i_{1}=1}^{k}SU(N_{f}')_{i_{1}}\prod_{i_{2}=1}^{k'}SU(N_{f})_{i_{2}}. 
\end{align}
From (\ref{d1d5a_modes})-(\ref{d5d5_modes}) 
we see that only orbifold actions on $(0,4)$ Fermi multiplets from D1-D5 strings and from D1-D5$'$ strings are non-trivial, 
that is $(\xi^{i}_{a}, \widetilde{\xi}^{a}_{i})$ transforming as $\bm{2}$ under the $SU(2)_{C}'$ and 
$(\zeta^{i}_{\tilde{a}},\widetilde{\zeta}^{\tilde{a}}_{i})$ trasforming as $\bm{2}$ under the $SU(2)_{H}'$. 
Let $\tilde{a}_{i_{1}}$ $=$ $1,\cdots, N_{f}'$ and 
$a_{i_{2}}$ $=$ $1,\cdots, N_{f}$ stand for $i_{1}$-th $SU(N_{f}')$ flavor indices 
and $i_{2}$-th $SU(N_{f})$ flavor indices respectively. 
Under the gauge group (\ref{zkzk_g}) and flavor symmetry group (\ref{zkzk_gf}) 
these Fermi multiplets transform as 
\begin{align}
\label{fm_d1d5zk}
\xi_{a}^{i}: \qquad (i;a)&=(i_{1}, i_{2};  a_{i_{2}+1}) \nonumber\\
\widetilde{\xi}^{a}_{i}: \qquad (i;a)&=(i_{1}, i_{2}; a_{i_{2}-1})  \nonumber\\
\zeta_{\tilde{a}}^{i}: \qquad (i;\tilde{a})&=(i_{1}, i_{2}; \widetilde{a}_{i_{1}+1}) \nonumber\\
\widetilde{\zeta}^{\tilde{a}}_{i}: \qquad (i;\tilde{a})&=(i_{1}, i_{2}; \widetilde{a}_{i_{1}-1}). 
\end{align}
We give a quiver diagram for 
adding $N_{f}$ D5- and $N_{f}'$ D5$'$-branes to D1-branes on $\mathbb{Z}_{k}$ $\times$ $\mathbb{Z}_{k'}$ 
in Figure \ref{fig04quiver_1}. 
Each of $k'$ blue square boxes which are vertically aligned represents the $SU(N_{f})$ flavor symmetry 
while each of $k$ green square boxes which are horizontally aligned represents the $SU(N_{f}')$ flavor symmetry. 
The blue and green edges are $\mathcal{N}=(0,4)$ fundamental hypermultiplets $(H^{i}_{a}, \widetilde{H}^{a}_{i})$
and twisted hypermultiplets $(T^{i}_{\tilde{a}}, \widetilde{T}_{i}^{\tilde{a}})$. 
The red lines connecting between gauge nodes and $SU(N_{f})$ boxes 
are the Fermi multiplets $(\xi_{a}^{i}, \widetilde{\xi}_{i}^{a})$ 
and those connecting between gauge nodes and $SU(N_{f}')$ boxes 
are the Fermi multiplets $(\zeta^{a}_{i}, \widetilde{\zeta}^{i}_{\tilde{a}})$. 
The red dotted line is the neutral $\mathcal{N}=(0,4)$ Fermi multiplets 
$(\gamma_{\tilde{a}}^{a}, \widetilde{\gamma}_{a}^{\tilde{a}})$ which arise from D5-D5$'$ strings.

\begin{figure}
\begin{center}
\includegraphics[width=9.5cm]{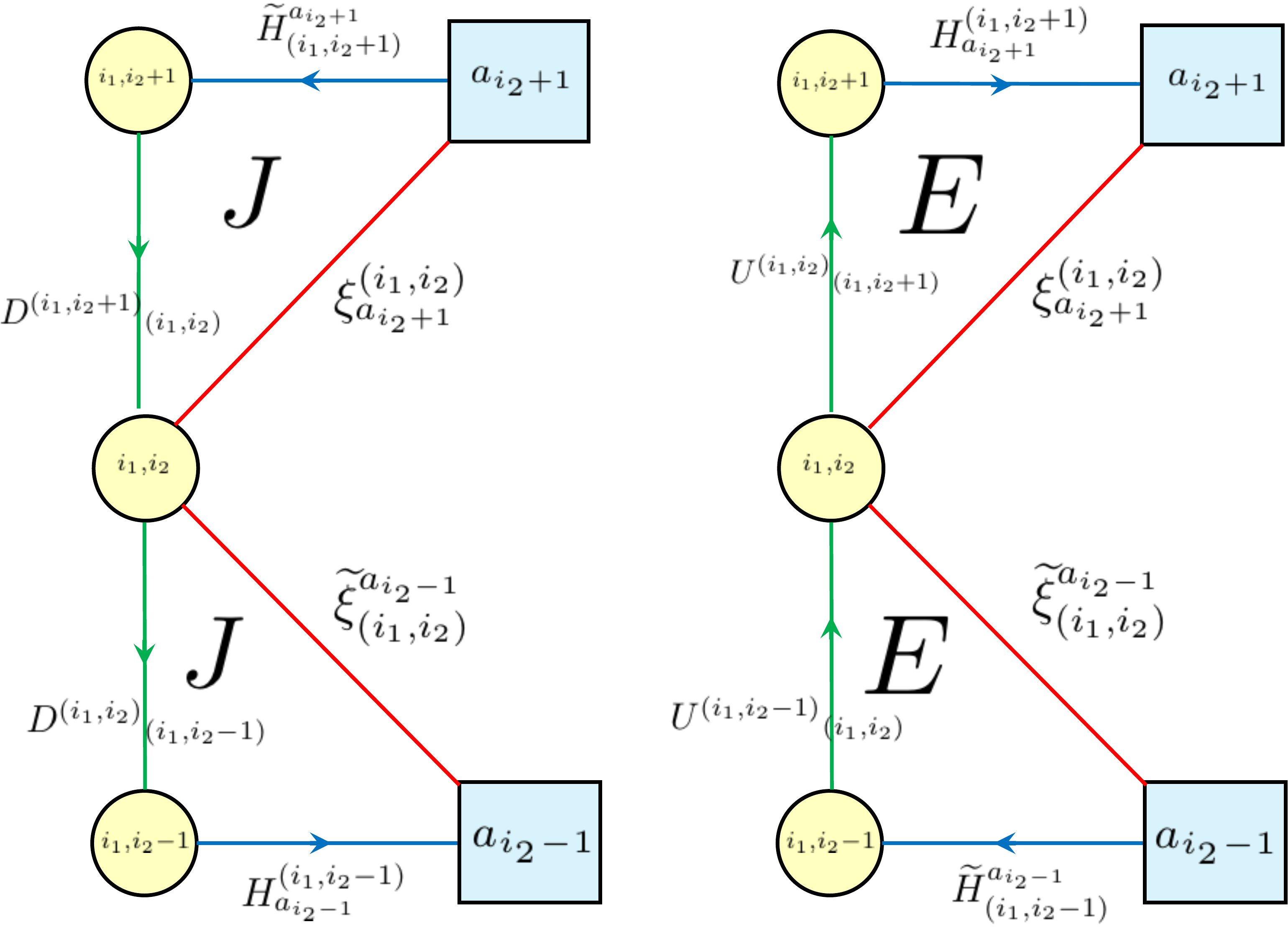}
\caption{$E$- and $J$-terms associated to $(\xi, \widetilde{\xi})$. 
A circular node labeled by a pair of two integers $(i_{1}, i_{2})$ 
corresponds to $(i_{1}, i_{2})$-th gauge factor. 
A blue square box labeled by an integer $a_{i_{2}}$ 
corresponds to $a_{i_{2}}$-th $SU(N_{f})$ flavor factor. }
\label{figje1}
\end{center}
\end{figure}

\begin{figure}
\begin{center}
\includegraphics[width=8.5cm]{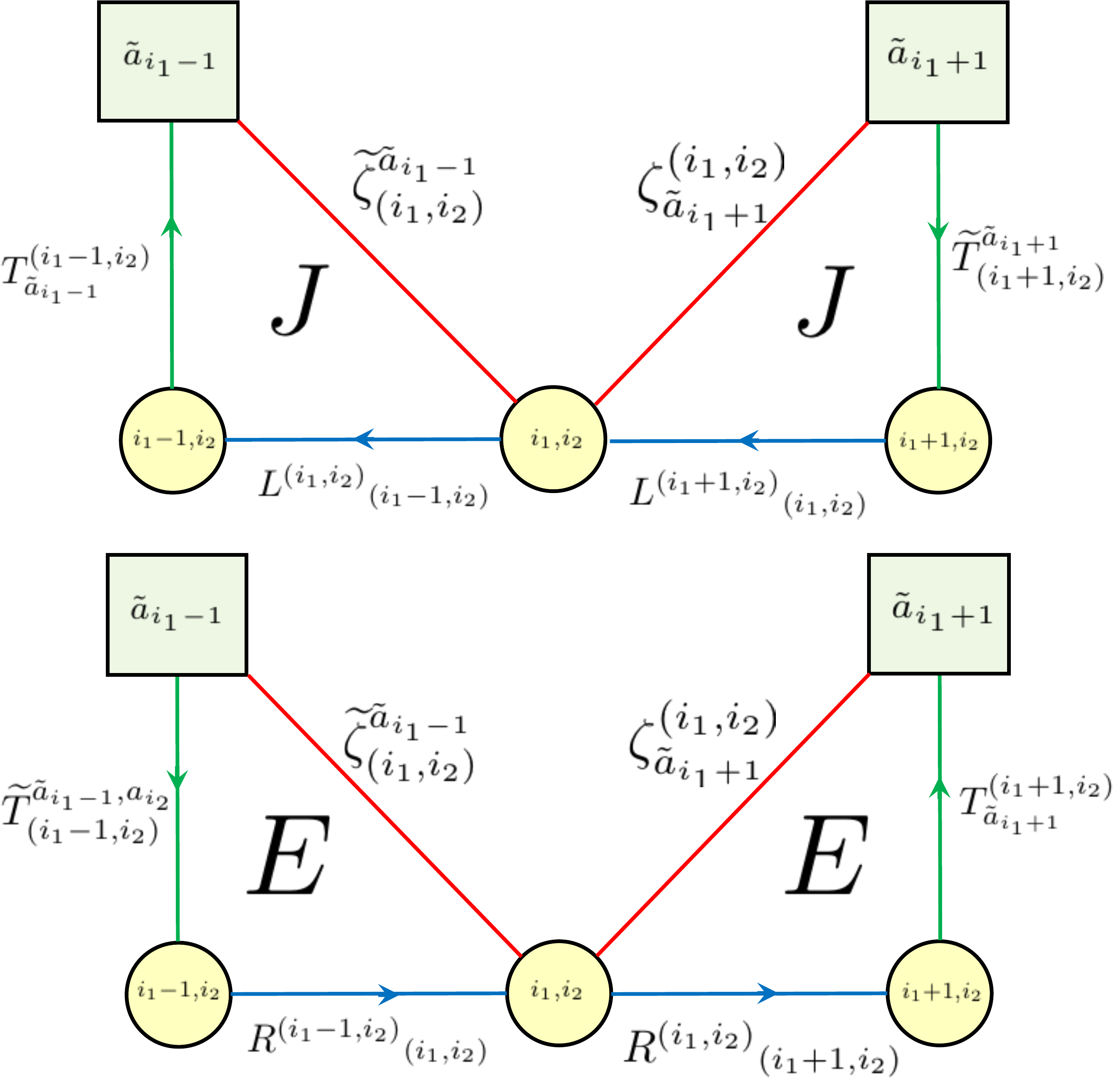}
\caption{$E$- and $J$-terms associated to $(\zeta,\widetilde{\zeta})$. 
A circular node labeled by a pair of two integers $(i_{1}, i_{2})$ 
corresponds to $(i_{1}, i_{2})$-th gauge factor. 
A green square box labeled by an integer $\tilde{a}_{i_{1}}$ 
corresponds to $\tilde{a}_{i_{1}}$-th $SU(N_{f})'$ flavor factor. }
\label{figje2}
\end{center}
\end{figure}

The $E$- and $J$-terms (\ref{je_d1d5}) 
associated to $(\xi^{i}_{a}, \widetilde{\xi}^{a}_{i})$ 
and the $E$- and $J$-terms (\ref{je_d1d5'}) 
associated to $(\zeta^{i}_{\tilde{a}},\widetilde{\zeta}^{\tilde{a}}_{i})$ become
\begin{align}
\label{je_d1d5kk}
\begin{array}{lcc}
&J&E \\
\xi^{(i_{1},i_{2})}_{a_{i_{2}+1}}:&
\widetilde{H}_{(i_{1},i_{2}+1)}^{a_{i_{2}+1}}\cdot {D^{(i_{1},i_{2}+1)}}_{(i_{1},i_{2})}=0&
{U^{(i_{1},i_{2})}}_{(i_{1},i_{2}+1)}\cdot H^{(i_{1},i_{2}+1)}_{a_{i_{2}+1}}=0\\
\widetilde{\xi}_{(i_1, i_2)}^{a_{i_{2}-1}}:&
{D^{(i_1, i_2)}}_{(i_1, i_2-1)}\cdot H^{(i_{1}, i_{2}-1)}_{a_{i_{2}-1}} =0&
-\widetilde{H}_{(i_{1}, i_{2}-1)}^{a_{i_{2}-1}}\cdot {U^{(i_{1}, i_{2}-1)}}_{(i_{1},i_{2})}=0\\
&& \\
\zeta^{(i_{1},i_{2})}_{\tilde{a}_{i_{1}+1}}:&
\widetilde{T}_{(i_{1}+1, i_{2})}^{\tilde{a}_{i_{1}+1}}\cdot {L^{(i_{1}+1, i_{2})}}_{(i_{1},i_{2})} =0&
{R^{(i_{1},i_{2})}}_{(i_{1}+1, i_{2})}\cdot T^{(i_{1}+1, i_{2})}_{\tilde{a}_{i_{1}+1}}=0\\
\widetilde{\zeta}_{(i_{1}, i_{2})}^{\tilde{a}_{i_{1}-1}}:&
{L^{(i_{1}, i_{2})}}_{(i_{1}-1, i_{2})}\cdot T^{(i_{1}-1, i_{2})}_{\tilde{a}_{i_{1}-1}}=0&
-\widetilde{T}_{(i_{1}-1, i_{2})}^{\tilde{a}_{i_{1}-1}, a_{i_{2}}}\cdot {R^{(i_{1}-1,i_{2})}}_{(i_{1}, i_{2})}=0\\
\end{array}. 
\end{align}
These interactions can be read off from the quiver diagram 
as in Figure \ref{figje1} and \ref{figje2}. 
%
%
%
%
%

%
%

\subsection{Examples}
\label{sec_d1d5kk_eg}

\subsubsection{D1-D5-KK$'$ system}
\label{sec_d1d5kk'_eg}
\begin{figure}
\begin{center}
\includegraphics[width=11.5cm]{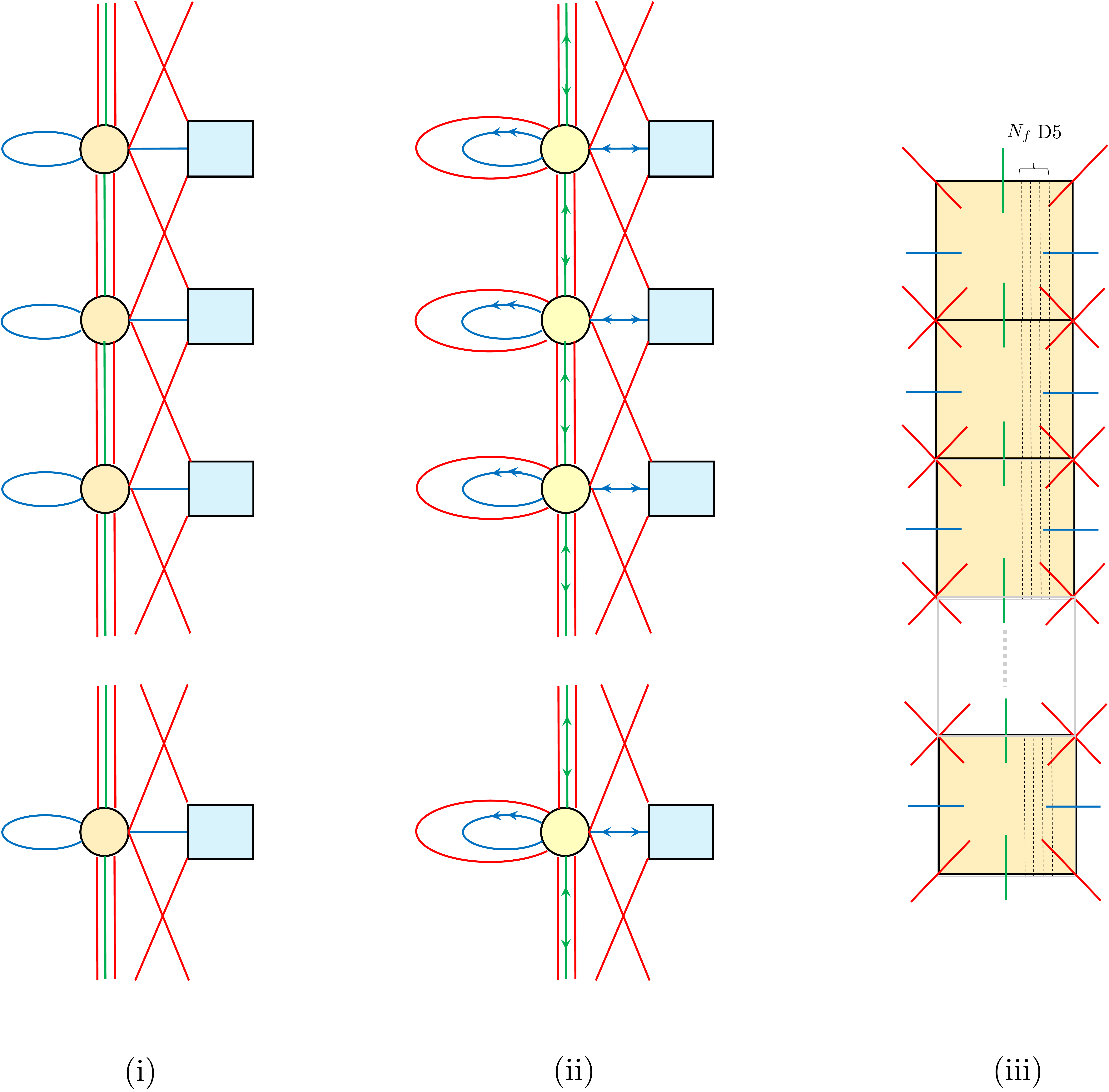}
\caption{
(i) $\mathcal{N}=(0,4)$ quiver diagram of D1-D5 brane system on $\mathbb{C}^{2}/\mathbb{Z}_{k'}$. 
This is a set of inner quiver and outer one. 
(ii) The corresponding $\mathcal{N}=(0,2)$ quiver. 
(iii) The T-dual $(1\times k')$ D3-brane box configuration with $N_{f}$ D5-branes. }
\label{figd1d5kk}
\end{center}
\end{figure}

When $k=1$ and $N_{f}'=0$, our brane setup reduces to D1-D5 brane system on $\mathbb{C}^{2}/\mathbb{Z}_{k'}$ 
\footnote{
This brane configuration has been studied in 
\cite{Sugawara:1999qp, Behrndt:1998nt, Kutasov:1998zh, Berenstein:1998rr, Sugawara:1999qp, Bena:2005ay}.}. 
The low-energy effective theory is $\mathcal{N}=(0,4)$ gauge theory which 
is encoded by a set of inner quiver  and outer quiver diagram shown in Figure \ref{figd1d5kk}. 
The inner quiver diagram is the affine $\widehat{A}_{k'-1}$ diagram 
with $U(N)$ gauge nodes corresponding to the orbifold of $N$ D1-branes. 
The outer quiver diagram is the another affine $\widehat{A}_{k'-1}$ Dynkin diagram 
with $SU(N_{f})$ gauge nodes corresponding to the orbifold of $N_{f}$ D5-branes.

Each gauge node comes up with 
a $\mathcal{N}=(0,4)$ vector multiplet, consisting of a $\mathcal{N}=(0,2)$ gauge multiplet 
and an adjoint $\mathcal{N}=(0,2)$ Fermi multiplet $\Lambda$,  
and an adjoint $\mathcal{N}=(0,4)$ hypermultiplet $(L, R)$. 
Between gauge nodes there is a bi-fundamental $\mathcal{N}=(0,4)$ twisted hypermultiplet $(U, D)$ 
and Fermi multiplets $(\Delta, \nabla)$. 
As discussed in section \ref{subsec_d1zkzk}, 
these are $\mathcal{N}=(4,4)$ gauge theory encoded in the inner quiver. 
In addition, there are flavor nodes and links between the inner and outer quiver. 
They represent fundamental hypermultiplets $(H,\widetilde{H})$ 
and Fermi multiplets $(\xi,\widetilde{\xi})$. 
It is compatible with the quiver previously studied in \cite{Okuyama:2005gq, Haghighat:2013tka, Gadde:2015tra}. 

At low energy the D1-D5-KK$'$ system 
is described by $\mathcal{N}=(0,4)$ SCFT of central charge \cite{Bena:2005ay}
\begin{align}
\label{cc_d1d4kk}
c&=6N_{c} N_{f} k'. 
\end{align}
and the microscopic states correspond to the chiral primary operators in the CFT.

The T-dual configuration is $(1\times k')$ D3-brane boxes including $N_{f}$ D5-branes illustrated in Figure \ref{figd1d5kk}.

\subsubsection{Mirrors}
\label{sec_selfmirror}
We can also consider more general $\mathcal{N}=(0,4)$ quiver gauge theories 
in which both D5- and D5$'$-branes exist at generic singularity. 
In this case we can study two distinct $\mathcal{N}=(0,4)$ gauge theories 
which map to the other under the S-duality in Type IIB string theory. 

\begin{figure}
\begin{center}
\includegraphics[width=15.5cm]{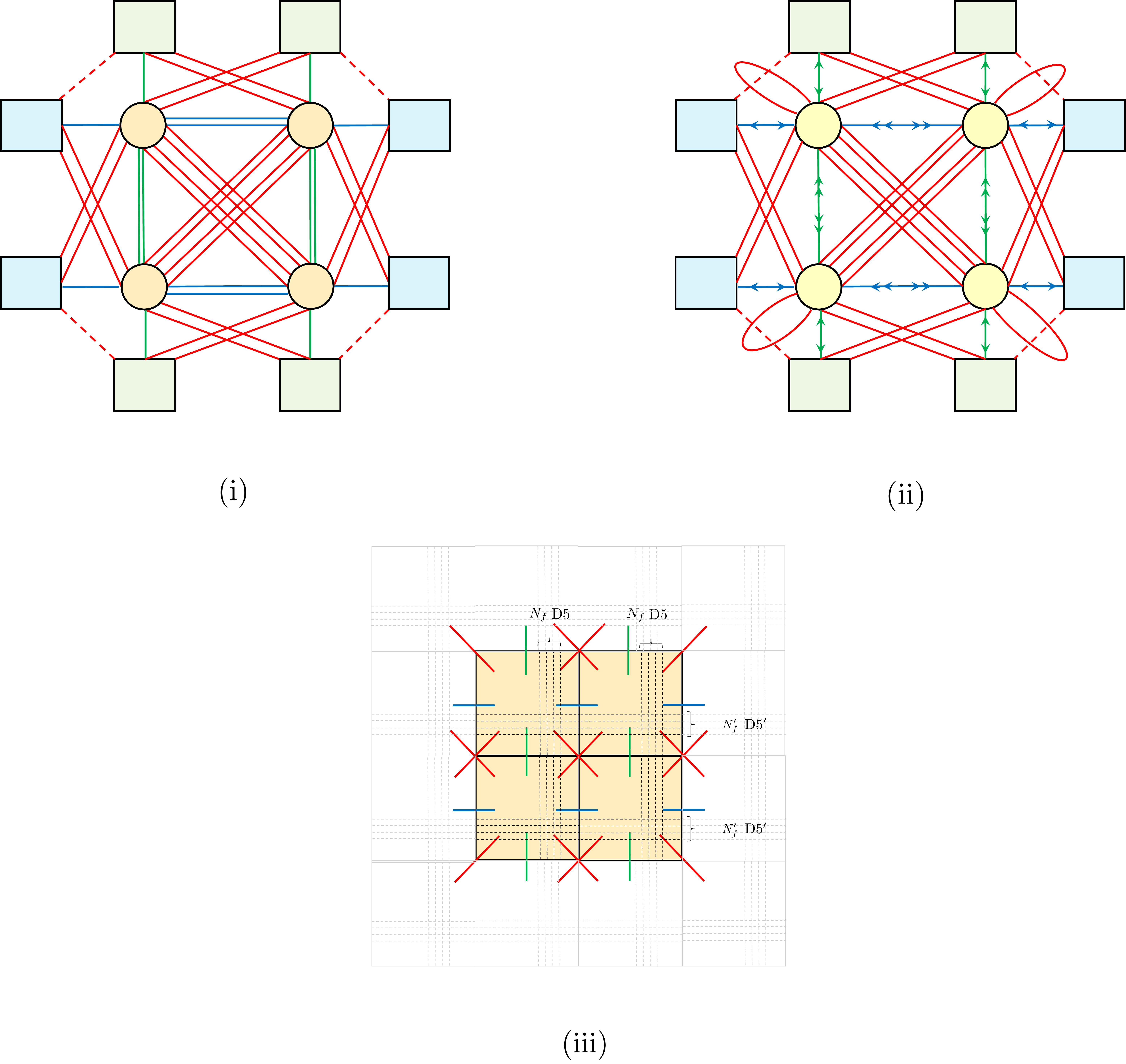}
\caption{(i) 
$\mathcal{N}=(0,4)$ quiver diagram for brane box model of (iii).
(ii) The corresponding $\mathcal{N}=(0,2)$ quiver. 
(iii) A grid of two NS5-branes and two NS5$'$-branes with $N_{f}$ flavor D5 and $N_{f}'$ D5$'$-branes 
which is T-dual of D1-D5-D5$'$ brane system on $\mathbb{C}^{2}/\mathbb{Z}_{2}$ $\times$ $\mathbb{C}^{2}/\mathbb{Z}_{2}$. }
\label{figquiver22d5d5}
\end{center}
\end{figure}

For example, let us consider $2\times 2$ D3-brane box model in 
which each of boxes intersecting with flavor $N_{f}$ D5- and $N_{f}'$ D5$'$-branes. 
Making use of our dictionary, we can easily obtain the quiver diagram illustrated in Figure \ref{figquiver22d5d5}. 
This is the generalized model of $2\times 2$ D3-brane box model in Figure \ref{figz2z2} 
involving $N_{f}$ flavor D5 and $N_{f}'$ D5$'$-branes. 
When $N_{f}=1$ and $N_{f}'=1$, 
this model is interesting in that 
it is invariant under the S duality in type IIB string theory, 
which indicates that the model is self mirror. 
We postpone further analysis of this issue to the future.

\section{Brane box model}
\label{sec_box}
Now that we have identified the matter content and interaction 
of $\mathcal{N}=(0,4)$ gauge theory for periodic D3-brane box configuration from T-dual configuration of D1-D5-KK system, 
we would like to further discuss the detail of the brane box model. 

\subsection{Brane box configuration and anomaly}
\label{sec_braneanomaly}

\subsubsection{Anomaly constraint}
\label{sec_NAanomaly}
As discussed in section \ref{sec_04anomaly}, $\mathcal{N}=(0,4)$ gauge theory must be free from gauge anomaly. 
Consequently only appropriate choices of gauge group and matter content are admitted. 
To see this constraint in the brane box construction, let us consider a box of $N$ D3-branes 
surrounded by eight adjacent boxes filled by D3-branes. 
Taking into account the $\mathcal{N}=(0,4)$ boundary conditions in subsection \ref{subsec_04BC1}, 
this leads to $\mathcal{N}=(0,4)$ $U(N)$ vector multiplet. 
Without any matter multiplets, the $\mathcal{N}=(0,4)$ $U(N)$ vector multiplet is anomalous. 

\begin{figure}
\begin{center}
\includegraphics[width=3.5cm]{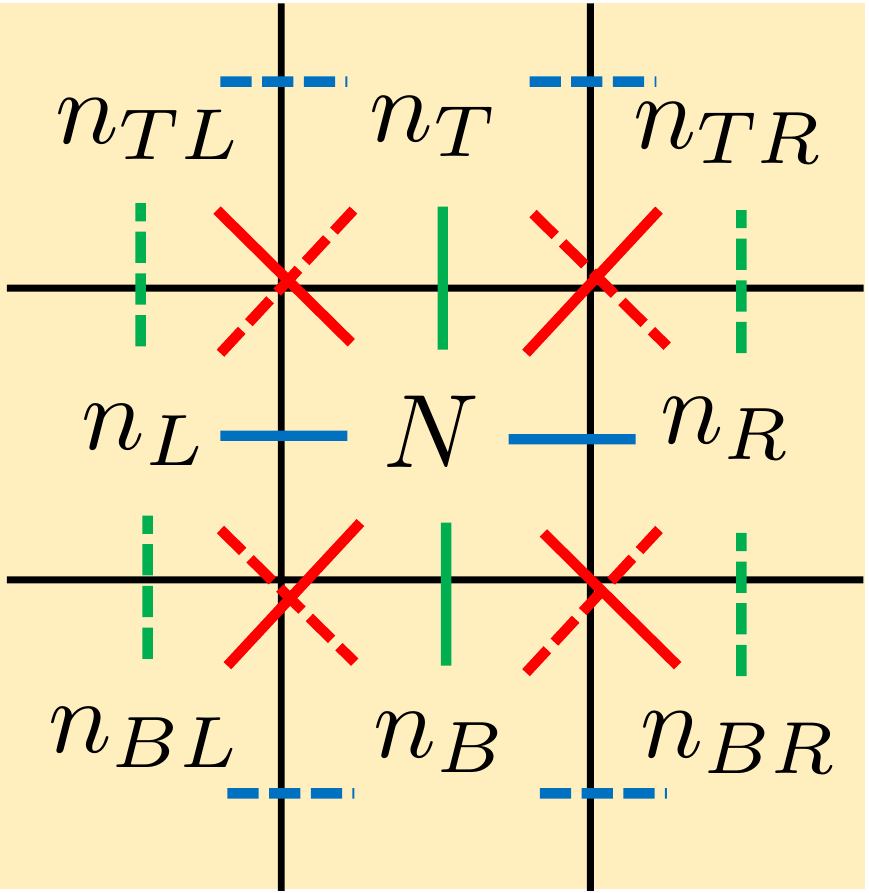}
\caption{A box of $N$ D3-branes surrounded by eight adjacent boxes filled by D3-branes.  
Blue, green and red solid edges correspond to hyper, twisted hyper and Fermi multiplets respectively. 
Blue and green dotted edges represent 3d hyper and twisted hypermultiplets respectively. 
Red dotted edges are neutral Fermi multiplets. }
\label{figsinglebox}
\end{center}
\end{figure}

In the absence of flavor 5-branes, 
an anomaly free theory is obtained by filling the adjacent D3-branes in $(x^{2}, x^{6})$ plane.  
Let $n_{TL}$, $n_{T}$, $n_{TR}$, $n_{L}$, $n_{R}$, 
$n_{BL}$, $n_{B}$, $n_{BR}$ be the numbers of D3-branes displaced in the neighboring infinite regions 
in top-left, top, top-right, 
left, right, bottom-left, bottom and bottom-right (see Figure \ref{figsinglebox}). 
Applying the brane rules for matter multiplets as above,
the horizontally aligned $n_{L}$ and $n_{R}$ D3-branes introduce $n_{L}$ and $n_{R}$ 
$\mathcal{N}=(0,4)$ bi-fundamental hypermultiplets, 
the vertically aligned $n_{T}$ and $n_{B}$ D3-branes provide 
$n_{T}$ and $n_{B}$ $\mathcal{N}=(0,4)$ bi-fundamental twisted hypermultiplets, 
and the diagonally aligned $n_{TL}$, $n_{TR}$, $n_{b}$ and $n_{BR}$ D3-branes 
lead to $n_{TL}$, $n_{TR}$, $n_{b}$ and $n_{BR}$ bi-fundamental Fermi multiplets. 
According to the anomaly contribution (\ref{t_Anom2a}), the
condition set by the ${\bf f}^{2}_{\mathfrak{su}(N)}$ gauge anomaly is given by 
\begin{align}
\label{brane_anomaly}
N&=\frac12 (n_{L}+n_{R}+n_{T}+n_{B})
-\frac{1}{4} (n_{TL}+n_{TR}+n_{BL}+n_{BR}). 
\end{align}
In the absence of D5- or D5$'$-branes, 
this constraint on the occupation numbers of D3-branes in the adjacent boxes 
is required for gauge anomaly cancellation.

\subsubsection{Junction tetravalent Fermi multiplet}
\label{sec_quadrant}
\begin{figure}
\begin{center}
\includegraphics[width=5cm]{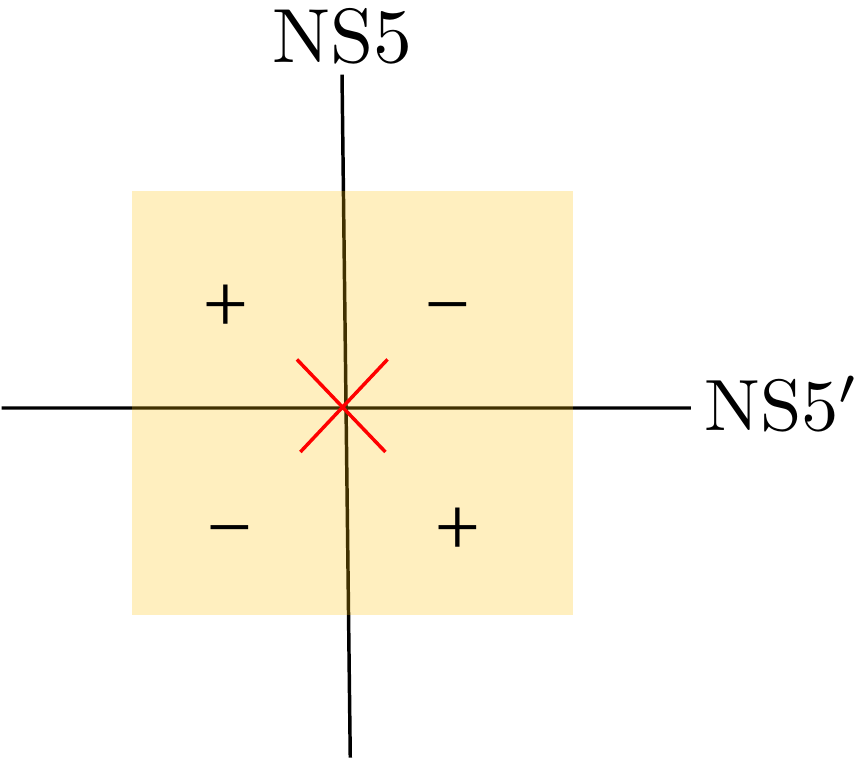}
\vspace*{0.5cm}
\caption{
The tetravalent Fermi multiplet living at NS5-NS5$'$ junction. 
This is required to cancel the Abelian gauge anomaly in such a way that 
it couples to the Abelian parts of quadrant of the gauge or/and global symmetries. 
It has the same charges along diagonal boxes. 
}
\label{figxfm0}
\end{center}
\end{figure}

Although the condition (\ref{brane_anomaly}) fixes the 
${\bf f}_{\mathfrak{su}(N)}^{2}$ gauge anomaly, 
the Abelian part is still anomalous 
because the Abelian gaugino has no contribution to the gauge anomaly. 
To see this precisely, 
let us compute the Abelian gauge anomaly polynomial by setting all the numbers of D3-branes to be one. 
We denote the field strength for the gauge and flavor symmetries which 
are associated to the multiplicities of D3-branes in Figure \ref{figsinglebox} as 
\begin{align}
\label{f_u1}
\begin{array}{c|c|c|c|c|c|c|c|c}
n_{TL}&n_{T}&n_{TR}&n_{L}&N&n_{R}&n_{BL}&n_{B}&n_{BR}\\ \hline 
\bf{a}&\bf{b}&\bf{c}&\bf{d}&\bf{s}&\bf{e}&\bf{f}&\bf{g}&\bf{h} \\
\end{array}
\end{align}
The two-dimensional fields which are illustrated as solid lines in Figure \ref{figsinglebox} have the following contributions to 
the Abelian gauge anomaly polynomial:
\begin{align}
\label{gauge_AN}
\mathcal{I}^{\textrm{2d bdy}}&=
\underbrace{
-2({\bf s}-{\bf e})^{2}
-2({\bf d}-{\bf s})^{2}
}_{\textrm{$(0,4)$ hypermultiplets}}
\underbrace{
-2({\bf s}-{\bf b})^{2}
-2({\bf g}-{\bf s})^{2}
}_{\textrm{$(0,4)$ twisted hypermultiplets}}
+
\underbrace{
({\bf s}-{\bf f})^{2}
+({\bf s}-{\bf h})^{2}
+({\bf a}-{\bf s})^{2}
+({\bf c}-{\bf s})^{2}
}_{\textrm{Fermi multiplets}}
\nonumber\\
&={\bf a}^{2}
-2{\bf b}^{2}
+{\bf c}^{2}
-2{\bf d}^{2}
-2{\bf e}^{2}
+{\bf f}^{2}
-2{\bf g}^{2}
+{\bf h}^{2}
\nonumber\\
&
-2{\bf a}\cdot {\bf s}
+4{\bf b}\cdot {\bf s}
-2{\bf c}\cdot {\bf s}
+4{\bf d}\cdot {\bf s}
+4{\bf e}\cdot {\bf s}
-2{\bf f}\cdot {\bf s}
+4{\bf g}\cdot {\bf s}
-2{\bf h}\cdot {\bf s}
-4{\bf s}^{2}. 
\end{align}
The non-vanishing terms which include the field strength ${\bf s}$ of the Abelian gauge field, 
provide the anomalous contributions. 
In addition, 
the three-dimensional bulk matter fields 
which are illustrated as dotted blue and green lines 
obeying the boundary conditions lead to the following contributions 
to the Abelian global anomaly:
\begin{align}
\label{3d_AN}
\mathcal{I}^{\textrm{3d matter}}=&
\underbrace{
-({\bf a}-{\bf b})^{2}
-({\bf b}-{\bf c})^{2}
-({\bf f}-{\bf g})^{2}
-({\bf g}-{\bf h})^{2}
}_{\textrm{N b.c. of 3d hypermultiplets}}
\underbrace{
-({\bf d}-{\bf a})^{2}
-({\bf e}-{\bf c})^{2}
-({\bf f}-{\bf d})^{2}
-({\bf h}-{\bf e})^{2}
}_{\textrm{N b.c. of 3d twisted hypermultiplets}}
\nonumber\\
=&-2{\bf a}^2
+2{\bf a}\cdot {\bf b}
-2{\bf b}^{2}
+2{\bf b}\cdot {\bf c}
-2{\bf c}^{2}
+2{\bf a}\cdot {\bf d}
-2{\bf d}^{2}
+2{\bf c}\cdot {\bf e}
\nonumber\\
&
-2{\bf e}^{2}
+2{\bf d}\cdot {\bf f}
-2{\bf f}^{2}
+2{\bf f}\cdot {\bf g}
-2{\bf g}^{2}
+2{\bf e}\cdot {\bf h}
+2{\bf g}\cdot {\bf h}
-2{\bf h}^{2}. 
\end{align}
Furthermore, there are four neutral Fermi multiplets 
illustrated as red dotted lines in Figure \ref{figsinglebox} which contribute to the Abelian global anomaly
\begin{align}
\label{nFM_AN}
\mathcal{I}^{\textrm{neutral Fermi}}=&
({\bf b}-{\bf d})^2
+({\bf b}-{\bf e})^2
+({\bf d}-{\bf g})^2
+({\bf e}-{\bf g})^2
\nonumber\\
=&2{\bf b}^2
-2{\bf b}\cdot {\bf d}
+2{\bf d}^2
-2{\bf b}\cdot {\bf e}
+2{\bf e}^2
-2{\bf d}\cdot {\bf g}
-2{\bf e}\cdot {\bf g}
+2{\bf g}^{2}. 
\end{align}
Collecting the contributions 
(\ref{gauge_AN})-(\ref{nFM_AN}), 
we have 
\begin{align}
\label{miss_AN}
&\mathcal{I}^{\textrm{2d bdy}}+
\mathcal{I}^{\textrm{3d matter}}+
\mathcal{I}^{\textrm{neutral Fermi}}
\nonumber\\
&=
-{\bf a}^2
+2{\bf a}\cdot {\bf b}
-2{\bf b}^2
+2{\bf b}\cdot {\bf c}
-{\bf c}^2
+2{\bf a}\cdot {\bf d}
-2{\bf b}\cdot {\bf d}
-2{\bf d}^2
-2{\bf b}\cdot {\bf e}
+2{\bf c}\cdot {\bf e}
\nonumber\\
&-2{\bf e}^{2}
+2{\bf d}\cdot {\bf f}
-{\bf f}^2
-2{\bf d}\cdot {\bf g}
-2{\bf e}\cdot {\bf g}
+2{\bf f}\cdot {\bf g}
-2{\bf g}^2
+2{\bf e}\cdot {\bf h}
+2{\bf g}\cdot {\bf h}
-{\bf h}^2
\nonumber\\
&
-2{\bf a}\cdot {\bf s}
+4{\bf b}\cdot {\bf s}
-2{\bf c}\cdot {\bf s}
+4{\bf d}\cdot {\bf s}
+4{\bf e}\cdot {\bf s}
-2{\bf f}\cdot {\bf s}
+4{\bf g}\cdot {\bf s}
-2{\bf h}\cdot {\bf s}
-4{\bf s}^{2}. 
\end{align}
The first and second lines are the Abelian global anomaly 
while the last line is the (mixed) Abelian gauge anomaly. 
It is quite remarkable that 
the remaining Abelian gauge anomaly 
and Abelian global anomaly
can be completely canceled by 
taking into account the additional tetravalent Fermi multiplets which are charged under the quadrant of D3-branes.  
They can give rise to anomaly contributions 
\begin{align}
\label{xfm_AN}
\mathcal{I}^{\textrm{tetravalent Fermi}}
&=
({\bf a}+{\bf s}-{\bf b}-{\bf d})^{2}
+({\bf b}+{\bf e}-{\bf c}-{\bf s})^{2}
+({\bf d}+{\bf g}-{\bf s}-{\bf f})^{2}
+({\bf s}+{\bf h}-{\bf e}-{\bf g})^{2}
\end{align}
This beautifully cancels the anomaly polynomial (\ref{miss_AN}) !
This anomaly computation strongly indicates that 
there are additional Fermi multiplets living at the NS5-NS5$'$ intersection in the brane box configuration 
\footnote{
We thank D. Gaiotto as we have noticed this resolution in a discussion with him. 
As in our model, such Fermi multiplets living at the NS5-NS5$'$ intersection is also considered 
in \cite{Costello:2018fnz}. 
}. 
As in Figure \ref{figxfm0}, the charges are equal on diagonal boxes and opposite on boxes which share a line.
They will only couple to the Abelian parts of gauge and/or global symmetries for quadrant of D3-branes 
separated by NS5- and NS5$'$-branes. 

For non-Abelian gauge and/or global symmetries, 
we can easily check that 
the Abelian part of the anomaly is cancelled under the balancing condition of equation (\ref{miss_AN})
\footnote{This is compatible with the rule which is discussed in section \ref{sec_2d3d}.}. 
In that case, the tetravalent Fermi multiplets is associated to the determinant representation of 
the gauge and/or global symmetries.

\subsection{Brane box with flavor 5-branes}
\label{sec_d3box_d5}
Let us consider a non-periodic array of D3-brane boxes. 
For non-periodic brane box configuration 
a pair of linking numbers can be introduced. 
It is defined as the number of 5-branes of the opposite kind that are to the left (bottom) of the given 5-branes plus 
net number of D3-branes ending on the 5-brane from the right (top) minus the number ending from the left (bottom). 
It is important to note that 
the linking numbers can be read off for each segment of 5-branes.

\begin{figure}
\begin{center}
\includegraphics[width=13.5cm]{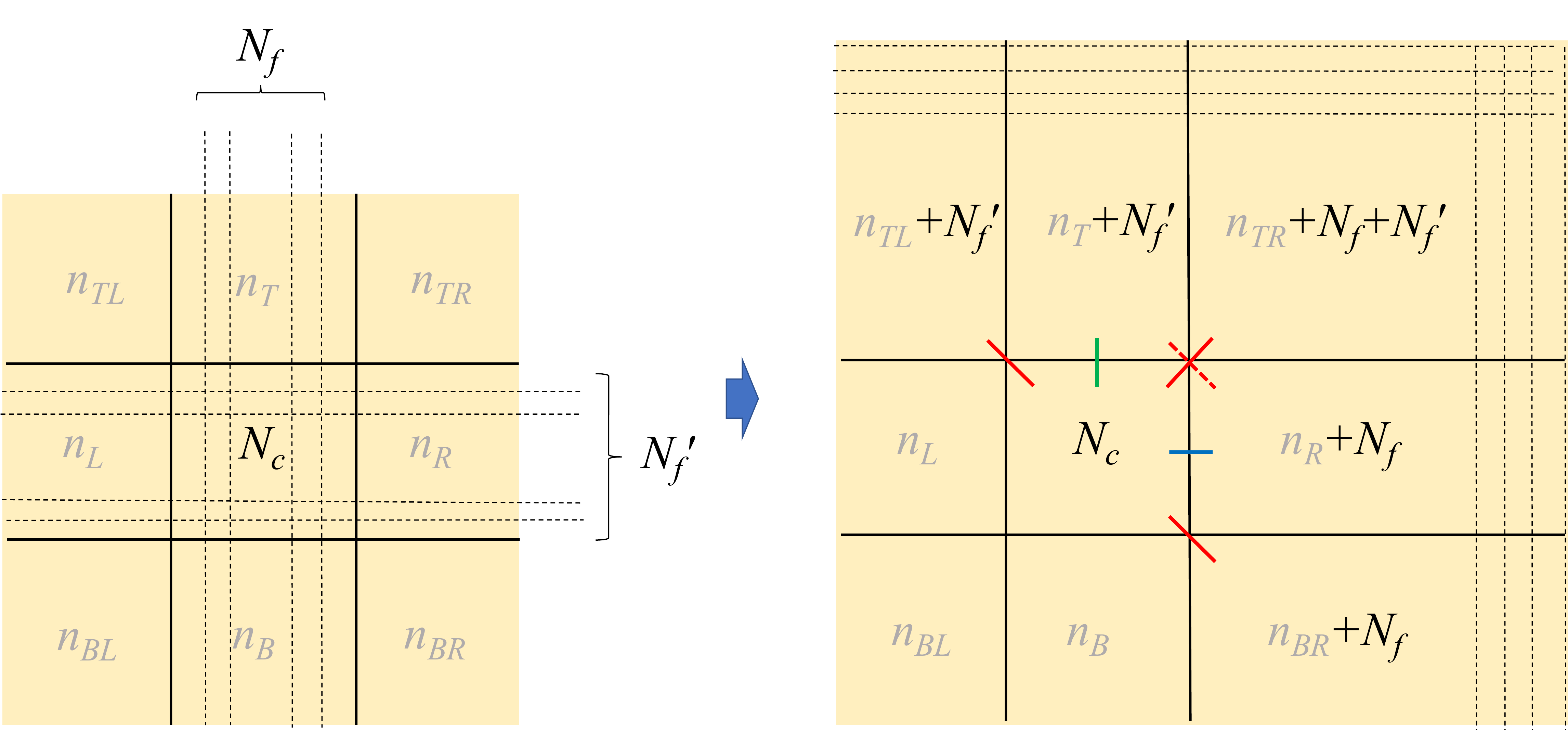}
\vspace*{0.5cm}
\caption{
A creation of D3-branes when $N_{f}$ D5- and $N_{f}'$ D5$'$ branes cross the right NS5-brane and the upper NS5$'$-branes. 
Additional $N_{f}$ D3-branes are created in the adjacent boxes at the top-right, right and bottom right 
and additional $N_{f}'$ D3-branes are created in the adjacent boxes at the top-right, right and bottom right. 
}
\label{figbranemove1}
\end{center}
\end{figure}

To consider the effect of adding D5- and D5$'$-branes in the box configuration, 
we take a box of $N_{c}$ D3-branes surrounded by two NS5-branes and two NS5$'$-branes which
intersect with $N_{f}$ D5-branes and $N_{f}'$ D5$'$-branes (see Figure \ref{figbranemove1}). 
By keeping the linking numbers of 5-branes, 
we rearrange 5-branes in such a way that 
all the D5-branes are located on the right and the D5$'$-branes are located on the top. 
The resulting configuration is illustrated in Figure \ref{figbranemove1}.

Moving $N_{f}$ vertical lines of D5-branes in the box to the right, 
additional $N_{f}$ D3-branes are created in the adjacent boxes at the top-right, right and bottom right. 
Now we can read off matter multiplets from the edges between the $N_{c}$ D3-branes and the $N_f$ D3-branes which are created. 
Three types of new edges connecting to top-right, right and bottom-right boxes show up.
These are $N_{f}$ fundamental Fermi multiplets $\xi$, 
$N_{f}$ fundametnal $\mathcal{N}=(0,4)$ hypermultiplets $(H,\widetilde{H})$ 
and $N_{f}$ fundamental Fermi multiplets $\widetilde{\xi}$, respectively. 

Similarly, moving $N_{f}'$ horizontal D5$'$-branes to the top leads to
$N_{f}'$ D3-branes which are created in the adjacent boxes at the top-left, top, and top-right. 
Correspondingly, three types of edges give rise to
$N_{f}'$ fundamental Fermi multiplets $\zeta$, 
$N_{f}'$ fundametnal $\mathcal{N}=(0,4)$ twisted hypermultiplets $(T, \widetilde{T})$,
and $N_{f}'$ fundamental  Fermi multiplets $\widetilde{\zeta}$, respectively.

When $N_{f}$ D5-branes and  $N_{f}'$ D5$'$-branes intersect in the central box of $N_{c}$ D3 branes, 
neutral $N_{f}N_{f}'$ Fermi multiplets can be read off from dotted diagonal edge between the top box and the right box. 
They are coupled to $N_{f}$ hypers and $N_{f}'$ twisted hypermultiplets.

The above brane rearrangement supports the analysis of the T-dual brane configuration 
in subsection \ref{sec_d1d5d5'}. 
In fact, the corresponding quiver diagram is again given in Figure \ref{fig04quiver_1}. 
Therefore the quiver diagram of Figure \ref{fig04quiver_1} obtained from 
the computation from D1-D5-D5$'$ brane system on $\mathbb{Z}_{k}\times \mathbb{Z}_{k}'$ can be also derived 
by the Hanany-Witten move in the non-periodic box model. 
We present a dictionary between the quiver and the brane box model as follows:
\begin{align}
\label{dic_box}
\begin{array}{c|c}
\textrm{$\mathcal{N}=(0,4)$ quiver}&\textrm{D3-brane box}\\ \hline 
& \\ 
\textrm{$U(N)_{i,j}$ vector} 
\begin{minipage}{1truecm}
\centering
\includegraphics[width=0.25truecm,clip]{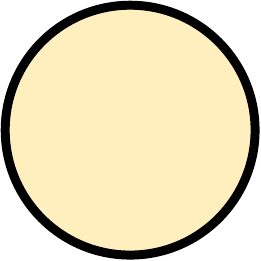}
\end{minipage}
&\textrm{box  $(i,j)$ of $N$ D3-branes} \\ 
\textrm{bi-fundamental hyper}
\begin{minipage}{2truecm}
\centering
\includegraphics[width=1.25truecm,clip]{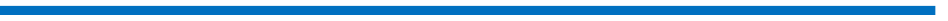}
\end{minipage}
& \textrm{horizontal edge between box $(i,j)$ and $(i\pm1, j)$} \\
\textrm{bi-fundamental twisted hyper}
\begin{minipage}{1truecm}
\centering
\includegraphics[width=0.04truecm,clip]{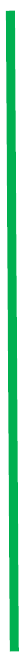}
\end{minipage}
&\textrm{vertical edge between boxes $(i,j)$ and $(i, j\pm 1)$}\\
\textrm{bi-fundamental Fermi} 
\begin{minipage}{2truecm}
\centering
\includegraphics[width=0.65truecm,clip]{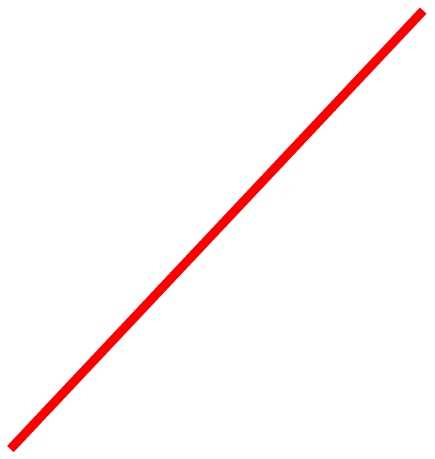}
\end{minipage}
& \textrm{diagonal edge between boxes $(i,j)$ and $(i+1,j-1)$}\\
&\\
\textrm{bi-fundamental Fermi}
\begin{minipage}{2truecm}
\centering
\includegraphics[width=0.65truecm,clip]{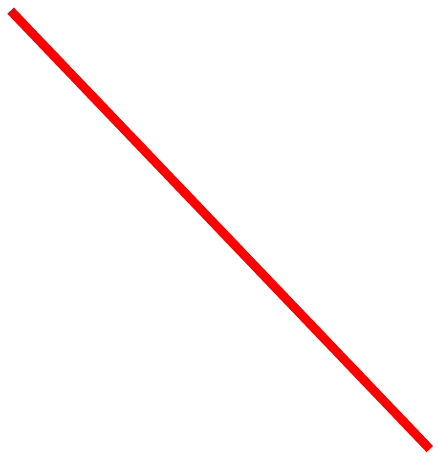}
\end{minipage}
& \textrm{diagonal edge between boxes $(i,j)$ and $(i-1, j-1)$}\\ 
\textrm{tetravalent Fermi}
\begin{minipage}{2truecm}
\centering
\includegraphics[width=0.25truecm,clip]{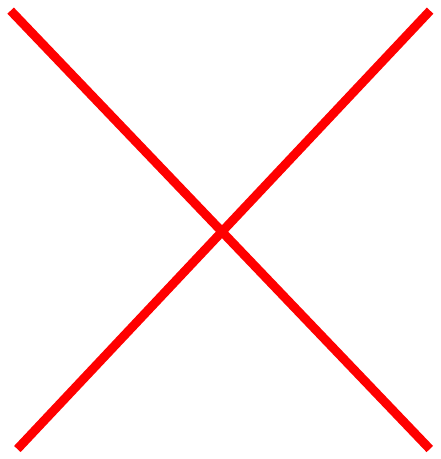}
\end{minipage}
& \textrm{NS5-NS5$'$ junction}\\ 
& \\
\hline
 & \\
\textrm{$N_{f}$ fundamental hyper}
\begin{minipage}{2truecm}
\centering
\includegraphics[width=1.25truecm,clip]{fig04hm.pdf}
\end{minipage}
& \textrm{vertical dotted lines of $N_{f}$ D5-branes}\\
& \\
\textrm{A pair of $N_{f}$ fundamental Fermi}
\begin{minipage}{2truecm}
\centering
\includegraphics[width=0.35truecm,clip]{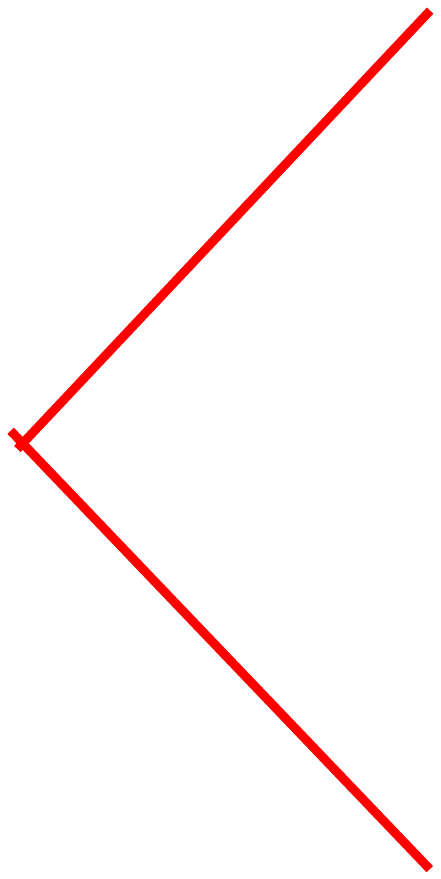}
\end{minipage} \\
& \\ \hline
& \\ 
\textrm{$N_{f}'$ fundamental twisted hyper}
\begin{minipage}{1truecm}
\centering
\includegraphics[width=0.04truecm,clip]{fig04thm.pdf}
\end{minipage}
& \textrm{horizontal dotted lines of $N_{f}'$ D5$'$-branes}\\
& \\
\textrm{A pair of $N_{f}'$ fundamental Fermi}
\begin{minipage}{2truecm}
\centering
\includegraphics[width=0.75truecm,clip]{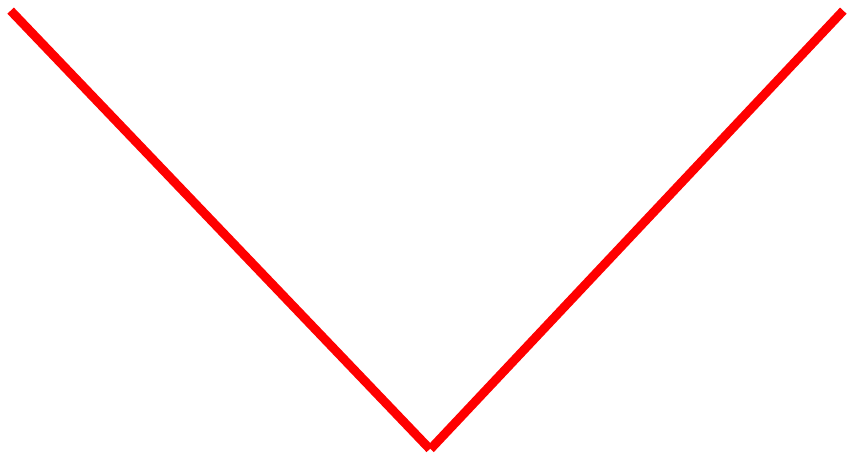}
\end{minipage}
& \\
& \\ \hline 
& \\
\textrm{$N_{f}N_{f}'$ neutral Fermi}
\begin{minipage}{2truecm}
\centering
\includegraphics[width=0.65truecm,clip]{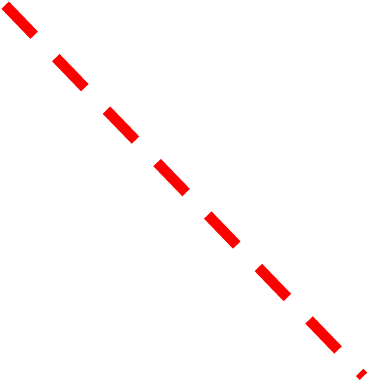}
\end{minipage}
&\textrm{intersecting points of $N_{f}$ D5- and $N_{f}'$ D5$'$-branes} \\
\end{array}
\end{align}

\subsection{2d-3d coupled system}
\label{sec_2d3d}
Let us go back to the issue of $\mathcal{N}=(0,4)$ boundary condition in section \ref{sec_bdyanomaly}  
and consider the 2d-3d coupled system realized in the brane box configuration. 

As we have discussed, the boundary gauge anomaly must be cancelled and it is encoded by the linking numbers. 
Remembering that 
Neumann boundary condition for a gauge field is realized by NS5$'$-brane, 
we would like to require that 
\begin{quote}
\textit{
linking numbers of NS5$'$-branes defining the boundary 
should share the same linking numbers. 
}
\end{quote}
In addition to the above requirement, 
we further assume that 
\begin{quote}
\textit{
linking numbers of NS5$'$-branes defining the boundary 
is not larger than the numbers of D3-branes terminating on the boundary NS5$'$-brane. 
}
\end{quote}
This condition guarantees that 
the numbers of adjacent D3-branes are positive. 

This suggests that the $\mathcal{N}=(0,4)$ boundary conditions in 3d $\mathcal{N}=4$ gauge theory 
can be labeled by the linking numbers of the boundary 5-branes. 
Now that we have a recipe to read off the $\mathcal{N}=(0,4)$ matter content from the brane box configuration, 
we can determine what types of boundary degrees of freedom may appear in 
the boundary conditions for given linking numbers. 
It can be seen that 
the above simple requirement and our identification of matter multiplets 
consistently provides the appropriate boundary degrees of freedom which cancel the gauge anomaly.

\begin{figure}
\begin{center}
\includegraphics[width=11cm]{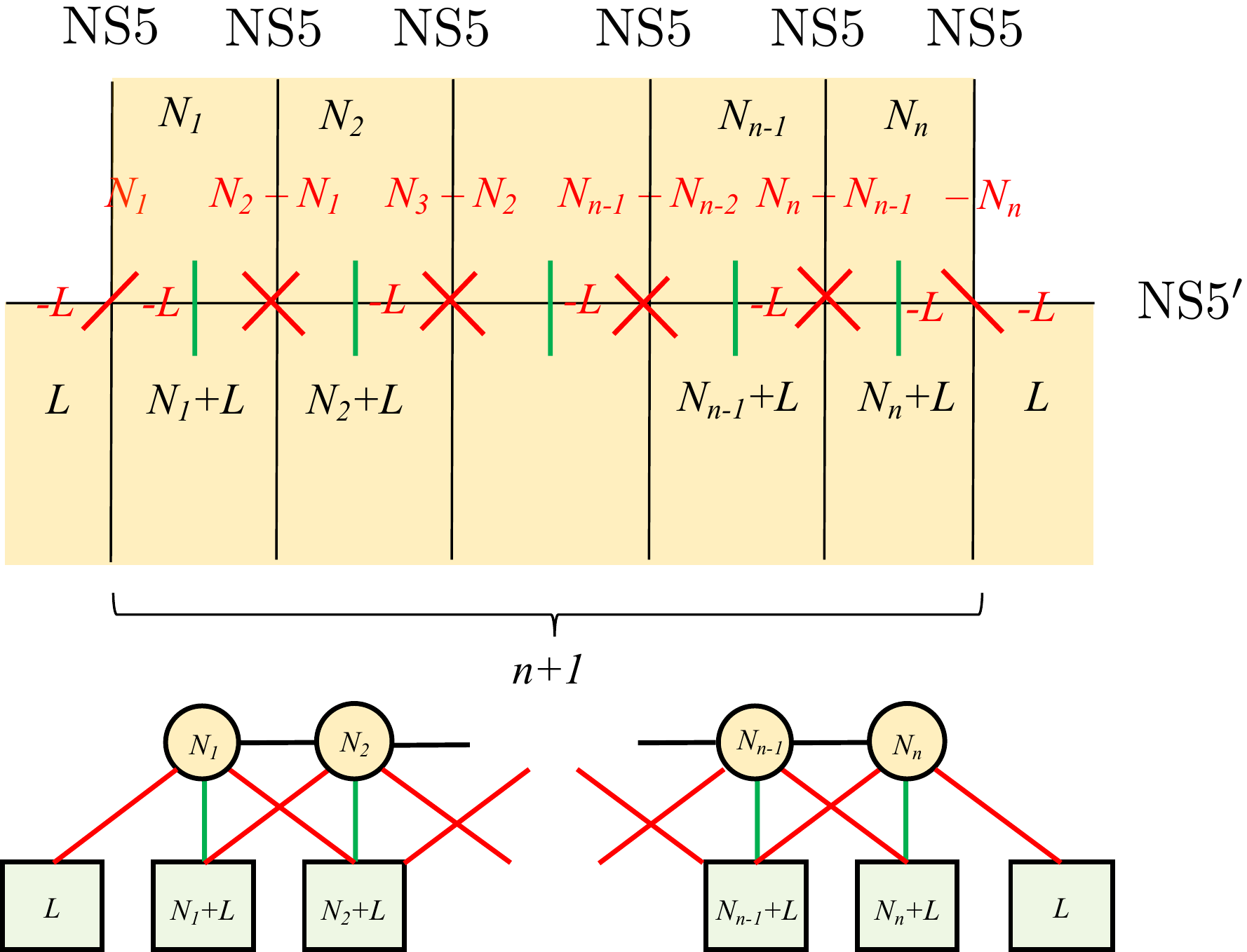}
\caption{Brane construction of interface in  
3d $\mathcal{N}=4$ $\prod_{i=1}^{n} U(N_{i})$ linear quiver gauge theory with bi-fundamental hypermultiplets. 
The numbers with red color indicate the linking numbers. 
The quiver diagram below illustrates the 2d-3d coupled system 
where the green and red edges are boundary twisted hyper and Fermi multiplets respectively. }
\label{fig2d3d}
\end{center}
\end{figure}

Let us start with the brane construction of 
3d $\mathcal{N}=4$ $\prod_{i=1}^{n} U(N_{i})$ linear quiver gauge theory with bi-fundamental hypermultiplets as in Figure \ref{figanomaly}. 
Using the above rules, we can choose a non-positive linking number $-L$ $\le 0$ for the NS5$'$-brane in each segment. 
These linking numbers can be realized when a certain set of D3-branes exist across the boundary NS5$'$-brane 
as shown in Figure \ref{fig2d3d}. 

Using the dictionary (\ref{dic_box}), 
we can read off the matter fields which couple to the 3d bulk gauge fields. 
For gauge nodes $U(N_{i})$ with $i=2,\cdots, n-1$, 
there are $(N_{i}+L)$ $\mathcal{N}=(0,4)$ fundamental twisted hypermultiplets 
and $(N_{i-1}+L)$ $+$ $(N_{i+1}+L)$ $\mathcal{N}=(0,2)$ fundamental Fermi multiplets. 
Gauge nodes $U(N_{1})$ and $U(N_{n})$ are the ends of the quiver and the numbers of Fermi multiplets are smaller.
For $U(N_{1})$ the number of Fermi multiplets is $L$ $+$ $(N_{2}+L)$, 
while for $U(N_{n})$ the number of Fermi multiplets is $(N_{n-1}+L)$ $+$ $L$. 
Recalling (\ref{t_Anom2a}), 
we find ${\bf f}_{\mathfrak{su}(N_{i})}^{2}$ anomaly contributions from the 2d boundary fields
\begin{align}
\label{quiver_bdy_AN}
\mathcal{I}_{\textrm{bdy}, L}
&=(N_{2}-2N_{1})
\Tr ({\bf s}_{1}^{2})
+\sum_{i=2}^{n-1}(-2N_{i}+N_{i-1}+N_{i+1})\Tr ({\bf s}_{i}^{2})
+(N_{n-1}-2N_{n})\Tr ({\bf s}_{n}^{2})
\nonumber\\
&=-\mathcal{I}^{(N_{1})-(N_{2})- \cdots (N_{n})}_{(\mathcal{N}, N)}. 
\end{align}
This shows that 
the set of boundary fields determined by the linking number of the boundary 5-brane and the dictionary (\ref{dic_box}) 
consistently produces a gauge anomaly free boundary condition.

\begin{figure}
\begin{center}
\includegraphics[width=17cm]{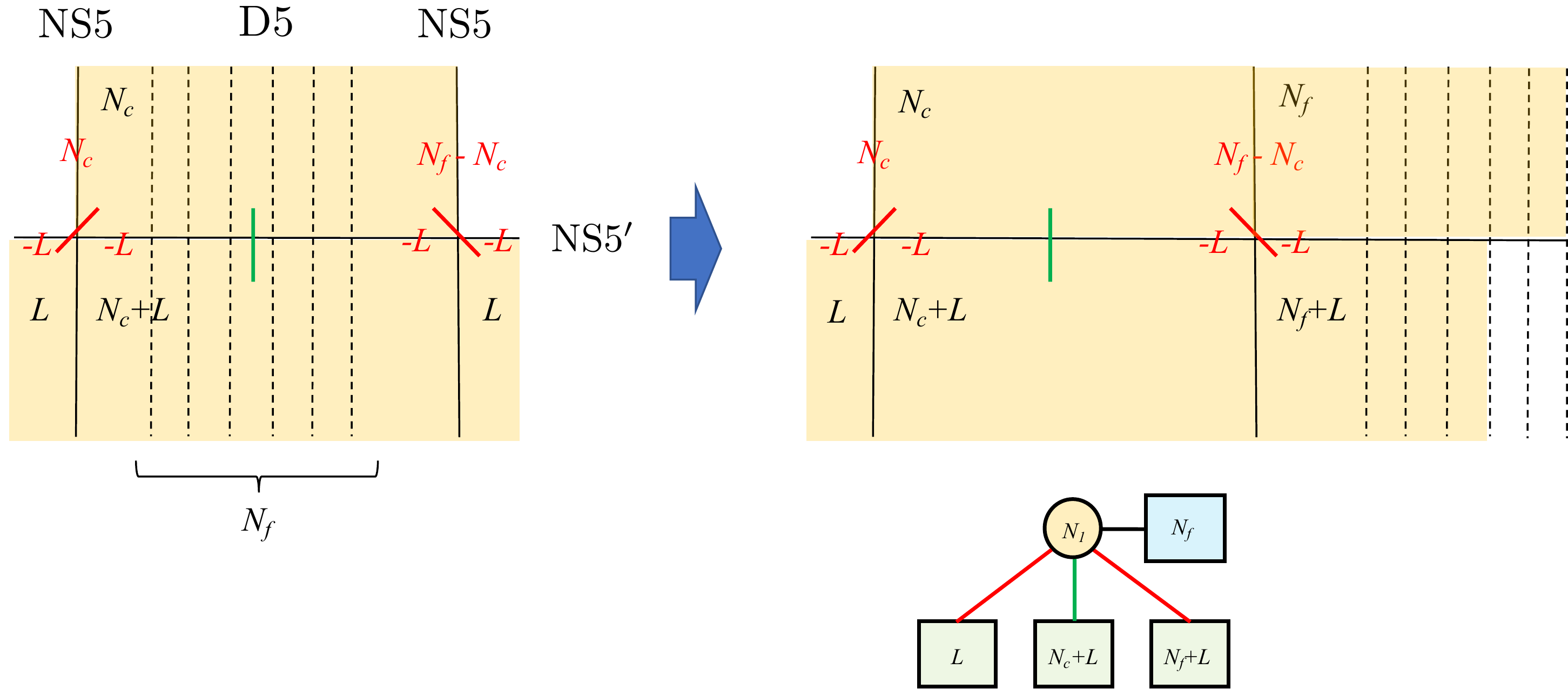}
\caption{Brane construction of interface in  
3d $\mathcal{N}=4$ $U(N_{c})$ gauge theory with $N_{f}$ hypermultiplets. 
After the brane creation, the required Fermi multiplets for gauge anomaly cancellation can be read off from the diagonal edge. }
\label{fig2d3d1}
\end{center}
\end{figure}

For 3d $\mathcal{N}=4$ $U(N_{c})$ gauge theory with $N_{f}$ hypermultiplets, 
we have ${\bf f}_{\mathfrak{su}(N)}^{2}$ gauge anomaly (\ref{ncnf_AN}). 
Making use of the rules above and rearranging the 5-branes, 
the boundary NS5$'$-brane with linking number $-L$ 
introduces $(N_{c}+L)$ twisted hypermultiplets and $L+(N_{f}+L)$ Fermi multiplets 
(see Figure \ref{fig2d3d1}). 
The ${\bf f}_{\mathfrak{su}(N)}^{2}$ anomaly contributions from these 2d boundary fields are given by 
\begin{align}
\label{ncnf_bdy_AN}
\mathcal{I}_{\textrm{bdy},L}
&=
(N_{f}-2N_{c})\Tr ({\bf s}^{2})
\nonumber\\
&=-\mathcal{I}_{(\mathcal{N},N)}^{(N_{c})-[N_{f}]}. 
\end{align}
Again this demonstrates that 
the rules above and the dictionary (\ref{dic_box}) consistently leads to 
gauge anomaly free boundary conditions with appropriate boundary degrees of freedom.

\begin{figure}
\begin{center}
\includegraphics[width=12cm]{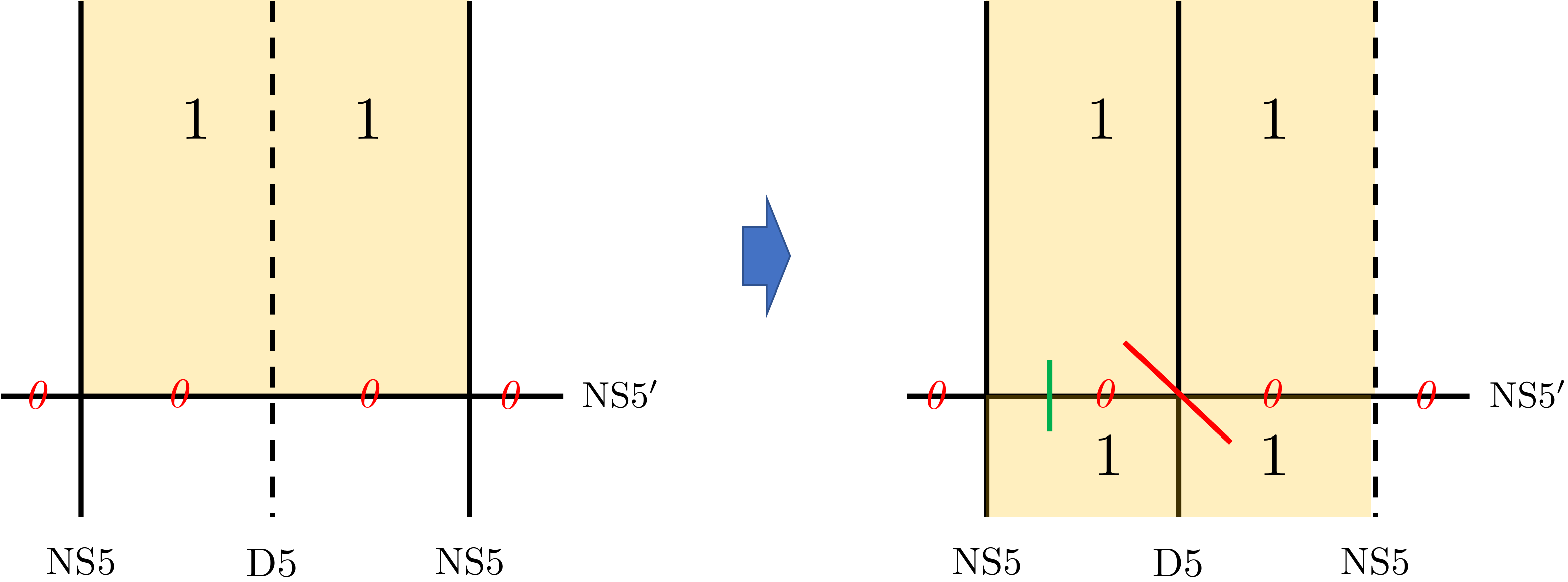}
\caption{The Neumann boundary condition of linking number $L=0$ for SQED with one hypermultiplet. 
The Abelian gauge anomaly is canceled by the tetravalent Fermi multiplets 
which live at the junction of NS5- and NS5$'$-branes. }
\label{fig2d3d2}
\end{center}
\end{figure}

We should note that the 
Abelian gauge anomaly again can be resolved by 
tetravalent Fermi multiplet living at the NS5-NS5$'$ junction. 
Let us consider the simplest example of SQED with one hypermultiplet. 
The bulk hypermultiplet obeying the Neumann boundary condition has boundary $U(1)$ gauge anomaly. 
Taking $L=0$ for the linking number of NS5$'$-brane 
and rearranging 5-branes, one can read off 
one charged twisted hypermultiplet as 
vertical edge, one charged Fermi multiplet as diagonal edge 
and tetravalent Fermi multiplet living at the NS5-NS5$'$ junction 
from Figure \ref{fig2d3d2}. 

We denote the field strengths 
for 3d gauge symmetry and 3d global symmetry 
by ${\bf f}$ and ${\bf a}$. 
Also we let ${\bf b}$ be 
the field strength for additional global symmetry under which 
the charged twisted hyper is charged and 
${\bf c}$ be the field strength for the other under which 
the charged Fermi is charged. 
The anomaly polynomial which is contributed from 
charged fields is evaluated as 
\begin{align}
\label{sqed1_AN}
\mathcal{I}^{\textrm{SQED}_{1}}_{\mathcal{N}, N, L=0}
&=
\underbrace{
-({\bf f}-{\bf a})^{2}
}_{
\begin{smallmatrix}
\textrm{N b.c. of}\\
\textrm{3d hyper}
\end{smallmatrix}
}
\underbrace{
-2({\bf f}-{\bf b})^{2}
}_{\textrm{$(0,4)$ twisted hyper}}
+
\underbrace{
({\bf f}-{\bf c})^{2}
}_{\textrm{Fermi}}
+
\underbrace{
({\bf b}-{\bf f})^{2}
+
({\bf f}+{\bf c}-{\bf a}-{\bf b})^{2}
}_{\textrm{tetravalent Fermi}}
\nonumber\\
&=2{\bf c}^{2}-2{\bf a}\cdot {\bf b}-2{\bf a}\cdot {\bf c}-2{\bf b}\cdot {\bf c}
\end{align}
This shows that the boundary condition $(\mathcal{N},N)$ with linking number $L=0$ for  
SQED with one hypermultiplet is free from gauge anomaly. 
The remaining global anomaly is beautifully 
resolved by further taking into account the anomaly contributions from 
3d hyper and twisted hypermultiplets 
which live across the boundary with the Neumann boundary condition 
and the uncharged boundary Fermi multiplet 
which can be read off from vertical, horizontal and diagonal edges respectively: 
\begin{align}
\label{sqed1_AN2}
\mathcal{I}_{\textrm{bdy},L=0}
&=
\underbrace{
-({\bf b}-{\bf c})^{2}
}_{
\begin{smallmatrix}
\textrm{N b.c. of}\\
\textrm{3d hyper}
\end{smallmatrix}
}
\underbrace{
-({\bf c}-{\bf a})^{2}
}_{\begin{smallmatrix}
\textrm{N b.c. of}\\
\textrm{3d twisted hyper}
\end{smallmatrix}
}
+
\underbrace{
({\bf a}-{\bf b})^{2}
}_{\textrm{neutral Fermi multiplet}}
\nonumber\\
&=-\mathcal{I}^{\textrm{SQED}_{1}}_{\mathcal{N}, N, L=0}. 
\end{align}

\subsection*{Acknowledgements}
We would like to thank Sebastian Franco, Davide Gaiotto, Rak-Kyeong Seong and Junya Yagi for useful discussions and comments. 
A.H. is supported by STFC Consolidated Grant ST/J0003533/1, and EPSRC Programme Grant EP/K034456/1. 
T.O. is supported in part by Perimeter Institute for Theoretical Physics and 
JSPS Overseas Research fellowships. 

\bibliographystyle{utphys}
\bibliography{ref}

\end{document}